\documentclass[iop]{emulateapj}

\usepackage{apjfonts}
\usepackage{natbib}
\usepackage{epsf}
\usepackage{color}
\usepackage{amsmath}
\usepackage{float}
\usepackage[colorlinks=true,citecolor=blue,urlcolor=blue,linkcolor=blue]{hyperref}
\newcommand{\hb}{\hbox{H$\beta$}}
\newcommand{\ha}{\hbox{H$\alpha$}}
\newcommand{\paa}{\hbox{Pa$\alpha$}}
\newcommand{\bra}{\hbox{Br$\alpha$}}
\newcommand{\ntwo}{\hbox{[\ion{N}{2}] $\lambda$6583 \AA}}
\newcommand{\othree}{\hbox{[\ion{O}{3}] $\lambda$5007 \AA}}
\newcommand{\spitzer}{\textit{Spitzer}}

\newcommand{\jwst}{\textit{JWST}}
\newcommand{\lir}{\hbox{L$_\mathrm{IR}$}}
\newcommand{\lpah}{\hbox{L$_\mathrm{PAH}$}}

\newcommand{\lpone}{\hbox{L$_\mathrm{6.2\micron}$}}
\newcommand{\lptwo}{\hbox{L$_\mathrm{7.7\micron}$}}
\newcommand{\lpthree}{\hbox{L$_\mathrm{11.3\micron}$}}

\newcommand{\lha}{\hbox{L$_{\mathrm{H} \alpha}$}}
\newcommand{\lpaa}{\hbox{L$_{\mathrm{Pa} \alpha}$}}
\newcommand{\ldhalpha}{\hbox{L$_{\mathrm{H} \alpha}$ + 0.020$\times$L$_{\mathrm{24~\micron}}$}}
\newcommand{\lfit}{\hbox{L$_\mathrm{fit}$}}
\newcommand{\lsun}{\hbox{L$_{\odot}$}}
\newcommand{\zsun}{\hbox{Z$_\odot$}}
\newcommand{\msun}{\hbox{M$_\odot$}}
\newcommand{\sfrunits}{\hbox{M$_\odot$ yr$^{-1}$}}

\newcommand{\zmetal}{\hbox{12 + log(O/H)$_{N2}$}}
\newcommand{\lsim}{\mathrel{\hbox{\rlap{\lower.55ex \hbox{$\sim$}} \kern-.3em \raise.4ex \hbox{$<$}}}}
\newcommand{\gsim}{\mathrel{\hbox{\rlap{\lower.55ex \hbox{$\sim$}} \kern-.3em \raise.4ex \hbox{$>$}}}}

\begin{document}

\submitted{Accepted for publication in ApJ}

\title{A New Star-Formation Rate Calibration from Polycyclic Aromatic Hydrocarbon Emission Features and Application to High Redshift Galaxies}

\author{\sc Heath V.\ Shipley\altaffilmark{1,2}, 
Casey Papovich\altaffilmark{1},
George H. Rieke\altaffilmark{3},
Michael J.\ I.\ Brown\altaffilmark{4},
John Moustakas\altaffilmark{5}}
\altaffiltext{1}{George P. and Cynthia Woods Mitchell Institute for Fundamental Physics and Astronomy, and Department of Physics \& Astronomy, Texas A \& M University, College Station, TX 77843-4242, USA}
\altaffiltext{2}{current address:  Department of Physics \& Astronomy, Tufts University, 574 Boston Avenue Suites 304, Medford, MA 02155, USA; heath.shipley@tufts.edu}
\altaffiltext{3}{Steward Observatory, University of Arizona, 933 North Cherry Avenue, Tucson, AZ 85719, USA}
\altaffiltext{4}{School of Physics \& Astronomy, Monash University, Clayton, Victoria 3800, Australia}
\altaffiltext{5}{Department of Physics \& Astronomy, Siena College, Loudonville, NY 12211, USA}

\begin{abstract}

\noindent  We calibrate the integrated luminosity from the polycyclic aromatic hydrocarbon (PAH) features at 6.2\micron, 7.7\micron\ and 11.3\micron\ in galaxies as a measure of the star-formation rate (SFR).  These features are strong (containing as much as 5-10\% of the total infrared luminosity) and suffer minimal extinction. Our calibration uses \spitzer\ Infrared Spectrograph (IRS) measurements of 105 galaxies at $0 < z < 0.4$, infrared (IR) luminosities of $10^9 - 10^{12} \lsun$, combined with other well-calibrated SFR indicators.  The PAH luminosity correlates linearly with the SFR as measured by the extinction-corrected \ha\ luminosity over the range of luminosities in our calibration sample.  The scatter is 0.14 dex comparable to that between SFRs derived from the \paa\ and extinction-corrected \ha\ emission lines, implying the PAH features may be as accurate a SFR indicator as hydrogen recombination lines.  The PAH SFR relation depends on gas-phase metallicity, for which we supply an empirical correction for galaxies with $0.2 < \mathrm{Z} \lsim 0.7$~\zsun.  We present a case study in advance of the \textit{James Webb Space Telescope} (\jwst), which will be capable of measuring SFRs from PAHs in distant galaxies at the peak of the SFR density in the universe ($z\sim2$) with SFRs as low as $\sim$~10~\sfrunits.  We use \spitzer/IRS observations of the PAH features and \paa\ emission plus \ha\ measurements in lensed star-forming galaxies at $1 < z < 3$ to demonstrate the ability of the PAHs to derive accurate SFRs.  We also demonstrate that because the PAH features dominate the mid-IR fluxes, broad-band mid-IR photometric measurements from \jwst\ will trace both the SFR and provide a way to exclude galaxies dominated by an AGN.

\end{abstract}


\section{INTRODUCTION}
\label{intro}

Star-formation is a fundamental property of galaxy formation and evolution.  Understanding the exact rate of star-formation and how it evolves with time and galaxy mass have deep implications for how galaxies form. Measuring accurate star-formation rates (SFRs) through cosmic time is therefore paramount for understanding galaxy evolution itself. Much work has been applied to calibrating emission from the UV, nebular emission lines, far-IR, X-ray and radio as tracers of the SFR in distant galaxies \citep[see the review by][]{KE2012}.

The epoch $1 < z < 3$ is particularly interesting as it corresponds to the peak cosmic SFR density in the Universe \citep[][and references therein]{MD2014}.  There are different measures of the instantaneous SFR \citep[see, e.g.,][]{Kennicutt1998}, which are useful for probing the SFR density evolution, but their utility depends on the redshift of the source and observational wavelengths available.  At low redshifts in this range, [\ion{O}{2}]~$\lambda $~3727\AA\ is frequently used to measure SFRs, but it is subject to large extinction corrections \citep{Kennicutt1998}, and hence large uncertainties.  Although the [\ion{O}{2}] line should be excited by star-formation, for galaxies with strong active galactic nuclei (AGN) such as quasars, their spectra generally show no [\ion{O}{2}] emission beyond that expected from the AGN \citep{Ho2005}, i.e., the quasar dominates the excitation and the line cannot be used to probe the host galaxies.  UV continuum emission is a more general probe, but is also subject to extinction and to ambiguities of interpretation in galaxies with AGN. Rest frame optical lines (e.g., \ha) shifted into the near-IR are observationally expensive and also suffer from ambiguities if there is an AGN.

Deep IR surveys with the {\it ISO}, \spitzer\ and {\it Herschel} Space Telescopes have proven to be a powerful tool to explore SFRs across this range, and have revealed that the majority of star-formation at redshifts of $z \sim 1 - 3$ occurs in dust enshrouded galaxies \citep[e.g.,][]{Elbaz2011,Murphy2011}; and that luminous IR galaxies (LIRGs, \lir\ = $10^{11} - 10^{12}$~\lsun) and ultraluminous IR galaxies (ULIRGs, \lir~$> 10^{12}$~\lsun) are much more prevalent during these earlier epochs than today \citep[][and references therein]{CE2001,LeFloc'h2005,PG2005,Magnelli2009,Murphy2011,Elbaz2011,Lutz2014}.  Not only do far-IR measurements reveal dust-obscured star-formation, but even in luminous quasars, this emission appears to be dominated by the power from star-formation in the host galaxies \citep{Rosario2013,Xu2015}.  The \textit{Atacama Large Millimeter Array} (\textit{ALMA}) can measure rest frame far-IR outputs for $z \gsim 2$, but at lower redshifts it becomes inefficient in this application.  Furthermore, no far-IR missions are planned for the immediate future, making it difficult to extend this method past the results from \spitzer\ and {\it Herschel}.  The \textit{James Webb Space Telescope} (\jwst) is a promising approach for $z \lsim 2$ if accurate SFRs can be derived from observations at rest wavelengths of $\sim$ 8\micron\ \citep[e.g.,][]{Rujopakarn2013}.

It has been argued that estimation of SFRs using the aromatic emission features\footnote{Lying in the range 3-19 \micron, attributed to polycyclic aromatic hydrocarbons and termed ``PAH" hereafter. Certain PAH bands are made up of several emission features (e.g., 7.7\micron, 8.6 \micron, 11.3\micron, 12.7\micron, and 17.0\micron\ PAH bands) and so we will use the general term ``feature'' to describe the PAH emission bands.} that dominate the 8 \micron\ emission of star forming galaxies is subject to substantial scatter \citep[e.g.,][]{Smith2007,Bendo2008}.  However, these features do track the total SFR at some level, and if they could be calibrated accurately, they would have numerous advantages. For example, \spitzer\ 24~\micron\ measurements provide perhaps the deepest probe available of SFRs for $1 < z < 3$ \citep{Elbaz2011}.  The luminosity in the mid-IR PAH emission bands is very high for galaxies with ongoing star-formation. The total PAH emission can contribute as much as 20\% of the total IR luminosity and the 7.7 \micron\ PAH band may contribute as much as 50\% of the total PAH emission \citep[e.g.,][]{Smith2007,Wu2010,Shipley2013}.  In addition, for galaxies with an AGN the PAH features have been shown to trace the SFR from the integrated light of the galaxy \citep[if the AGN is not the dominant source of the integrated light; e.g.,][]{Shipley2013}. Furthermore, the 11.3 \micron\ PAH feature even in the immediate vicinity of an AGN is not significantly suppressed \citep{DR2010}.

Interpreting existing deep \spitzer\ and future \jwst\ measurements of embedded star-formation depends upon developing confidence in the use of the PAH features for this purpose.  In this paper, we therefore discuss the utility of using the PAH emission to study the SFR in galaxies over a large range of total IR luminosity and metallicity to determine a robust calibration of the relation between PAH emission and the SFR.  There are previous efforts that calibrate the PAH emission as a SFR indicator.  Some of these works focus on high-luminosity objects and calibrate against the total IR luminosity, where for these objects there may be unknown contributions to the emission from AGN and/or unknown optical depth effects \citep{Pope2008,Lutz2008,FS2009}. Other studies have focused on the broadband emission in the mid-IR which contains contributions both from the continuum and PAH emission \citep{Calzetti2007,Fumagalli2014,Battisti2015}.  There is a need for a PAH SFR indicator calibrated over a range of bolometric luminosity using the luminosity in the PAH emission features themselves. Once this has been accomplished, it is also possible to understand and improve measurements of SFRs based on photometry in the 8 \micron\ region.

This paper presents one of the first efforts to calibrate the PAH emission itself against robust SFR measures of the integrated light for distant galaxies over a large range of total IR luminosity (Section \ref{SFR comparisons}). To do so, we take advantage of PAH emission feature measurements using a sample of galaxies with \spitzer/IRS spectroscopy out to z $< 0.4$.  The outline for the paper is as follows. In \S\ \ref{sample}, we define our main calibration sample and describe a high-redshift demonstration sample. In \S\ \ref{derived quantities}, we describe our analysis of the derived quantities. In \S\ \ref{results}, we present our SFR relations in terms of the PAH luminosity. In \S\ \ref{SFR comparisons}, we discuss previous calibrations of the PAH luminosity as a SFR indicator. In \S\ \ref{preview jwst}, we demonstrate our PAH SFR relations with the demonstration sample of high-redshift lensed galaxies.  In \S\ \ref{conclusions}, we present our conclusions. The $\Lambda$CDM cosmology we assume is H$_{0}=$ 70 km s$^{-1}$Mpc$^{-1}$, $\Omega_{m}=$ 0.3, and $\Omega_{\lambda}=$ 0.7 throughout this work (previous studies we reference throughout this work predominantly use this cosmology and we adopt it as well to be consistent as much as possible). We adopt a Kroupa initial mass function (IMF) throughout this work, where the IMF has the slope $\alpha =$ 2.3 for stellar masses $0.5 -100$~M$_\odot$ and a shallower slope $\alpha =$ 1.3 for the mass range 0.1 $-$ 0.5 M$_\odot$ The Kroupa IMF is consistent with observations of the Galactic field IMF \citep[e.g.,][]{Chabrier2003,Kroupa2003}.

\section{Sample and Data}
\label{sample}

\subsection{Calibration Sample}
\label{primary sample}

For proper calibration of the PAH features as a SFR indicator, we carefully selected galaxies that had full coverage of the \spitzer /IRS spectrum and \spitzer /MIPS 24~\micron\ observations.  We required the galaxies in our sample to have complete optical coverage of the important emission lines needed (specifically \hb, \othree, \ha\ and \ntwo) to measure extinction-corrected SFRs from the \ha\ emission line luminosity \citep{Kennicutt2009}, and to estimate the ionization state \citep{Kewley2001,Kauffmann2003} and the gas-phase metallicity \citep{Pettini2004}.  We also required that the galaxies be sufficiently distant that the \spitzer/IRS contains the majority of the integrated light in the spectroscopic slit.  In effect, this limits us to $z > 0.01$ or where robust aperture corrections are available.  Aperture corrections for each sample were performed in the same general way.  Photometry for the integrated light of the galaxy was obtained and used to estimate the correction factors applied to the optical and IR spectra from the smaller aperture sizes used for the spectral slits (see below for more detail).

We identified three galaxy samples from the literature that fulfill our selection criteria.  We use these three samples as our \textbf{calibration sample} to establish the PAH luminosity as a SFR indicator.  The calibration sample includes a broad range of galaxies allowing for a robust calibration of the PAH luminosity for vastly different systems.  The full calibration sample includes 286 galaxies with data from \citet{ODowd2009}, \citet{Shipley2013} and \citet{Brown2014} samples (specific steps taken for data reduction of the spectra can be found therein).  We discuss each sample briefly in the following sections.  Optical and IR aperture corrections were performed by both \citet{ODowd2009} and \citet{Brown2014} and can be found therein.  For \citet{Shipley2013}, IR aperture corrections were performed and can be found in the reference.  Optical aperture corrections were on average $\sim$10\% for the \citet{Shipley2013} sample for the optical line fluxes shown in Table \ref{Shipley optical fluxes}.  A few sources from \citet{Brown2014} are below our redshift requirement of $z > 0.01$.  We keep these galaxies in our primary and secondary calibration samples because \citet{Brown2014} achieved very accurate aperture corrections to construct their spectral energy distributions (SEDs), but this may introduce some bias for a few galaxies as the aperture corrections may not apply equally between the nuclear regions used for spectroscopy and the galaxy outskirts.

We excluded from our full calibration sample those objects with poor data quality, where measured line/feature fluxes are upper limits in one or more of the required emission features: \ha, the three brightest PAH features (6.2\micron, 7.7\micron\ and 11.3\micron\ features) or the MIPS 24~\micron\ band.  Also, other reasons made it necessary to exclude galaxies from the full calibration sample:  not all galaxies in \citet{Brown2014} have \spitzer\ IRS coverage (resulting from passive galaxies without significant mid-IR dust emission);  the sources in the \spitzer\ First Look Survey (FLS) from \citet{Shipley2013} do not have optical spectroscopy covering \ha.  We discuss these reasons in detail in the following sections for each sample.  This reduced our calibration sample to 227 galaxies.

Finally, we restricted our calibration sample to include only star-forming galaxies with no indications of AGN (which can contribute both to the \ha\ and 24~\micron\ emission) and we required all galaxies to have approximately solar gas-phase metallicity (see \S\ \ref{metals and pahs} below).  These restrictions reduced our calibration sample to 105 galaxies fitting all the requirements.  This constitutes our \textbf{primary calibration sample}.  In what follows, we focus the PAH SFR calibration on this primary calibration sample of 105 galaxies, which span redshift 0.0 $<$ z $<$ 0.4 and IR luminosity of 10$^9 <$ \lir /\lsun\ $< 10^{12}$.  We also consider a secondary calibration sample of 25 galaxies that satisfy all our selection criteria, but have sub-solar metallicity (see \S~\ref{metals and pahs} and \S~\ref{metal corrections}).

\subsubsection{\citet{ODowd2009} Sample}
\label{ODowd09}

The \citet{ODowd2009} sample consists of 92 galaxies that cover a range in total IR luminosity of \lir\ = $10^{9}-10^{11}$~\lsun\ that complements the lower luminosity galaxies in \citet{Brown2014} with a redshift of $0.03 < z < 0.22$ and varying AGN activity (from starbursts to AGN).  Compared to the solar oxygen abundance, 12 + log(O/H) = 8.69 \citep{Asplund2009}, the galaxies in the \citet{ODowd2009} sample span a narrow range around solar (\zmetal\ = 8.6 to 8.8 with one galaxy at 9.0).  This sample has complete coverage of the optical spectra (using the emission line fluxes from the SDSS DR7), IRS spectra and MIPS 24~\micron\ photometry taken from the published data.  63 galaxies from this sample fit our primary calibration sample criteria.  Zero galaxies fit the secondary calibration sample criteria.

\subsubsection{\citet{Shipley2013} Sample}
\label{Shipley13}

The \citet{Shipley2013} sample consists of 65 galaxies covering a range in total IR luminosity of \lir\ = $10^{10}-10^{12}$~\lsun\ with mostly higher IR luminosity compared to the \citet{Brown2014} and \citet{ODowd2009} samples.  The sample has a redshift of $0.02 < z < 0.6$ with varying AGN activity (from domination by starbursts to AGN).  Compared to the solar oxygen abundance the galaxies in the \citet{Shipley2013} sample span a range around solar (\zmetal\ = 8.4 to 8.8) but have metallicities that bridge the other two samples.  This sample has complete IR coverage of the IRS and MIPS 24~\micron\ photometry and complete optical spectroscopic data for 26 of the galaxies in the sample.  12 galaxies from this sample fit our primary calibration sample criteria.  Two galaxies fit the secondary calibration sample criteria.

\subsubsection{\citet{Brown2014} Sample}
\label{Brown14}

The \citet{Brown2014} study included 129 galaxies, covering a large range in total IR luminosity (\lir\ = $10^{9}-10^{12}$~\lsun) and a diverse range in morphology (from dwarf irregulars to ellipticals) with a redshift of $0.0 < z < 0.06$ and varying AGN activity (from starbursts to AGN), and is largely drawn from the integrated optical spectroscopic samples of \citet{MK2006} and \citet{Moustakas2010}.  Because of the lower redshift and mostly lower IR luminosity range of the \citet{Brown2014} sample, it contains lower metallicity galaxies (7.8 $<$ \zmetal\ $<$ 8.8) compared to the \citet{ODowd2009} and \citet{Shipley2013} samples.  The \citet{Brown2014} sample also includes clearly merging systems.  This extends the study of the PAH emission as a SFR indicator over a range of activity and metallicity.  This sample has complete optical spectra and MIPS 24~\micron\ photometry coverage and includes IRS spectra for 111 of the galaxies that come from several surveys \citep[see][and references therein; galaxies without IRS spectra are passive]{Brown2014}.  We use these for our calibration sample.  30 galaxies from this sample fit our primary calibration sample criteria.  23 galaxies fit the secondary calibration sample criteria.

Because the redshift distribution of the \citet{Brown2014} sample is lower compared to the other datasets used in the calibration sample, the aperture corrections for light lost outside the slit can be higher.   \citet{Brown2014} provided aperture corrections at 8\micron\ and 12\micron, and caution must be applied when these differ significantly.  For the \citet{Brown2014} galaxies in our primary calibration sample, most (23/30 galaxies) have 8\micron\ and 12\micron\ aperture corrections consistent with the overall sample (the aperture corrections are $\sim$1.37 and $\sim$1.34 for 8\micron\ and 12\micron, respectively).  The remaining seven galaxies have larger differences (the aperture corrections are $\sim$2.52 and $\sim$2.35 for 8\micron\ and 12\micron, respectively), but a visual inspection of the spectra show no obvious discontinuities and none of our results would change if we exclude these galaxies.

\subsection{Demonstration Sample of High Redshift Galaxies}
\label{application sample}

To demonstrate the PAH SFR calibration, we use \spitzer /IRS data for a sample of high-redshift ($1 < z < 3$) galaxies  that have been gravitationally lensed by foreground massive galaxies and/or galaxy clusters.  In many cases the gravitational lensing is substantial, boosting the flux from the background galaxies by factors of $\mu \simeq 3-30$.  Because of the lensing factors the sensitivity is what can be expected for ``typical'' distant galaxies observed with the \jwst\ Mid-IR Instrument (MIRI).  \jwst/MIRI will be able to observe the 6.2\micron\ and 7.7\micron\ PAH features for galaxies with SFRs as low as $\sim$10 \sfrunits\ to $z \lsim$~2 by covering a wavelength range from 4.9\micron\ to 28.8\micron\ with greater sensitivity than any previous IR mission\footnote{see:  http://ircamera.as.arizona.edu/MIRI/performance.htm}.

We use the sample and results reported in \citet{Rujopakarn2012}, which include line flux measurements (or limits) for \ha, \paa\ or \bra\ and the PAH features.  The \paa\ and \bra\ emission lines are ``gold standard" SFR indicators as they trace the ionization from the same star-forming populations as \ha, but \paa\ and \bra\ suffer substantially less extinction owing to their longer wavelengths in the rest-frame near-IR.  Another excellent extinction-free SFR indicator is \ha+24\micron\ \citep{Kennicutt2009}; \citet{Rujopakarn2012} provide this metric based on rest-frame 24 \micron\ fluxes estimated from spectral energy distribution (SED) template fitting.  In addition, we reprocess the \spitzer/IRS data for the 8 o'clock arc as \citet{Rujopakarn2012} did not report measurements for the PAH features in the mid-IR spectroscopy for this galaxy.  We used their values for the redshift and lensing magnification for each galaxy (and references therein), including the 8 o'clock arc to be consistent for the sample\footnote{The approximate magnification of a lensed galaxy introduces a large uncertainty in the estimated SFR as the actual magnification may differ by a factor of a few to that of the measured.  As a result, other studies may use a different magnification; which can cause the estimated SFRs to be over- or underestimated in comparison to ours.}.  We define these seven high-redshift lensed galaxies from \citet{Rujopakarn2012} as our \textbf{demonstration sample}.

We focus our comparison of the estimated PAH SFRs to the estimated \paa\ and \ha+24\micron\ SFRs in our demonstration of the PAH SFR calibration (see \S\ \ref{highz galaxies}).  Depending on the redshift and \spitzer/IRS channels observed, only one of these SFR indicators was available for each galaxy.  In general, \ha+24\micron\ is available for the entire sample and \paa\ is available for $z > 2$ galaxies, where the rest-frame 24 \micron\ fluxes come from longer wavelength MIPS (and in some cases \textit{Herschel}) data, or extrapolations from observed 24 \micron\ data.  The galaxies with measured \ha+24\micron\ fluxes only are A2218b \citep[][at $z=1.034$ and $\mu =6.1$]{Rigby2008} and A2667a \citep[][at $z=1.035$ and $\mu =17$]{Rigby2008}.  The galaxies with measured \ha+24\micron\ and \paa\ fluxes are the Clone \citep[][at $z=2.003$ and $\mu =27 \pm 1$]{Hainline2009,Lin2009}, A2218a \citep[][at $z=2.520$ and $\mu =22 \pm 2$]{Rigby2008,Papovich2009,Finkelstein2011}, A1835a \citep[][at $z=2.566$ and $\mu =3.5 \pm 0.5$]{Smail2005,Rigby2008}, cB58 \citep[][at $z=2.729$ and $\mu \sim30$]{Seitz1998,Teplitz2000}, and the 8 o'clock arc \citep[][at $z=2.731$ and $\mu \sim8$]{Allam2007,Finkelstein2009}.

For the 8 o'clock arc, we reprocessed \spitzer\ data taken from program 40443 (PI:  C.\ Papovich).  We reduced the spectroscopic data for this galaxy using the S17.2.0 version of the processing software from the basic calibrated data (BCD) for the IRS SL2 and LL1 modules by following the \spitzer\ Data Analysis Cookbook, Recipe 17\footnote{see:  http://irsa.ipac.caltech.edu/data/SPITZER/docs/\\
dataanalysistools/cookbook/23/}.  As in \citet{Rujopakarn2012}, we were able to de-blend the flux by performing PSF photometry to measure the flux of the arc.  We reduced the IRAC and MIPS 24~\micron\ data using S16.1.0 version of the processing software from the BCD images by following the data reduction steps outlined in \citet{Papovich2009}.  We used three apertures of 1.8$^{\prime \prime}$ each centered on the three components identified by \citet{Finkelstein2009} using the \spitzer\ APEX source extraction software to extract the IRAC flux from the 5.8\micron\ band.  We estimate that the aperture corrections are 30$-$50\%\footnote{see:  http://irsa.ipac.caltech.edu/data/SPITZER/docs/irac/\\
iracinstrumenthandbook/} with the contamination from a nearby LIRG at the 30$-$50\% level and the corrections roughly cancel each other.

To calibrate the absolute flux density of the IRS spectra for the 8 o'clock arc, we integrated the spectrum from the SL2 module with the IRAC 5.8\micron\ transmission function and the spectrum from the LL1 module with the MIPS 24~\micron\ transmission function. We took the ratio of the observed flux densities as measured directly from the IRAC and MIPS observations to the flux densities synthesized from the IRS spectra as an aperture correction.  We did this to account for contamination from a foreground LIRG or light that is lost outside the spectroscopic slits.  Figure \ref{8oclock spec} shows the extracted, flux calibrated IRS SL2 and LL1 spectra for the 8 o'clock arc.  We use these flux-corrected spectra to measure the \paa\ emission line from the IRS/SL2 spectrum and the PAH features from the IRS/LL1 spectrum.  We extracted line fluxes from the IRS data for the 8 o'clock arc and the galaxies in our calibration sample using the methods described in \S~ \ref{fits}.\\
\\
\\

\section{Derived Quantities}
\label{derived quantities}

\subsection{IRS Spectral Fitting}
\label{fits}
To determine flux estimates for the PAH features in the IRS spectra of our calibration sample, we used the PAHFIT spectral decomposition code \citep{Smith2007}, designed for \spitzer\ IRS data.  PAHFIT uses a $\chi^2$ minimization routine to fit a non-negative combination of multiple emission features and continua to the one-dimensional spectra of our sources.  The features included in PAHFIT are the dust emission features from PAHs (modeled as Drude profiles), thermal dust continuum, continuum from starlight, atomic and molecular emission lines (modeled as Gaussians), and dust extinction.   The PAH emission features at (e.g. 7.7\micron, 11.3\micron, and 17\micron) are blends of multiple components, but PAHFIT treats them as individual complexes.

We take the published PAH fluxes from \citet{ODowd2009} and \citet{Shipley2013}, which used the same PAHFIT input parameters as we use here for the \citet{Brown2014} sample (see references for specific parameters used).  We fit the features in the IRS spectrum for each galaxy in \citet{Brown2014} that had an IRS spectrum (111/129 galaxies).

For the demonstration sample, the published values for the \spitzer /IRS spectra were fit using PAHFIT with the same parameters as our calibration sample.  We use these flux measurements of \paa, \ha+24\micron\ and the PAH features to estimate the SFRs.

\begin{figure}[t!]	
\epsscale{1.1}
\plotone{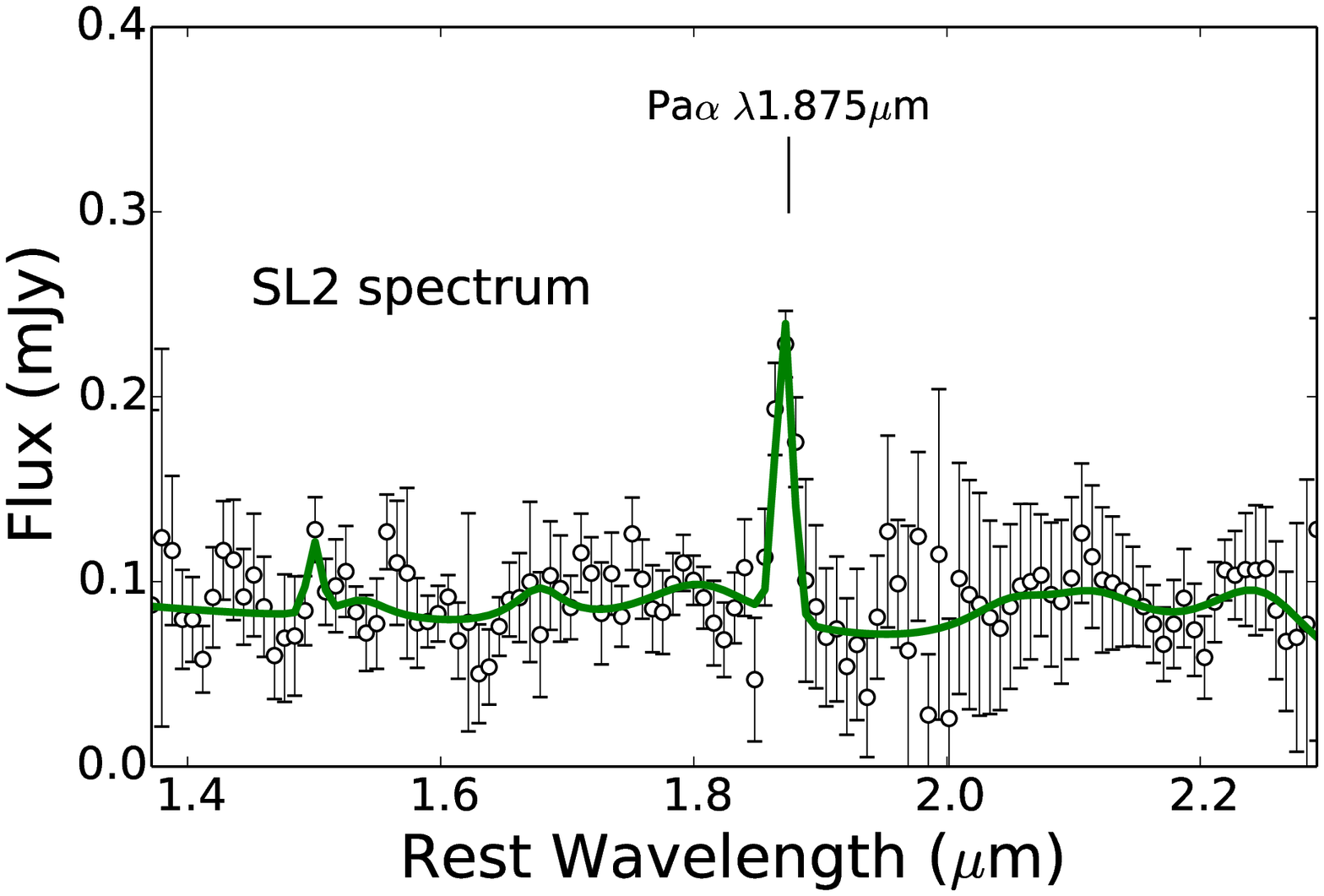}
\plotone{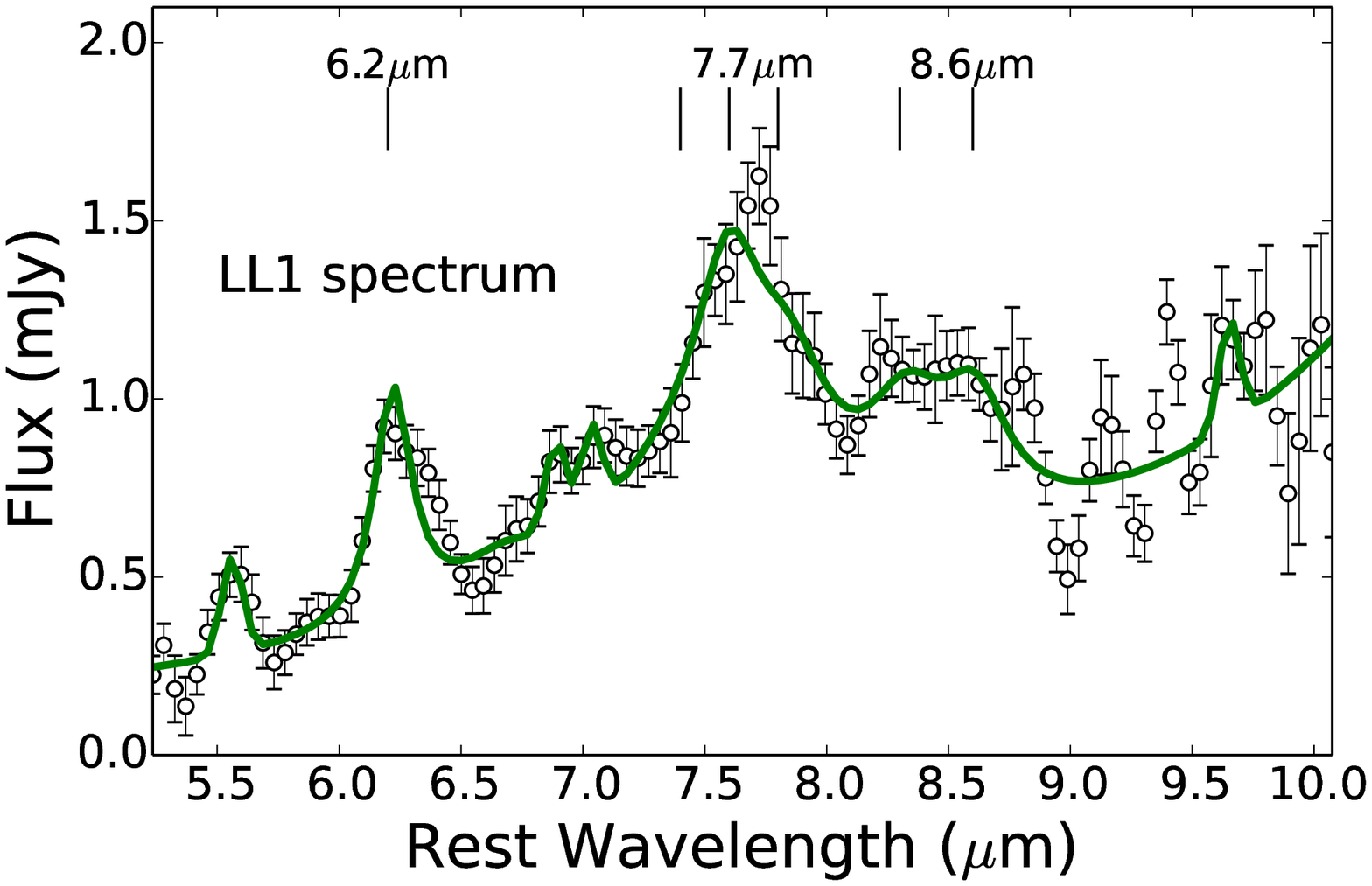}
\caption{IRS spectra of the strongly gravitationally lensed galaxy, the ``8 oÕclock arc".  The \underline{Top} panel shows the SL2 spectrum, and the \underline{Bottom} panel shows the LL1 spectrum.  In each panel, the data points and error bars show the measurements from the data, and the green curve shows the spectral decomposition from the best fit PAHFIT model to the data.}
\label{8oclock spec}
\end{figure}

We fit the 8 o'clock arc spectra with the same procedure.  It was important to have the flux measurement fit the same way for each sample to have an accurate comparison between our PAH SFRs and the SFR indicators (\paa\ and \ha+24\micron) available in \citet{Rujopakarn2012}.  Our measured flux values for the 8 o'clock arc are \lpaa\ = $7.4 \pm 0.8 \times 10^{-16}$ erg s$^{-1}$ cm$^{-2}$ and \lpone\ = $21.1 \pm 1.4 \times 10^{-15}$ erg s$^{-1}$ cm$^{-2}$, \lptwo\ = $80.0 \pm 11.2 \times 10^{-15}$ erg s$^{-1}$ cm$^{-2}$.  But as suggested in \citet{Rujopakarn2012}, the resulting spectrum (for the LL1) has signal-to-noise (S/N) too low for PAHFIT to fit accurately the continuum and emission features due to the contamination of the nearby sources.  We do see this to some extent as PAHFIT failed to converge and yield a more accurate fit to the PAH features in the \spitzer /IRS LL1 spectrum (Figure \ref{8oclock spec}).  Because of the galaxy's high redshift, $z=2.731$, not all PAH features are accessible, but the 6.2\micron\ feature indicates a SFR = 302 $\pm$ 20.0 \sfrunits\ and the 7.7\micron\ feature indicates a SFR =  262 $\pm$ 36.7 \sfrunits. In contrast, the \spitzer/IRS SL2 spectrum gives a consistent SFR from \paa, SFR = 265 $\pm$ 29.6 \sfrunits.

\subsection{Optical Spectral Fitting}
\label{optical fits}

\begin{figure}[t!]	
\epsscale{1.1}
\plotone{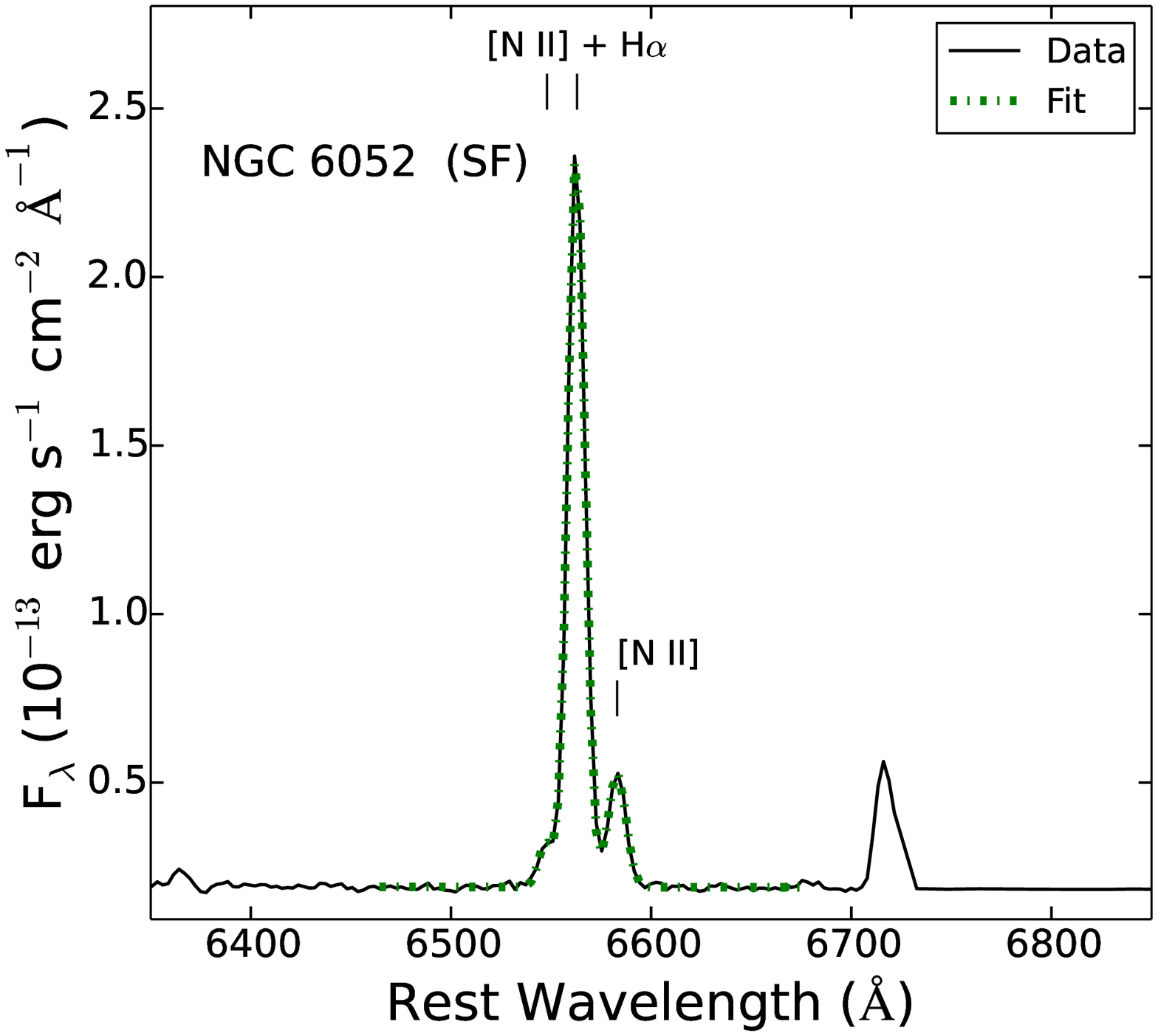}
\plotone{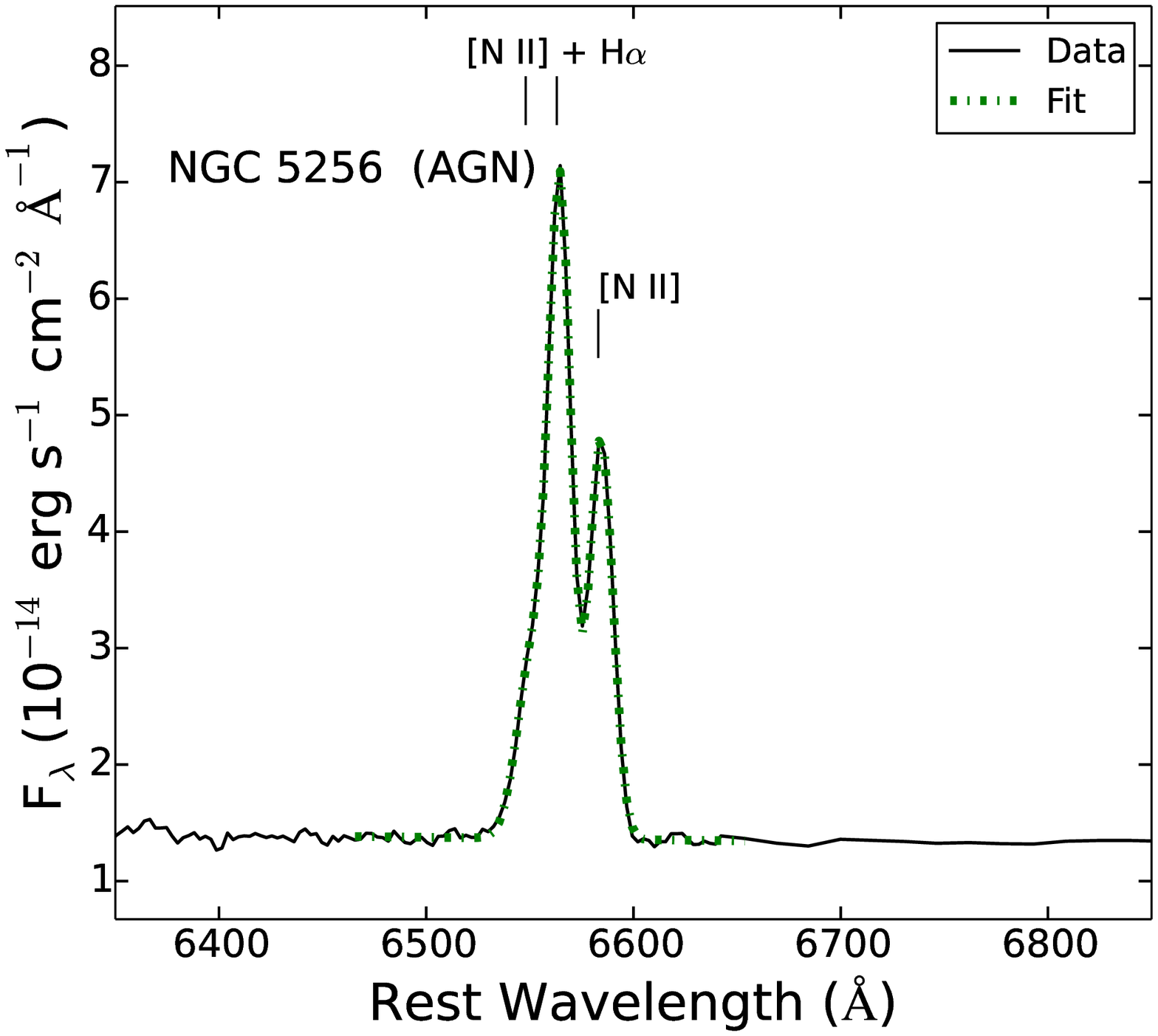}
\caption{Examples of our fits (green dot-dashed line) to the optical spectra (solid black line) for \citet{Brown2014} for the \ha\ and [\ion{N}{2}] emission lines (see \S\ \ref{optical fits} for fitting procedure).  Here, we show NGC 6052 (star-forming) and NGC 5256 (AGN) that is representative of the entire \citet{Brown2014} sample for each classification (see \S\ \ref{optical fits} for classification).}
\label{Brown spec fits}
\end{figure}

For the \citet{ODowd2009} sample, we used the updated line flux measurements (\hb, \othree, \ha\ and \ntwo) from SDSS DR7 \citep[MPA/JHU value-added galaxy catalog,][]{Brinchmann2004,Tremonti2004}.\footnote{see http://www.mpa-garching.mpg.de/SDSS/DR7/}  For \citet{Shipley2013}, we used the line flux measurements from the AGN Galaxy Evolution Survey \citep[AGES;][]{Kochanek2012}, reported in Table \ref{Shipley optical fluxes} and described in \citet{Moustakas2011} based on the continuum and emission-line fitting technique of \citet{Moustakas2010}.  For \citet{Brown2014}, we fit the emission lines as simple Gaussians using a least squares routine to obtain the best fit.  We fit the \ha\ and [\ion{N}{2}] doublet simultaneously using three Gaussians, with the ratio of the central wavelengths fixed.  Figure \ref{Brown spec fits} shows representative spectra and our fits for a star-forming galaxy and AGN in the \citet{Brown2014} sample.

The relationship between the \ntwo/\ha\ and \othree/\hb\ emission line ratios provide a test for ionization from a central AGN using an optical Baldwin, Phillips and Terlevich (BPT) classification \citep{Kewley2001,Kauffmann2003}.  We adopted the SDSS DR7 measured emission line fluxes for the \citet{ODowd2009} sample.  We note that a few of the galaxy classifications are changed from those in \citet{ODowd2009}, that used the SDSS DR4 flux measurements.  For the \citet{Shipley2013} sample, we used the measured fluxes in Table \ref{Shipley optical fluxes} for classification.  For the \citet{Brown2014} sample, we use their reported optical BPT classifications.

\begin{deluxetable*}{cccccc}		
\tablecaption{Optical Fluxes and BPT Classification for \citet{Shipley2013} Sample \vspace{-6pt}
\label{Shipley optical fluxes}}
\tablecolumns{6}
\tabletypesize{\footnotesize}
\tablewidth{0pc}
\tablehead{
\colhead{ID} & \colhead{\hb} & \colhead{\othree} & \colhead{\ha} & \colhead{\ntwo} & \colhead{BPT} \\
 & \colhead{($10^{-16}$)} & \colhead{($10^{-16}$)} & \colhead{($10^{-16}$)} & \colhead{($10^{-16}$)} & \colhead{Class}
}
\startdata
1 & 22.4 $\pm$ 0.33 & 41.8 $\pm$ 0.32 & 89.6 $\pm$ 0.44 & 16.0 $\pm$ 0.27 & SF \\ 
2 & $\ldots$ & $\ldots$ & $\ldots$ & $\ldots$ & $\ldots$ \\ 
3 & 0.99 $\pm$ 0.34 & $\ldots$ & $\ldots$ & $\ldots$ & $\ldots$ \\ 
4 & 9.88 $\pm$ 0.30 & 1.22 $\pm$ 0.20 & 63.2 $\pm$ 0.37 & 21.9 $\pm$ 0.28 & SF \\ 
5 & 60.5 $\pm$ 1.18 & 18.6 $\pm$ 1.00 & 414. $\pm$ 2.29 & 155. $\pm$ 1.79 & SF \\ 
6 & 1.70 $\pm$ 0.30 & 1.49 $\pm$ 0.32 & 11.9 $\pm$ 0.51 & 7.40 $\pm$ 0.47 & SF/AGN \\ 
7 & 51.7 $\pm$ 0.43 & 10.7 $\pm$ 0.33 & 258. $\pm$ 1.56 & 98.9 $\pm$ 0.62 & SF \\ 
8 & 1.83 $\pm$ 0.30 & 1.78 $\pm$ 0.26 & 13.8 $\pm$ 0.39 & 13.8 $\pm$ 0.35 & AGN \\ 
9 & 0.56 $\pm$ 0.12 & 0.49 $\pm$ 0.10 & 4.32 $\pm$ 0.20 & 3.32 $\pm$ 0.19 & SF/AGN \\ 
10 & 2.92 $\pm$ 0.24 & 2.38 $\pm$ 0.22 & 22.4 $\pm$ 0.51 & 12.7 $\pm$ 0.50 & SF/AGN \\ 
11 & 0.36 $\pm$ 0.09 & 0.24 $\pm$ 0.12 & 1.32 $\pm$ 0.24 & $\ldots$ & $\ldots$ \\ 
12 & 6.93 $\pm$ 0.26 & 7.52 $\pm$ 0.28 & 40.9 $\pm$ 0.75 & 17.0 $\pm$ 0.64 & SF/AGN \\ 
13 & $\ldots$ & $\ldots$ & $\ldots$ & $\ldots$ & $\ldots$ \\ 
14 & 0.82 $\pm$ 0.28 & $\ldots$ & 4.33 $\pm$ 0.76 & 2.64 $\pm$ 0.49 & $\ldots$ \\ 
15 & 10.0 $\pm$ 0.26 & 5.75 $\pm$ 0.25 & 65.6 $\pm$ 1.00 & 29.4 $\pm$ 0.77 & SF \\ 
16 & $\ldots$ & 1.36 $\pm$ 0.27 & $\ldots$ & $\ldots$ & $\ldots$ \\ 
17 & 4.52 $\pm$ 0.17 & 3.37 $\pm$ 0.16 & 38.0 $\pm$ 0.50 & 17.5 $\pm$ 0.39 & SF/AGN \\ 
18 & 21.7 $\pm$ 0.35 & 12.4 $\pm$ 0.34 & 120. $\pm$ 0.72 & 42.6 $\pm$ 0.59 & SF \\ 
19 & 2.14 $\pm$ 0.13 & 3.42 $\pm$ 0.09 & 11.0 $\pm$ 0.36 & 2.57 $\pm$ 0.27 & SF \\ 
20 & 2.42 $\pm$ 0.20 & 0.69 $\pm$ 0.18 & 14.4 $\pm$ 0.41 & 7.04 $\pm$ 0.44 & SF \\ 
21 & 1.51 $\pm$ 0.24 & 8.42 $\pm$ 0.25 & 7.22 $\pm$ 0.62 & 5.65 $\pm$ 0.60 & AGN \\ 
22 & 7.71 $\pm$ 0.35 & 13.9 $\pm$ 0.37 & 39.3 $\pm$ 1.65 & 25.4 $\pm$ 1.16 & SF/AGN \\ 
23 & 10.4 $\pm$ 0.75 & 4.74 $\pm$ 0.65 & 75.6 $\pm$ 2.34 & 30.5 $\pm$ 2.04 & SF \\ 
24 & 0.94 $\pm$ 0.26 & 12.3 $\pm$ 0.24 & $\ldots$ & $\ldots$ & $\ldots$ \\ 
25 & 5.66 $\pm$ 0.38 & 4.95 $\pm$ 0.40 & 28.1 $\pm$ 1.10 & 18.1 $\pm$ 1.19 & SF/AGN \\ 
26 & $\ldots$ & 4.77 $\pm$ 0.42 & $\ldots$ & $\ldots$ & $\ldots$ \\ 
27 & 1.46 $\pm$ 0.25 & 4.45 $\pm$ 0.27 & 7.11 $\pm$ 1.15 & 6.72 $\pm$ 0.95 & AGN \\ 
28 & 3.43 $\pm$ 0.13 & 1.66 $\pm$ 0.15 & 22.5 $\pm$ 0.82 & 6.31 $\pm$ 0.71 & SF \\ 
29 & 2.01 $\pm$ 0.28 & 5.83 $\pm$ 0.31 & $\ldots$ & $\ldots$ & $\ldots$ \\ 
30 & 2.42 $\pm$ 0.24 & 2.01 $\pm$ 0.18 & $\ldots$ & $\ldots$ & $\ldots$ \\ 
31 & 4.92 $\pm$ 0.21 & 1.83 $\pm$ 0.16 & $\ldots$ & $\ldots$ & $\ldots$ \\ 
32 & 1.76 $\pm$ 0.10 & 0.58 $\pm$ 0.10 & $\ldots$ & $\ldots$ & $\ldots$ \\ 
33 & 1.21 $\pm$ 0.29 & $\ldots$ & 7.17 $\pm$ 0.64 & 2.78 $\pm$ 0.44 & $\ldots$ \\ 
34 & 1.81 $\pm$ 0.14 & 0.92 $\pm$ 0.11 & $\ldots$ & $\ldots$ & $\ldots$ \\ 
35 & 2.00 $\pm$ 0.47 & $\ldots$ & $\ldots$ & $\ldots$ & $\ldots$ \\ 
36 & 1.01 $\pm$ 0.16 & $\ldots$ & $\ldots$ & $\ldots$ & $\ldots$ \\ 
37 & 0.47 $\pm$ 0.08 & 1.53 $\pm$ 0.09 & 2.97 $\pm$ 0.29 & 1.67 $\pm$ 0.19 & AGN \\ 
38 & $\ldots$ & 0.81 $\pm$ 0.13 & $\ldots$ & $\ldots$ & $\ldots$ \\ 
39 & $\ldots$ & $\ldots$ & $\ldots$ & $\ldots$ & $\ldots$ \\ 
40 & 0.51 $\pm$ 0.14 & $\ldots$ & 6.08 $\pm$ 0.62 & 5.79 $\pm$ 0.54 & $\ldots$ \\ 
41 & $\ldots$ & 0.71 $\pm$ 0.20 & 3.08 $\pm$ 0.33 & 5.53 $\pm$ 0.36 & $\ldots$ \\ 
42 & 1.16 $\pm$ 0.19 & 1.27 $\pm$ 0.21 & 10.3 $\pm$ 0.81 & 9.67 $\pm$ 0.73 & AGN \\ 
43 & $\ldots$ & $\ldots$ & $\ldots$ & $\ldots$ & $\ldots$ \\ 
44 & 1.58 $\pm$ 0.27 & 9.40 $\pm$ 0.34 & $\ldots$ & $\ldots$ & $\ldots$ \\ 
45 & 1.22 $\pm$ 0.22 & 2.93 $\pm$ 0.26 & $\ldots$ & $\ldots$ & $\ldots$ \\ 
46 & 2.75 $\pm$ 0.26 & 3.15 $\pm$ 0.20 & 12.1 $\pm$ 0.32 & 2.55 $\pm$ 0.21 & SF \\ 
47 & 7.01 $\pm$ 0.29 & 3.68 $\pm$ 0.22 & 40.5 $\pm$ 0.51 & 14.9 $\pm$ 0.55 & SF \\ 
48 & 5.45 $\pm$ 0.17 & 1.62 $\pm$ 0.19 & 59.2 $\pm$ 2.79 & 26.0 $\pm$ 2.99 & SF \\ 
49 & 4.16 $\pm$ 0.71 & 18.0 $\pm$ 1.29 & 20.9 $\pm$ 4.05 & 13.4 $\pm$ 4.39 & AGN \\ 
50 & $\ldots$ & $\ldots$ & $\ldots$ & $\ldots$ & $\ldots$ \\ 
\enddata
\vspace{-6pt}
\tablecomments{Column 1 is the galaxy identification defined in \citet{Shipley2013}.  Columns 2$-$5 are the measured fluxes for the emission lines \hb, \othree, \ha\ and \ntwo, respectively (in units of erg s$^{-1}$ cm$^{-2}$).  Column 6 is the BPT classification from the measured fluxes.}
\end{deluxetable*}		

\subsection{Extinction-corrected \ha\ Emission Using Rest-Frame MIPS 24~\micron\ Flux}
\label{ha contribution}

We estimate extinction-corrected \ha\ line luminosities using the sum of the observed \ha\ emission line and a fraction of the rest 24~\micron\ monochromatic luminosity as defined by \citet{Kennicutt2009},
\begin{equation}
\mathrm{L}_{\ha}^{\mathrm{corr}}\ (\mathrm{erg\ s}^{-1}) = \mathrm{L}_{\ha} + 0.020\ \mathrm{L}_{24~\micron}\ (\mathrm{erg\ s}^{-1})
\end{equation}
For the \citet{Brown2014} subsample this was straightforward considering the SEDs for their galaxies were given in the rest-frame.  For the \citet{ODowd2009} and \citet{Shipley2013} subsamples we used the best fit SEDs (see \S\ \ref{LIR}) in the rest-frame for each galaxy and again estimated the rest 24~\micron\ flux densities from the SEDs for each galaxy.  This was necessary for these two samples to be consistent with each other as a significant amount of the galaxies' IRS spectra at rest-frame 24~\micron\ are shifted out of the observed IRS spectrum or dominated by noise due to the redshift of the galaxies (49/142 galaxies for the combined samples).  For the \citet{ODowd2009} and \citet{Shipley2013} samples, we synthesized MIPS 24~\micron\ broadband flux densities by computing the (rest-frame) MIPS bandpass-weighted average flux density from the IR SED that best fit the long-wavelength photometry\footnote{see http://irsa.ipac.caltech.edu/data/SPITZER/docs/\\
dataanalysistools/cookbook/14/}.

\begin{figure}[t!]	
\epsscale{1.1}
\plotone{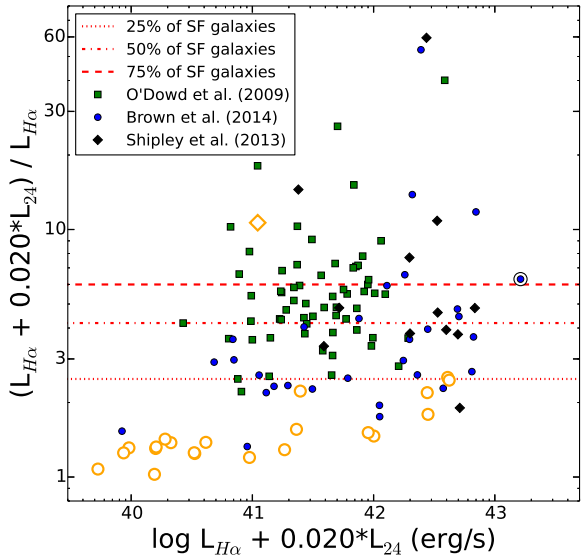}
\caption{The relation between the extinction-corrected \ha\ luminosity and the dust correction factor, defined as the ratio of the extinction-corrected \ha\ to observed (uncorrected) \ha.  The filled symbols show the primary calibration sample.  The horizontal lines show the 25, 50, and 75 percentiles of the distribution.  The median correction factor is 4.2 and the interquartile range (25$-$75 percentiles) is 2.5 $-$ 5.9.  The open orange symbols show the secondary calibration sample of low metallicity galaxies as defined in \S\ \ref{metals and pahs}.  The blue point enclosed by a black circle denotes II Zw 096, see \S\ \ref{zwicky}.}
\label{Ha contribution}
\end{figure}

In most cases the \ha\ extinction corrections are substantial.  Figure \ref{Ha contribution} shows the relation between the corrected \ha\ luminosity and the ratio of observed \ha\ luminosity (uncorrected for dust extinction) to the extinction-corrected \ha\ luminosity.  Most of the galaxies in our primary and a few in the secondary calibration samples have corrections of more than a factor of 2 to the \ha\ luminosity.  The median correction factor is 4.2 and the interquartile range (25$-$75 percentiles) is 2.5 $-$ 5.9.

\begin{figure}[t!]	
\epsscale{1.1}
\plotone{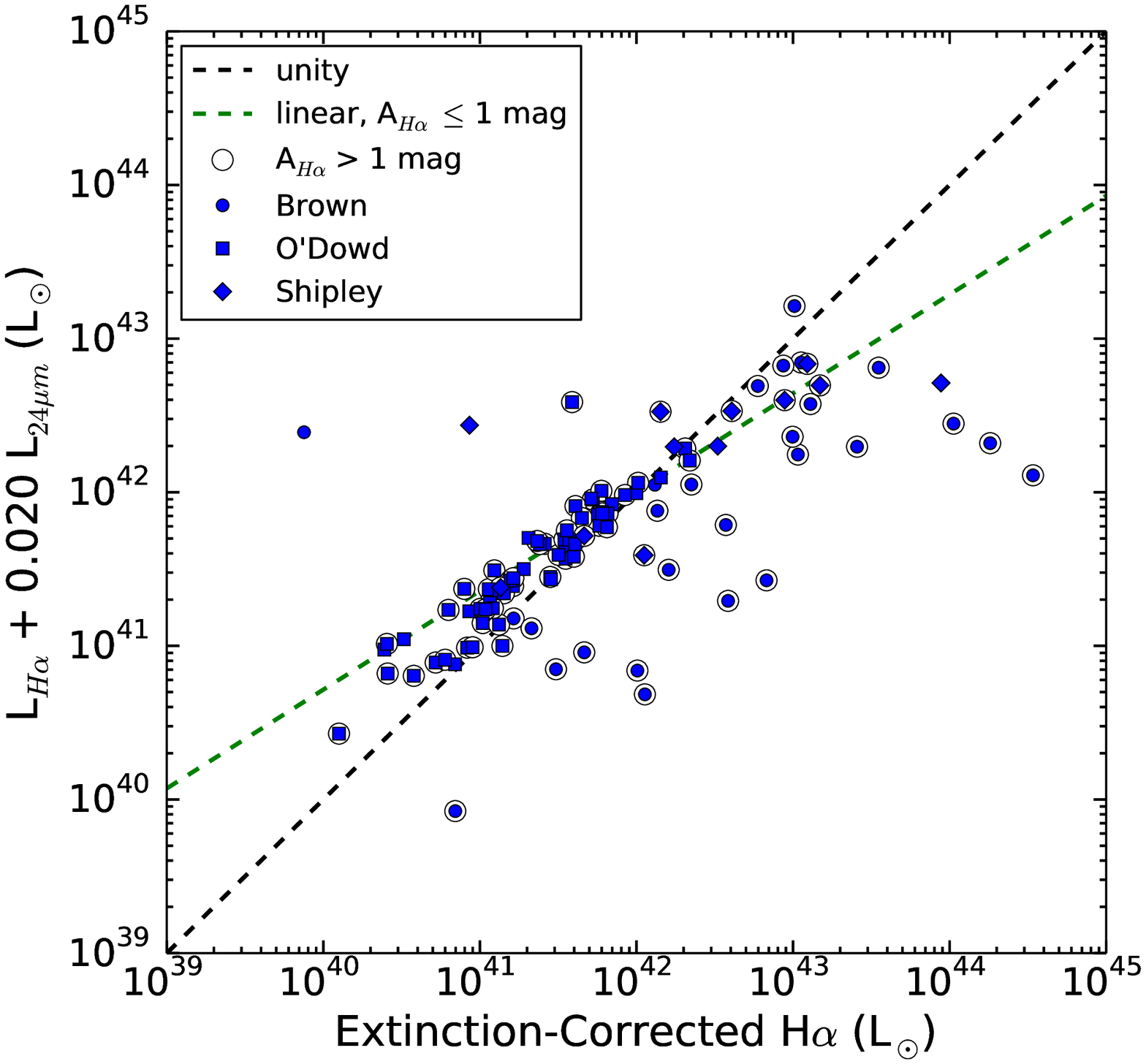}
\caption{Extinction-corrected \ha\ luminosities corrected from the Balmer decrement (ratio of \ha /\hb) compared to the extinction-corrected \ha\ luminosities corrected from the 24~\micron\ luminosity for the primary calibration sample.  We denote galaxies with A$_{\ha} > 1$~mag (open circles).  We performed unity (black dashed-line) and linear (green dashed-line) fits to the galaxies with A$_{\ha} \leq 1$~mag.  For many sources the results are consistent.  But, there is a sample of highly attenuated objects, A$_{\ha} > 1$~mag, where we see big deviations (see \S\ \ref{ha contribution} for further explanation).  Three galaxies from the primary calibration sample do not appear in the figure due to unreliable \hb\ flux measurements.}
\label{balmer}
\end{figure}

In Figure \ref{balmer}, we compare our extinction corrections from the monochromatic 24~\micron\ luminosity to extinction-corrected \ha\ luminosities from the Balmer decrement, measured from the ratio of the \ha\ and \hb\ emission lines.  Most of our primary calibration sample consists of dusty star-forming galaxies (i.e.\ LIRGs) that heavily attenuate the \hb\ emission line and to a lesser extent the \ha\ emission line.  We indicate galaxies with A$_{\ha} > 1$~mag that, in essence, results in an attenuation factor of $\gsim$2.5 for \ha\ and a factor $\gsim$3.6 for \hb.  For many sources the results are consistent.  But, there is a sample of highly attenuated objects, A$_{\ha} > 1$~mag, where we see big deviations (at the higher extinction-corrected \ha\ luminosities the Balmer decrement corrections saturate).  The likely explanation for this attenuation is that the interstellar medium (ISM) at \hb\ is optically thick, so we are only seeing the ``skin" of the star-forming region (down to where $\tau \sim 1$ for \hb), whereas we can see deeper into the star-forming region for \ha.  Balmer absorption could be another issue.  Figure \ref{balmer} suggests it is a significant contributor, since there are many galaxies where correcting with the Balmer decrement overcorrects \ha.  Only about half the sources in the primary calibration sample for A$_{\ha} > 1$~mag allow us to extinction correct using the Balmer decrement and $\sim 10 \%$ of the sample have unreliable corrections due to \hb\ being heavily attenuated.  For this reason, we adopt the \ldhalpha\ method as the best estimator for the total \ha\ emission for our primary calibration sample.

\subsection{Gas-Phase Metallicities}
\label{metals and pahs}

The PAH emission from a galaxy is known to depend on the gas-phase metallicity.  \citet{Engelbracht2005} demonstrated this dependence using 8~\micron-to-24~\micron\ ratios of galaxies above and below a metallicity of 12 + log(O/H) $=$ 8.2, where they attributed the smaller ratios for galaxies with low metallicity to a decrease in the mid-IR emission bands (attributed to PAHs).  \citet{Calzetti2007} showed star-forming galaxies with low metallicities have less than expected mid-IR dust emission (presumably from PAH emission) based on a linear fit to their calibration of extinction corrected \paa\ to the 8~\micron\ dust emission.  This trend is not seen in their calibration of extinction corrected \paa\ to \ha\ + 24~\micron\ luminosities, illustrating the PAH emission dependence on metallicity.

\begin{figure}[t!]	
\epsscale{1.1}
\plotone{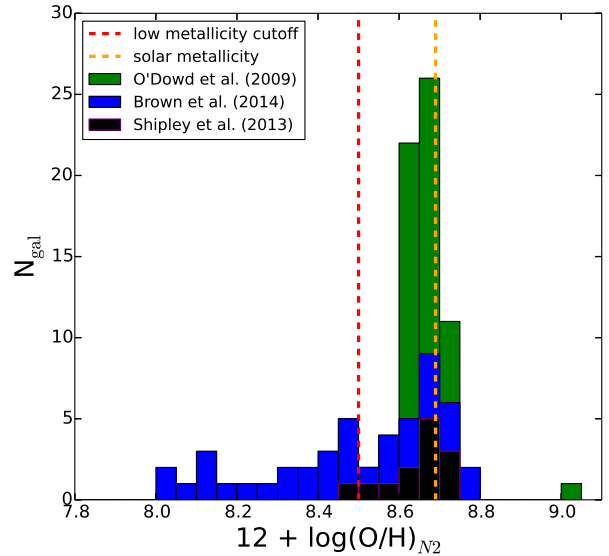}
\caption{Histogram of the metallicities for all the star-forming galaxies in the calibration sample, where the metallicities are \zmetal.  For the calibration of the SFR relations using PAH luminosity we used galaxies with \zmetal\ $\geq$ 8.5 (red dashed line) for our primary calibration sample.  The resulting cut being 0.65 of the solar abundance \citep[12 + log(O/H) = 8.69,][orange dashed line]{Asplund2009}.}
\label{metallicity}
\end{figure}

We calibrate the relation between metallicity, PAH emission, and SFRs using the data in our samples.  For all galaxies in the calibration sample, we estimate the gas-phase oxygen abundance using the $N2$ index, $N2 \equiv$ log \ntwo /\ha, using the relation from \citet{Pettini2004},
\begin{equation}
12 + \mathrm{log(O/H)} = 8.90 + 0.57 \times N2
\end{equation}
We use \zmetal\ because the $N2$ index is available for all galaxies in our calibration sample.  Figure \ref{metallicity} shows the distribution of oxygen abundance for the galaxies in our primary and secondary calibration samples.  Most sources fall in a tight distribution ($\pm0.2$ dex) around the solar abundance \citep[12 + log(O/H) = 8.69,][]{Asplund2009} with a long tail to lower abundances.  This is unsurprising because of the well-known mass-metallicity relation (Tremonti et al. 2004).  Most sources in the calibration samples have $L_\mathrm{IR} > 10^{10}$ \lsun, and likely have higher stellar mass, and therefore higher abundance \citep[e.g., based on][galaxies with stellar mass $> 10^9$ \msun\ have abundances $>$ 0.5 \zsun]{Tremonti2004}.  We use star-forming galaxies with \zmetal\ $\geq$ 8.5 for the primary calibration sample.

\subsection{Total IR luminosity}
\label{LIR}

We estimate the total IR luminosity (\lir\ = L$_{\mathrm{8-1000}~\micron}$) using the MIPS 24~\micron\ flux densities \citep[also 70~\micron\ and 160~\micron, where available in][]{Shipley2013} and the method in \citet{Shipley2013}.  We use the \citet{Rieke2009} IR SEDs because for star-forming galaxies the total IR luminosities derived from these templates using the 24~\micron\ flux density only are closest to the IR luminosities derived when more IR photometric bands are available.  These results help us understand the range of total IR luminosity over which our PAH SFRs are calibrated; our calibration may not extend to higher IR luminosities, where the PAH features tend to be suppressed.  We were able to determine the contributing total IR luminosity range for the \citet{Brown2014} sample by using the values reported in \citet{Haan2013} for the majority of the galaxies.

\section{The PAH SFR Calibration}
\label{results}

Previous studies have shown that the \ha\ luminosity is a robust indicator of the SFR once corrected for dust extinction but this correction is non-trivial \citep[e.g.,][]{Kennicutt1998}.  More recent work has shown that the \ha\ luminosity summed with a fraction of the monochromatic 24~\micron\ light provides a linear fit against extinction corrected \ha\ and \paa\ emission \citep{Calzetti2007, Kennicutt2007, Kennicutt2009,Treyer2010}.  Here we use the extinction-corrected \ha\ from the 24~\micron\ monochromatic emission as our SFR calibrator, using \citet{Kennicutt2009},
\begin{multline} \label{K09 SFR}
\mathrm{SFR}\ (\sfrunits) = \\
5.5\times 10^{-42}[\mathrm{L}(\ha) + 0.020\ \mathrm{L}(24~\micron)]\ (\mathrm{erg\ s}^{-1})
\end{multline}
where the constant of proportionality is appropriate for the Kroupa IMF we have adopted.

\subsection{PAH SFR Relations}
\label{SFR relations}

We compare the relation between the luminosity in the PAH features to the extinction-corrected \ha\ luminosity.  Because different PAH features are accessible at different wavelengths depending on the redshift of the source and wavelength capabilities of the telescope/instrument, we calibrate both the individual features at 6.2\micron, 7.7\micron\ and 11.3\micron, as well as sums of these features.  It is also worth noting that \ha\ probes very young star-formation from OB stars, whereas it is likely that the PAH features cover a broader range of ages (presumably including contributions from A and early F stars) and that different SFR indicators are sensitive to slightly different effects from the star-formation age range measured.

\begin{figure*}[t!]	
\epsscale{1.0}
\plotone{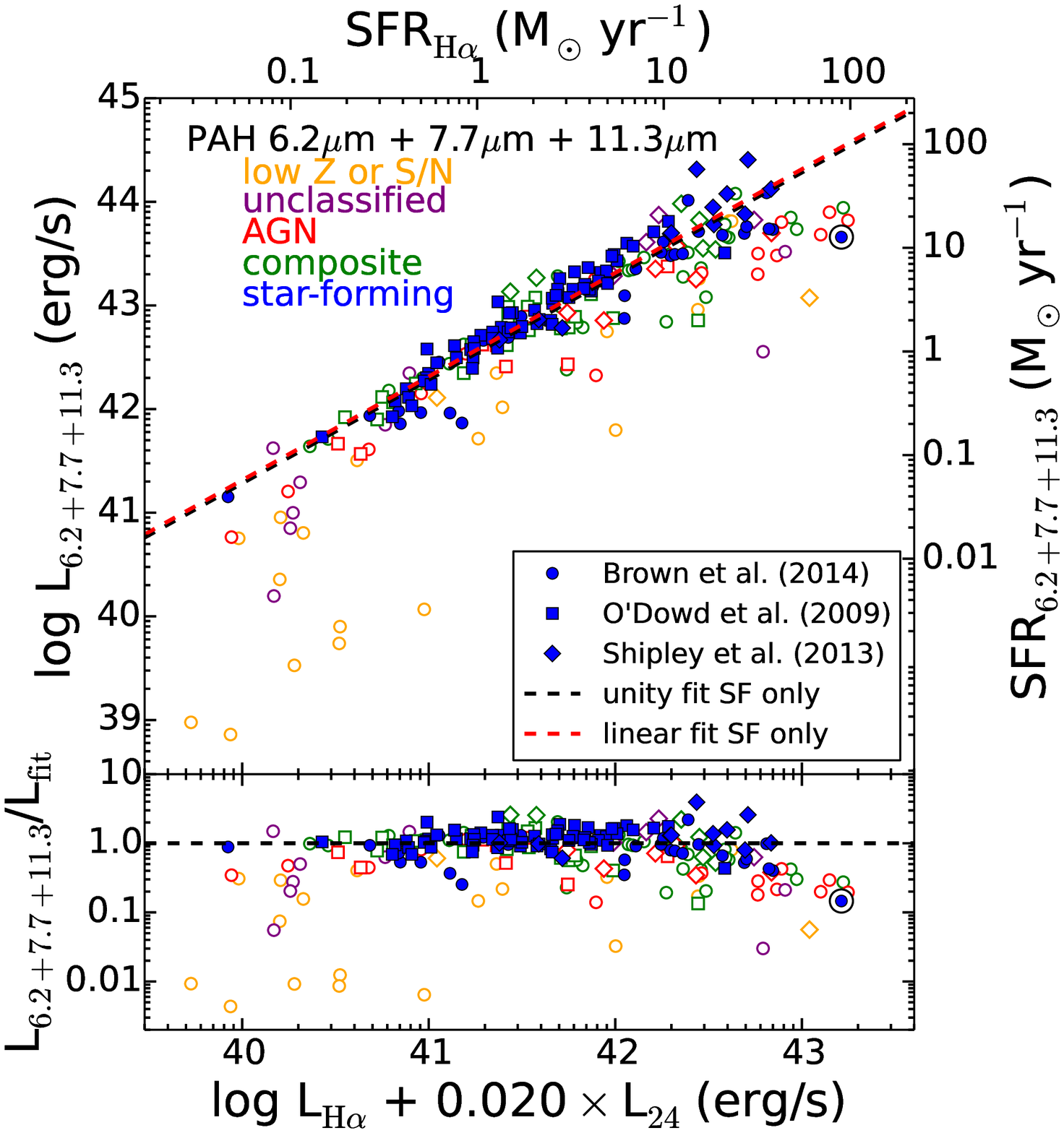}
\caption{The top panel is the extinction-corrected \ha\ luminosity (\ldhalpha) versus total PAH luminosity (\lpone\ + \lptwo\ + \lpthree\ PAH features).  We fit a unity relation (dashed black line) to the primary calibration sample (filled blue points) and use this line (that we define as \lfit) to show the ratio of the total PAH luminosity, for each galaxy, to the unity relationship as a function of extinction-corrected \ha\ luminosity (bottom panel).  We classify galaxies as star-forming (filled blue points), composite (unfilled green points) or an AGN (unfilled red points) based on the location of their emission line ratios on a BPT diagram.  Unclassified galaxies (unfilled purple points) did not have all lines required for BPT classification and the star-forming galaxies with low metallicity or S/N (unfilled orange points) are not used for the PAH SFR calibration (filled blue points; see text for more detail).  The blue point enclosed by a black circle denotes II Zw 096, see \S\ \ref{zwicky}.}
\label{ha24 pah wAGN}
\end{figure*}

Figure \ref{ha24 pah wAGN} shows the correlation between the total PAH luminosity\footnote{\lpah\ we use to denote the PAH luminosity for the total luminosity of the 6.2\micron, 7.7\micron\ and 11.3\micron\ features.  All other instances of PAH luminosity will be denoted by the included PAH features (e.g. \lpone).  However in figures, we label explicitly which features are used.}
\begin{equation}
\lpah = \lpone\ + \lptwo\ + \lpthree
\end{equation}
and the extinction-corrected \ha\ luminosity for the 227 galaxies in our calibration sample.  For star-forming galaxies with high metallicity (primary calibration sample), the total PAH luminosity correlates linearly with the extinction-corrected \ha\ luminosity.  Our derived relations between the extinction-corrected \ha\ luminosity and the total PAH luminosity are
\begin{multline}\label{lpah unity}
\mathrm{log}\ \lpah\ = \\
(1.30 \pm 0.03) + \mathrm{log}(\ldhalpha) 
\end{multline}
for the unity relation and 
\begin{multline}\label{lpah linear}
\mathrm{log}\ \lpah\ = (1.30 \pm 0.03)\ + \\
(1.00 \pm 0.03)\ \mathrm{log}(\ldhalpha) 
\end{multline}
for the linear relation.  We note that the forms of Equations \ref{lpah unity} and \ref{lpah linear} are essentially identical, and this is because the  linear fit yields a slope consistent with unity to five significant digits ($1.0003\pm 0.03$).  Using Equation \ref{K09 SFR}, this yields the following calibrations for the total PAH luminosity as a SFR indicator,
\begin{multline}
\mathrm{log\ SFR\ (M}_{\odot} \mathrm{\ yr}^{-1}) = \\
(-42.56 \pm 0.03) + \mathrm{log\ } \lpah\ (\mathrm{erg\ s}^{-1})
\end{multline}
and using the linear fit we find,
\begin{multline}
\mathrm{log\ SFR\ (M}_{\odot} \mathrm{\ yr}^{-1}) = (-42.56 \pm 0.03)\ + \\
(1.00 \pm 0.03)\ \mathrm{log\ } \lpah\ (\mathrm{erg\ s}^{-1})
\end{multline}
showing there is statistical uncertainty of $\approx$0.03 dex and scatter of $\approx$0.135 dex on the total PAH luminosity unity SFR relationship (where we remind the reader that this applies for galaxies of roughly solar metallicity and luminosities of $10^{9} < \lir /\lsun < 10^{12}$).

We observe small offsets from the extinction-corrected \lha\ and \lpah\ calibration between the different calibration subsamples.  We measure average offsets of \lpah /\lfit\ = 0.06, 0.06, -0.10 for the \citet{ODowd2009}, \citet{Shipley2013} and \citet{Brown2014} subsamples, respectively.  This may imply there is an additional systematic uncertainty of 0.1 dex (i.e., 20\%) on the PAH SFR calibration from differences in aperture corrections or data processing and analysis.

The scatter about this relation for the primary calibration sample is very small, 0.14 dex, at least over the luminosity range of our sample.  This implies that the total PAH emission correlates linearly with the SFR.  The scatter is similar to that measured for other star-formation indicators measured against each other \citep{Kennicutt2009}, indicating that the total PAH emission is of similar accuracy in determining SFRs.

\begin{figure*}[t!]	
\epsscale{1.1}
\plottwo{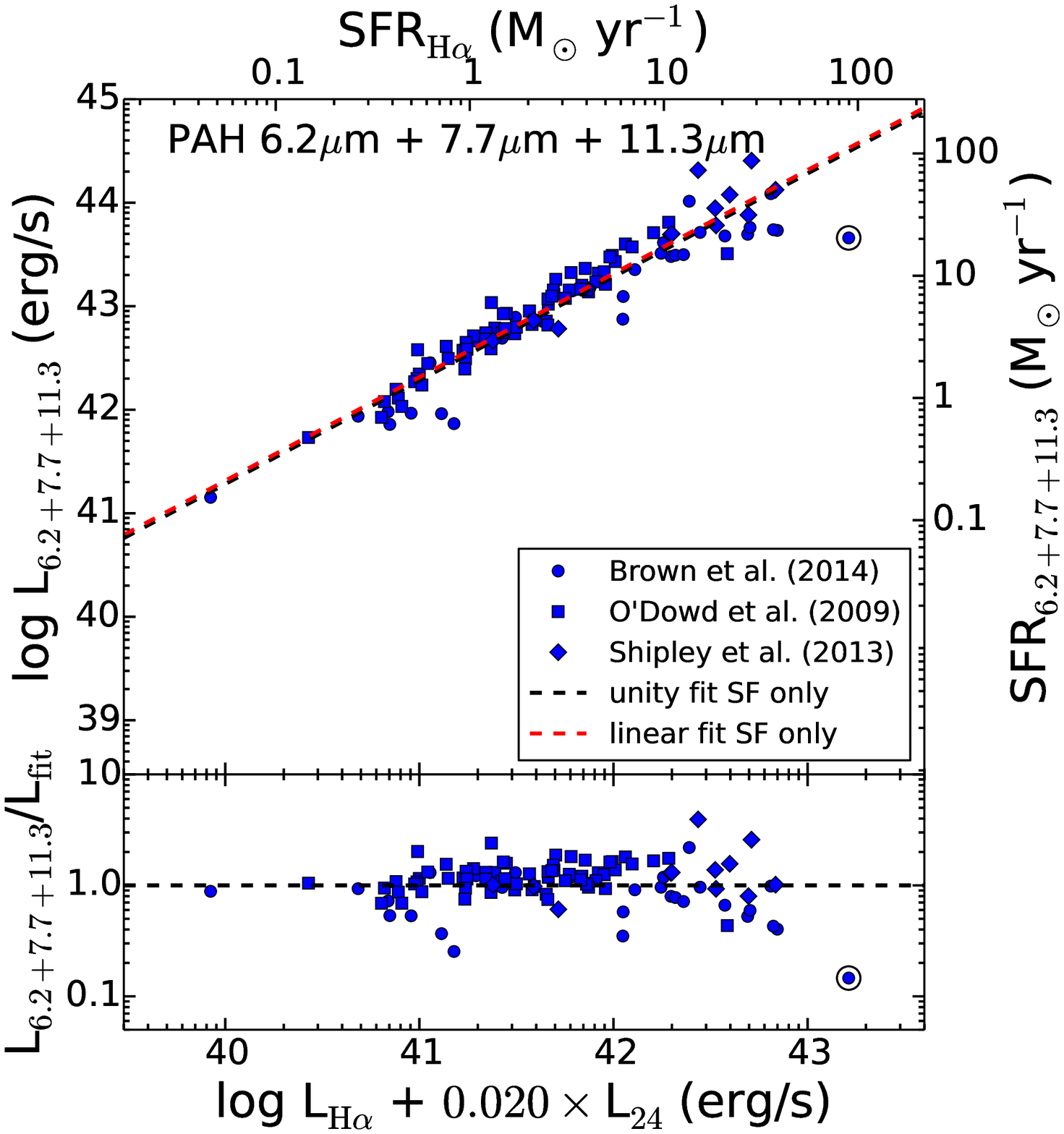}{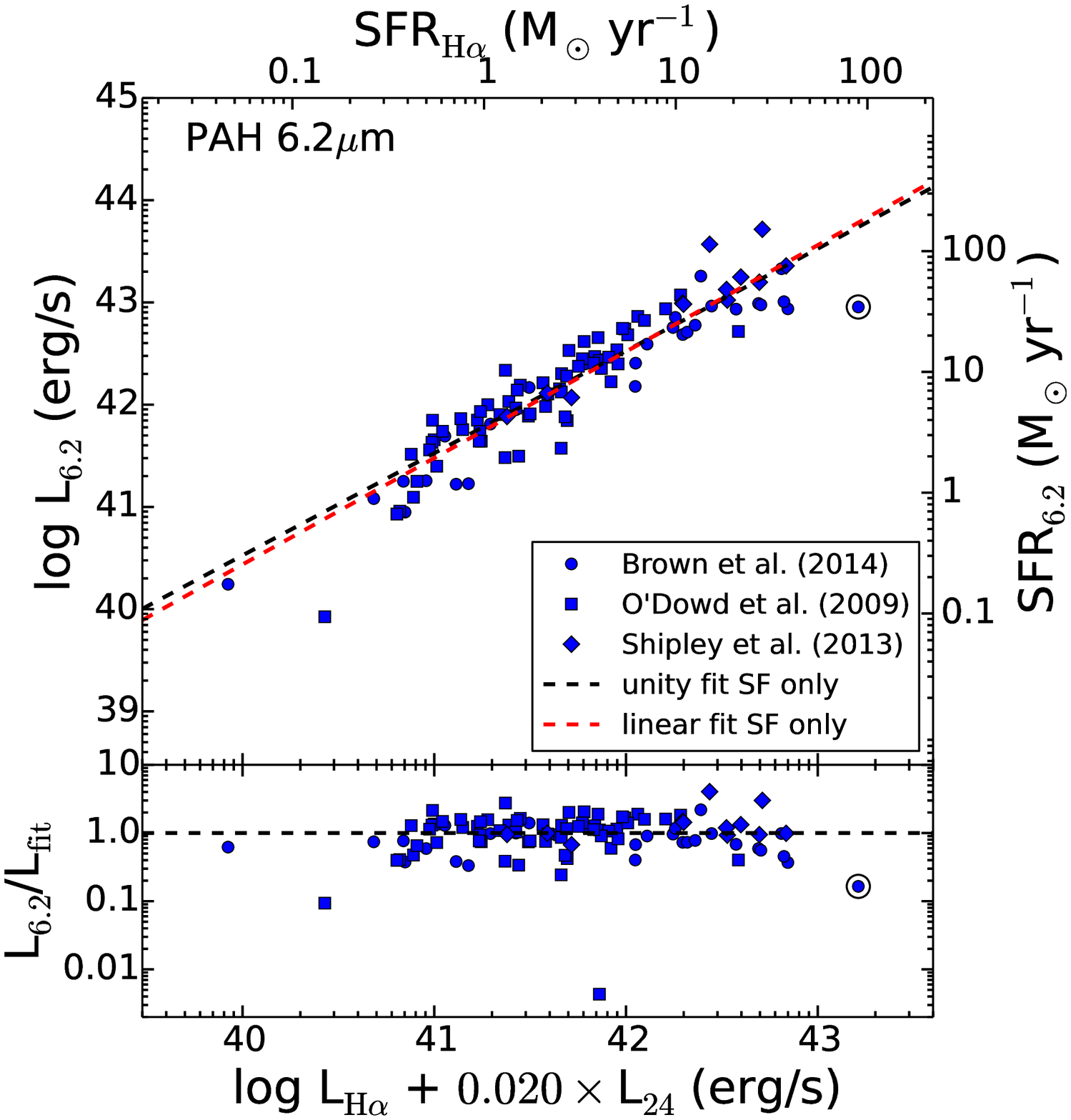}
\plottwo{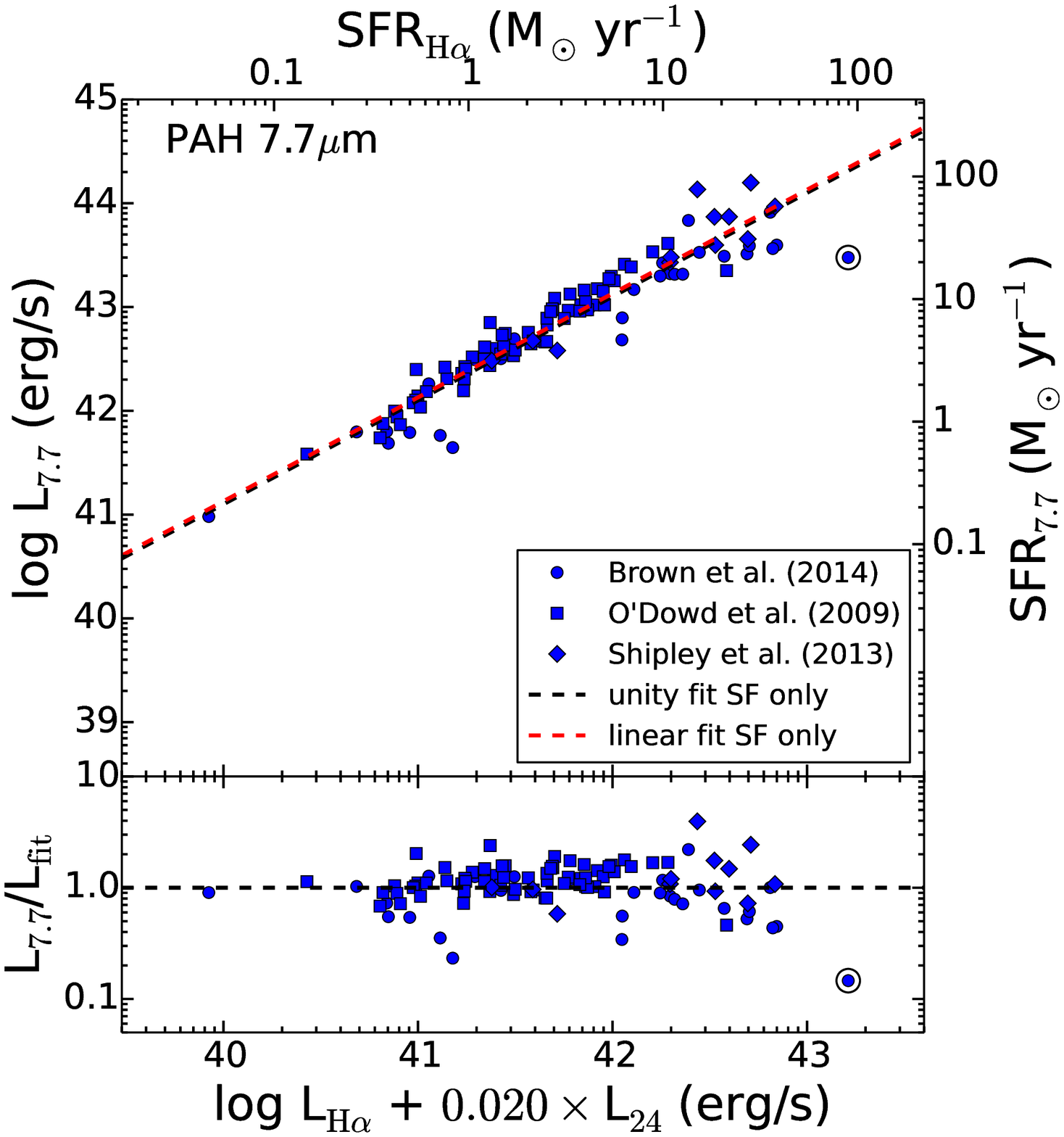}{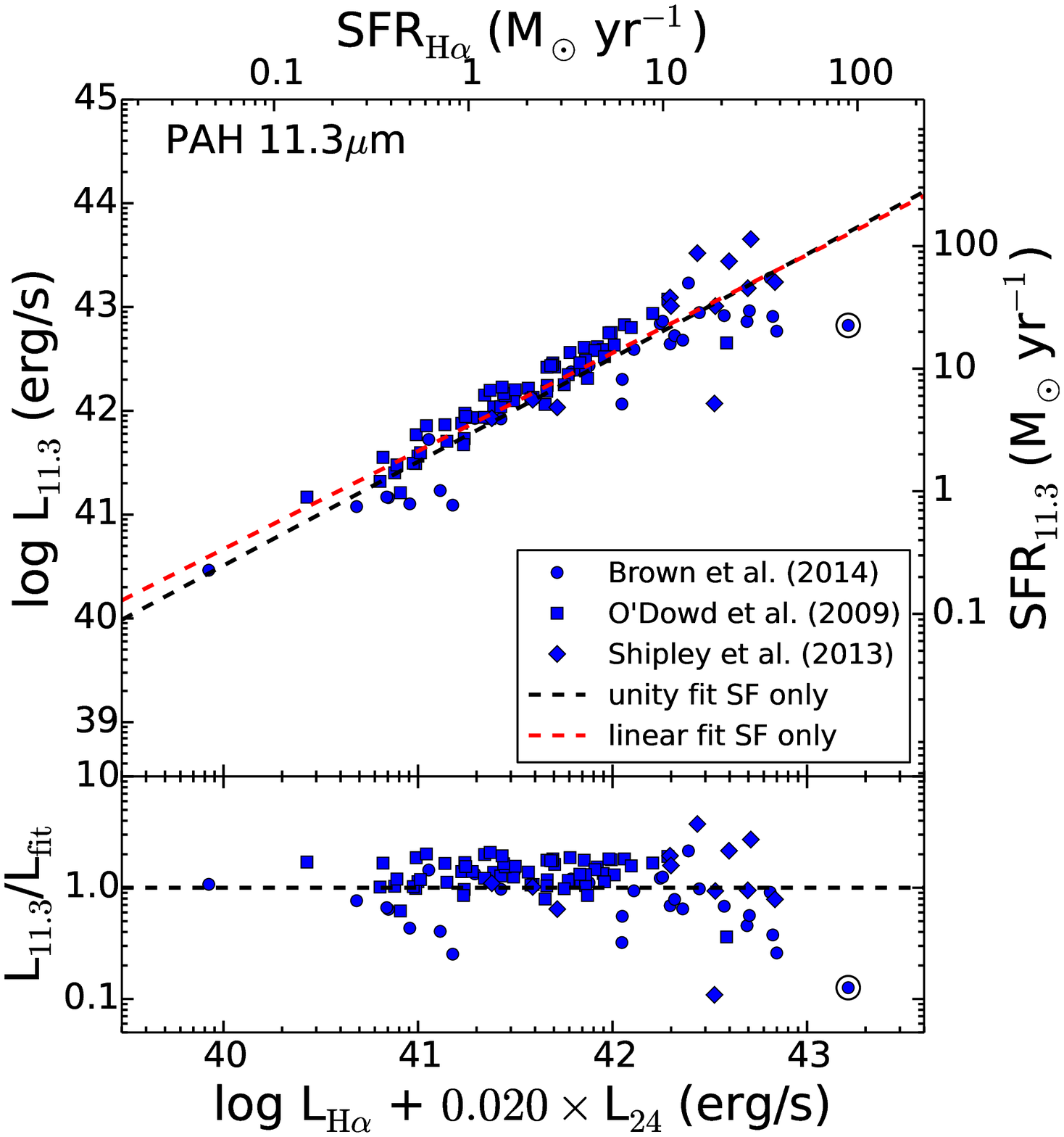}
\caption{The extinction-corrected \ha\ luminosity \citep[\ldhalpha,][]{Kennicutt2009} versus \lpah\ (top left) and for the 6.2\micron\ (top right), 7.7\micron\ (bottom left) and 11.3\micron\ (bottom right) PAH features separately.  This illustrates that the small scatter persists for the individual PAH features ($\sigma_\mathrm{MAD} \lsim 0.2$ dex).  We fit a unity relation to the primary calibration sample (solid black line) and use this line as \lfit\ to determine residuals of expected \lpah\ values (bottom for each panel).  We used this fit to determine a SFR relation for \lpah\ and each PAH feature shown individually (see \S\ \ref{SFR relations}).  The blue point enclosed by a black circle denotes II Zw 096, see \S\  \ref{zwicky}.}
\label{ha24 pah onlySF}
\end{figure*}

Figure \ref{ha24 pah wAGN} shows that it was necessary to remove the low metallicity (and low S/N; S/N $<$ 5) galaxies from our primary calibration sample.  The low metallicity (and low S/N) galaxies have low \lpah/\lfit\ ratios (Figure \ref{ha24 pah wAGN}, bottom), where \lfit\ is defined as the unity relation fit to the primary calibration sample for the expected luminosity of the PAH luminosity for a given \ldhalpha\ luminosity.  We mark the galaxy II Zw 096 as it is an outlier in the figures (Figures \ref{ha24 pah wAGN}, \ref{ha24 pah onlySF} and \ref{metallicity residual}) but do not omit it as it does not significantly affect our calibration of the PAH features (we discuss this interesting galaxy in \S~\ref{zwicky}).

\begin{deluxetable}{ccc}		
\tablecaption{Unity PAH Luminosity SFR Relations \vspace{-6pt} 
\label{Unity SFRs}}
\tablecolumns{3}
\tablewidth{0pc}
\setlength{\tabcolsep}{36pt}
\tablehead{
\colhead{PAH Feature(s)} & \colhead{C} & \colhead{$\sigma_{\mathrm{MAD}}$}
}
\startdata
6.2 + 7.7 + 11.3 & -42.56 $\pm$ 0.03 & 0.135 \\ 
6.2 & -41.73 $\pm$ 0.08 & 0.214 \\ 
7.7 & -42.37 $\pm$ 0.05 & 0.131 \\ 
11.3 & -41.80 $\pm$ 0.07 & 0.187 \\ 
6.2 + 7.7 & -42.47 $\pm$ 0.04 & 0.135 \\ 
6.2 + 11.3 & -42.09 $\pm$ 0.04 & 0.160 \\ 
7.7 + 11.3 & -42.48 $\pm$ 0.04 & 0.120 \\ 
\enddata
\vspace{-6pt}
\tablecomments{The calibration sample consisted of 105 galaxies used to determine the PAH SFR relations (see \S\ \ref{primary sample}).  Column 1 defines which PAH features were used in the SFR relation.  Column 2 gives the values for the unity SFR relations (defined as log SFR [\sfrunits] = C + log L$_{\mathrm{PAH}, \lambda}$ [erg s$^{-1}$] ).  Column 3 is the 1$\sigma$ scatter in the SFR relation for the galaxies used in the fit (using the median absolute deviation, MAD ($\sigma_\mathrm{MAD}$) to represent the scatter).}
\end{deluxetable}		

\begin{figure*}[t!]	
\epsscale{1.1}
\plottwo{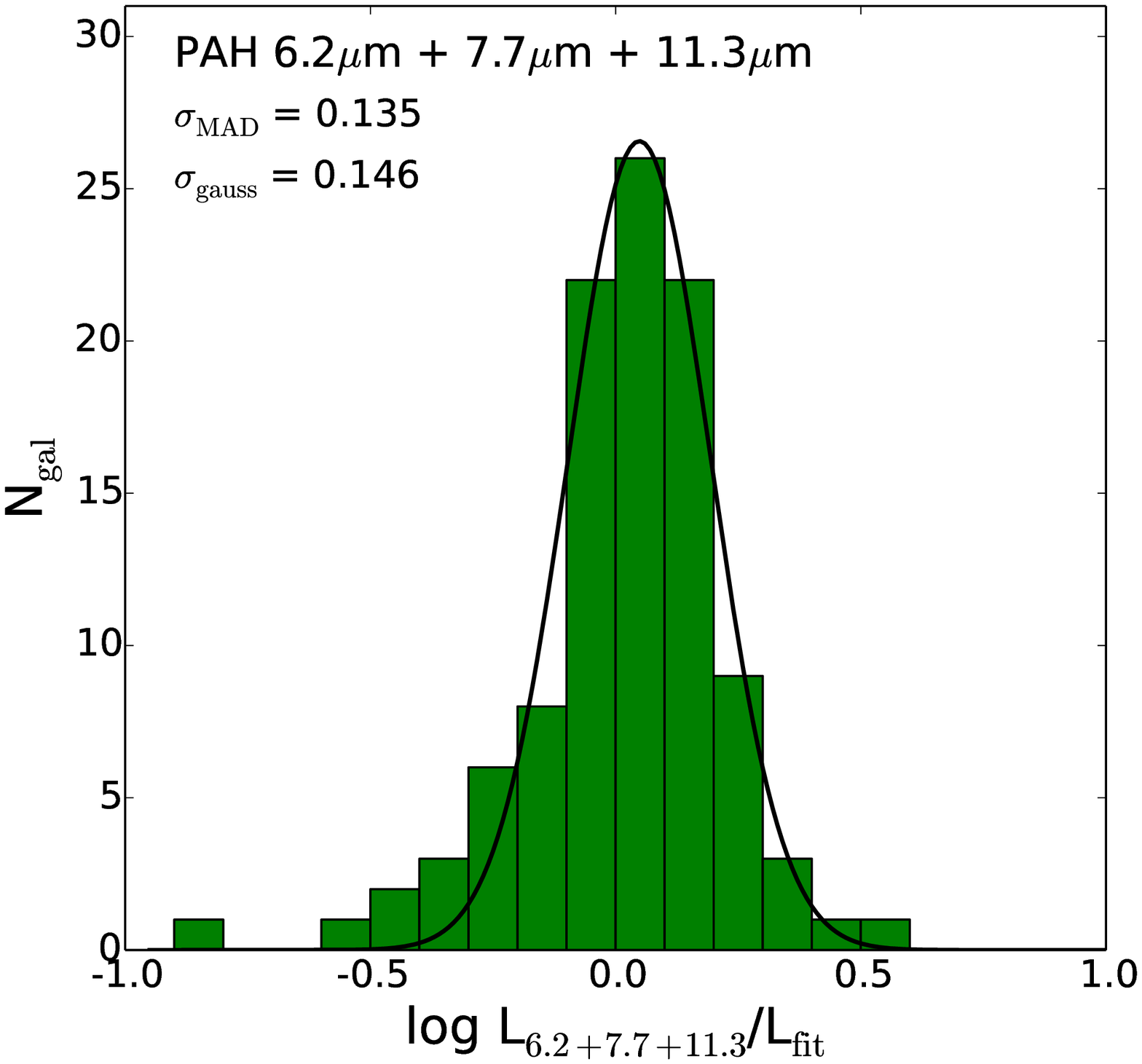}{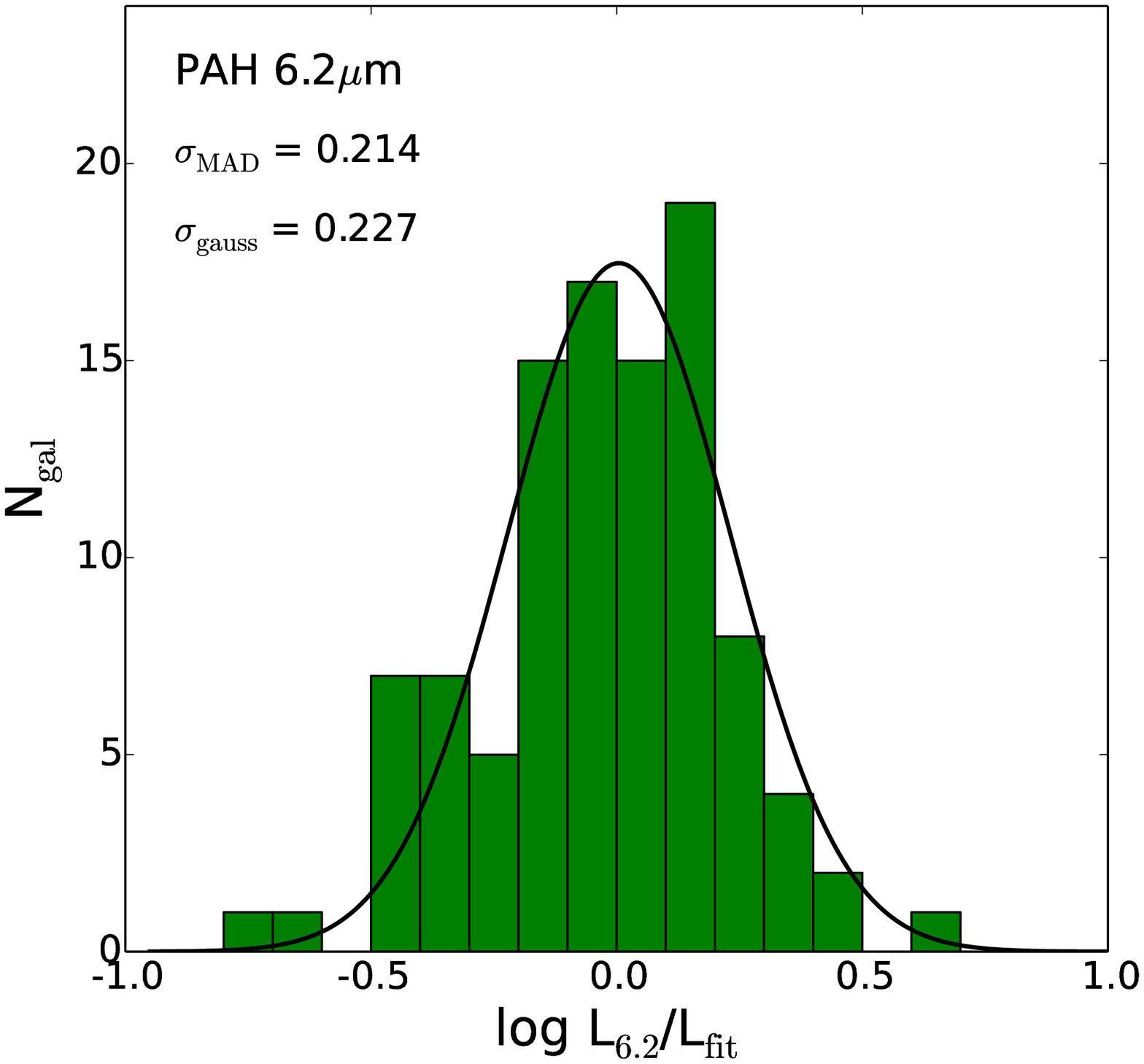}
\plottwo{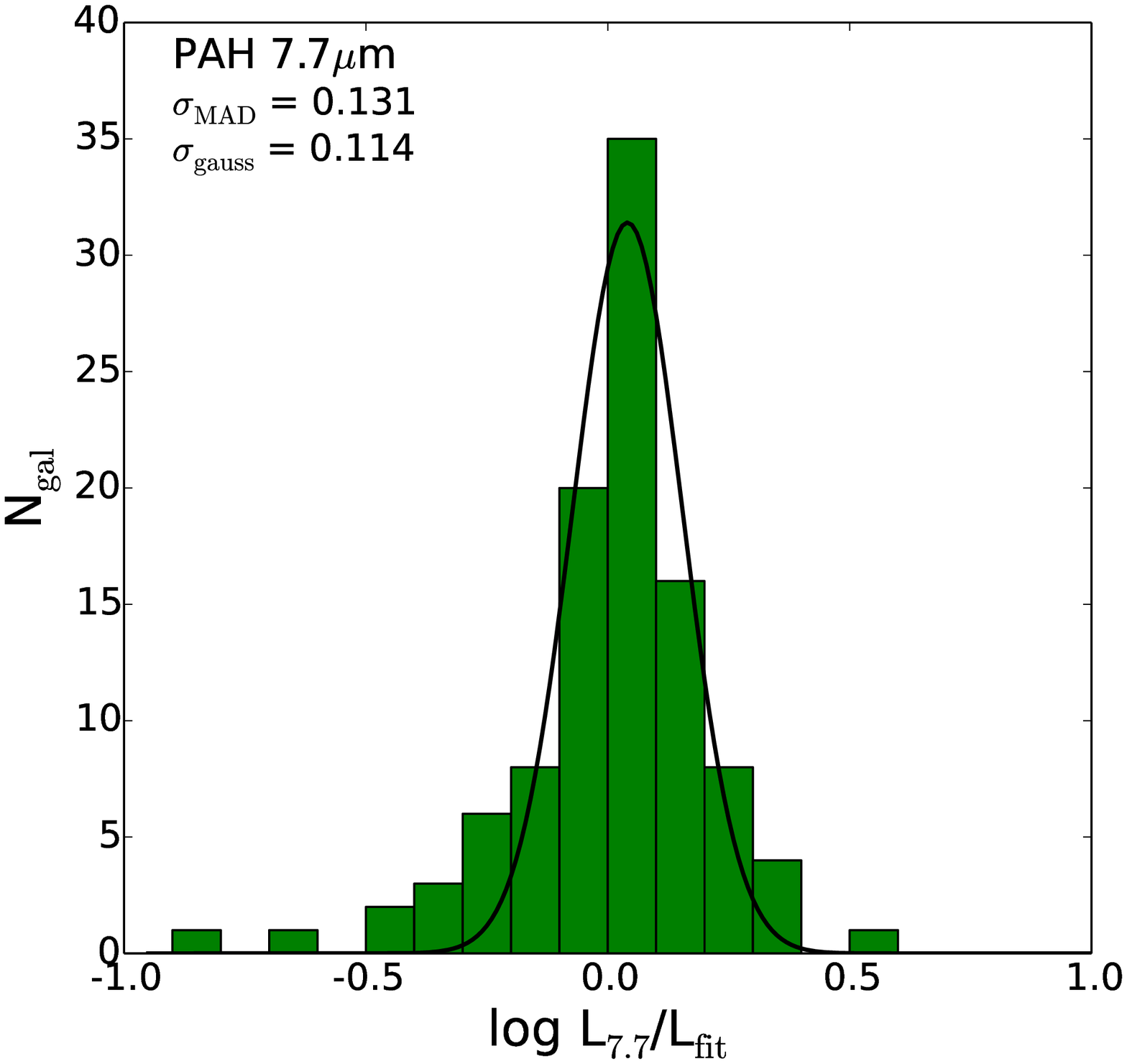}{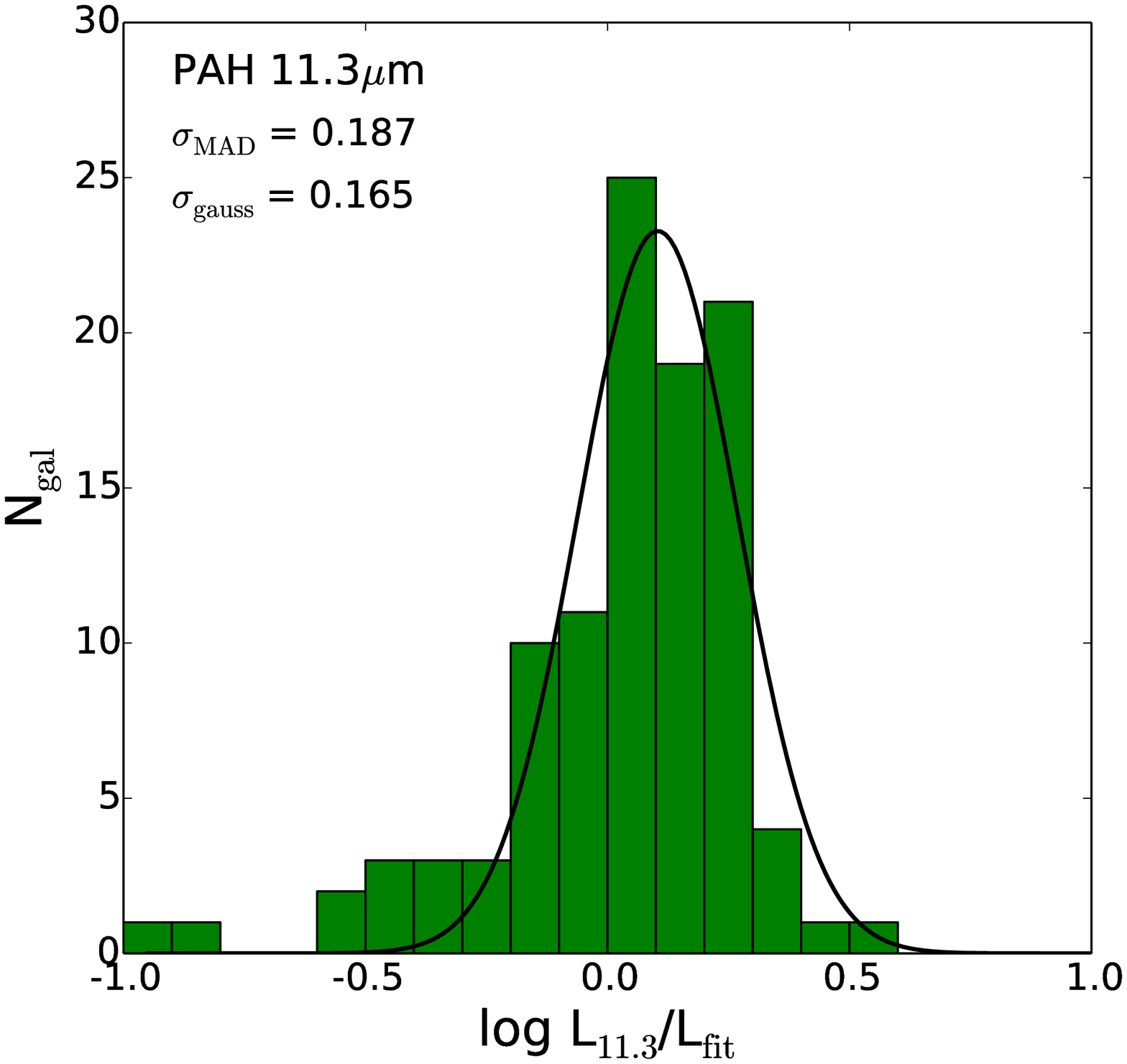}
\caption{A histogram of the ratio of \lpah /\lfit , where \lfit\ is the unity relation to the star-forming galaxies shown in Figure \ref{ha24 pah onlySF}, demonstrating the scatter in the sample and consistency of \lpah\ as a SFR indicator in comparison to extinction-corrected \ha\ for the 6.2\micron\ (top right), 7.7\micron\ (bottom left), 11.3\micron\ (bottom right) PAH features individually and \lpah\ (top left).  The bin size used is 0.1 log \lpah /\lfit\ and N$_{\mathrm{gal}}$ represents the number of galaxies in each bin.  We fit a Gaussian distribution to determine the standard deviation of the scatter ($\sigma_{\mathrm{gauss}}$, shown in plot) and calculated the median absolute deviation ($\sigma_{\mathrm{MAD}}$) to represent accurately the scatter of L$_{\mathrm{PAH}, \lambda}$ as a SFR indicator.}
\label{resfit wL7}
\end{figure*}

We also calibrate the luminosity from the individual PAH features and combinations of them as SFR indicators as in some cases only one or two PAH features may be useable to get a SFR, such as for high redshift galaxies (see \S\ \ref{highz galaxies}).  Figure \ref{ha24 pah onlySF} shows the relations between the extinction-corrected \ha\ luminosity and the individual PAH features and the total PAH luminosity for the primary calibration sample.  Figure \ref{resfit wL7} shows the distribution of residuals between the PAH luminosity and the fit, and shows both the scatter computed from the median absolute deviation \citep[MAD,][]{Beers1990} and a Gaussian fit, both of which are consistent.  The scatter is smallest, $\sigma_{\mathrm{MAD}}$ = 0.14 dex (i.e., 40\%) whenever the 7.7\micron\ feature is involved in the fit.  This implies this feature may provide the most robust measure of the SFR.  The relation against the 6.2\micron\ feature has the highest scatter, $\sigma_{\mathrm{MAD}}$ = 0.21 dex (i.e., 60\%), which may indicate more variation in galaxies between the luminosity in this feature and the SFR.

Our derived relations between the extinction-corrected \ha\ luminosity and the individual PAH luminosities are
\begin{equation}
\begin{split}
\mathrm{log}\ \lpone = \\
(0.47 \pm & 0.08) + \mathrm{log}(\ldhalpha) \\
\mathrm{log}\ \lptwo = \\
(1.11 \pm & 0.05) + \mathrm{log}(\ldhalpha) \\
\mathrm{log}\ \lpthree = \\
(0.54 \pm & 0.07) + \mathrm{log}(\ldhalpha)
\end{split}
\end{equation}

for the unity relations and the linear relations are
\begin{equation}
\begin{split}
\mathrm{log}\ \lpone = (-1.20 \pm 0.09)\ + \\
(1.04 \pm 0.04)\ \mathrm{log}(\ldhalpha) \\
\mathrm{log}\ \lptwo = (1.12 \pm 0.06)\ + \\
(1.00 \pm 0.03)\ \mathrm{log}(\ldhalpha) \\
\mathrm{log}\ \lpthree = (2.87 \pm 0.08)\ + \\
(0.94 \pm 0.03)\ \mathrm{log}(\ldhalpha)
\end{split}
\end{equation}

\begin{deluxetable}{ccccc}	
\tablecaption{Linear PAH Luminosity SFR Relations \vspace{-6pt} 
\label{Linear SFRs}}
\tablecolumns{5}
\tabletypesize{\small}
\tablewidth{0pc}
\setlength{\tabcolsep}{18pt}
\tablehead{
\colhead{PAH Feature(s)} & \colhead{A} & \colhead{B} & \colhead{$\sigma_{\mathrm{AB}}$} & \colhead{$\sigma_{\mathrm{MAD}}$}
}
\startdata
6.2 + 7.7 + 11.3 & -42.56 $\pm$ 0.03 & 1.00 $\pm$ 0.03 & -0.03 & 0.138 \\ 
6.2 & -40.06 $\pm$ 0.09 & 0.96 $\pm$ 0.04 & -0.07 & 0.233 \\ 
7.7 & -42.38 $\pm$ 0.06 & 1.00 $\pm$ 0.03 & -0.03 & 0.138 \\ 
11.3 & -44.14 $\pm$ 0.08 & 1.06 $\pm$ 0.03 & -0.04 & 0.166 \\ 
6.2 + 7.7 & -42.05 $\pm$ 0.04 & 0.99 $\pm$ 0.03 & -0.03 & 0.137 \\ 
6.2 + 11.3 & -42.52 $\pm$ 0.05 & 1.01 $\pm$ 0.03 & -0.04 & 0.144 \\ 
7.7 + 11.3 & -42.85 $\pm$ 0.04 & 1.01 $\pm$ 0.03 & -0.03 & 0.139 \\ 
\enddata
\vspace{-6pt}
\tablecomments{The calibration sample consisted of 105 galaxies used to determine the PAH SFR relations (see \S\ \ref{primary sample}).  Column 1 defines which PAH features were used in the SFR relation.  Columns 2 and 3 give the values for the linear SFR relations (defined as log SFR [\sfrunits] = A + B log L$_{\mathrm{PAH}, \lambda}$ [erg s$^{-1}$] ).  Column 4 is the 1$\sigma$ measurement in the covariance of the linear fit (see \S\ \ref{error on SFRs}).  Column 5 is the 1$\sigma$ scatter in the SFR relation for the galaxies used in the fit (using the median absolute deviation, MAD ($\sigma_\mathrm{MAD}$) to represent the scatter).}
\end{deluxetable}		

In every instance, the linear relations are consistent with a unity slope between the PAH luminosities and the SFR (within 2$\sigma$ of unity).  Again, using the relation from \citet{Kennicutt2009} given in Equation \ref{K09 SFR}, this yields the following calibrations for the individual PAH luminosities as SFR indicators,

\begin{equation}\label{unity single SFRs}
\begin{split}
\mathrm{log\ SFR\ (M}_{\odot} \mathrm{\ yr}^{-1}) = \\
(-41.73 \pm 0.08)& + \mathrm{log\ } \lpone\ (\mathrm{erg\ s}^{-1}) \\
\mathrm{log\ SFR\ (M}_{\odot} \mathrm{\ yr}^{-1}) = \\
(-42.37 \pm 0.05)& + \mathrm{log\ } \lptwo\ (\mathrm{erg\ s}^{-1}) \\
\mathrm{log\ SFR\ (M}_{\odot} \mathrm{\ yr}^{-1}) = \\
(-41.80 \pm 0.07)& + \mathrm{log\ } \lpthree\ (\mathrm{erg\ s}^{-1})
\end{split}
\end{equation}
and using the linear fits we find,
\begin{equation}
\begin{split}
\mathrm{log\ SFR(M}_{\odot} \mathrm{\ yr}^{-1}) = (-40.06 \pm 0.09)\ + \\
(0.96 \pm 0.04)\ \mathrm{log\ } \lpone\ & (\mathrm{erg\ s}^{-1}) \\
\mathrm{log\ SFR(M}_{\odot} \mathrm{\ yr}^{-1}) = (-42.38 \pm 0.06)\ + \\
(1.00 \pm 0.03)\ \mathrm{log\ } \lptwo\ & (\mathrm{erg\ s}^{-1}) \\
\mathrm{log\ SFR(M}_{\odot} \mathrm{\ yr}^{-1}) = (-44.14 \pm 0.08)\ + \\
(1.06 \pm 0.03)\ \mathrm{log\ } \lpthree\ & (\mathrm{erg\ s}^{-1})
\end{split}
\end{equation}

All of these SFR relations coefficients as well as the scatter for each can be found in Tables \ref{Unity SFRs} (unity relation defined as log SFR [\sfrunits] = C + log L$_{\mathrm{PAH}, \lambda}$ [erg s$^{-1}$] ) and \ref{Linear SFRs} (linear relation defined as log SFR [\sfrunits] =  A + B log L$_{\mathrm{PAH}, \lambda}$ [erg s$^{-1}$] ), which includes all possible PAH feature combinations.  We note there are two galaxies \citep[galaxies SSGSS 52 and 83 from the sample of][]{ODowd2009} for the 6.2\micron\ calibration that are well off the relation.  We attribute this to the ``problematic 6.2\micron\ feature fits" discussed in \citet{ODowd2009}.  These galaxies did not affect our calibration of the 6.2\micron\ feature, and the uncertainties decrease only slightly when excluding them from the PAH SFR relations.

\subsection{Uncertainties for Derived SFR Relations}
\label{error on SFRs}

We determined uncertainties for our PAH SFR relations using a robust linear fit to the 105 galaxies in the primary calibration sample that utilizes a least squares fitting routine.  This routine determines residuals for the fit of the luminosities and estimates a 1$\sigma$ uncertainty on the mean of the estimated values returned for the fit (slope and intercept for the linear fits).  We then propagate these uncertainties for the derived PAH SFR relations.  The slope and intercept are covariant and we report the covariance matrix for these fits in Table \ref{Linear SFRs}, where the covariance matrix has the following form.

\begin{equation} \label{covariance}
cov = \dbinom{\sigma_\mathrm{A}^2\ \sigma_\mathrm{AB}}{\sigma_\mathrm{AB}\ \sigma_\mathrm{B}^2}
\end{equation}
The fact that the off diagonal elements of the covariance matrix are negative means that A and B are not independent and are anti-correlated.  The diagonal elements are the variance on parameters A and B.

\subsection{Correction to PAH Luminosity for Low Metallicity Galaxies}
\label{metal corrections}

\begin{figure}[t!]	
\epsscale{1.1}
\plotone{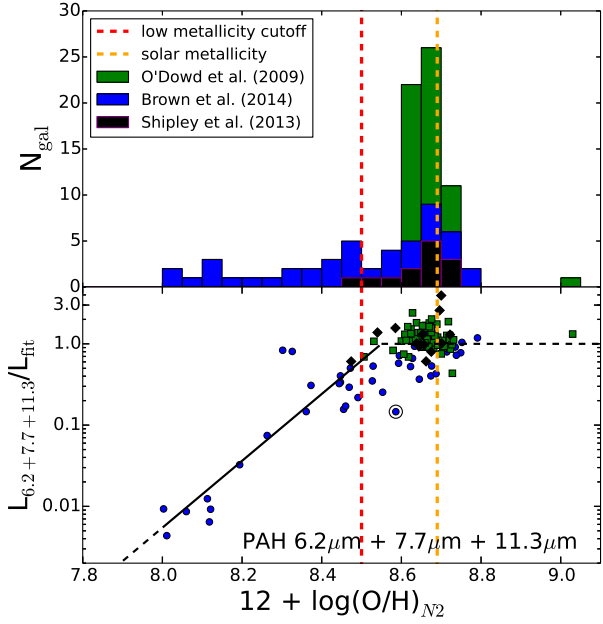}
\caption{The top panel shows the same metallicity distribution for the star-forming galaxies in the calibration sample as Figure \ref{metallicity}.  The bottom panel shows the \lpah /\lfit\ ratios for the star-forming galaxies as a function of oxygen abundance.  This shows how the low metallicity galaxies result in weaker observed PAH emission than expected from the correlation.  For the corrections to the PAH luminosity we used galaxies with \zmetal\ $\leq$ 8.55.  We determined a correction to the PAH luminosity by fitting a line (solid black line) for the galaxies below Z$_{0}$ (Z$_0 = \zsun - 8.55 = -0.14$) and extrapolations (dashed black lines) beyond the sample limit (low end) and the fit (high end, see \S\  \ref{metal corrections} for more details).  The blue point enclosed by a black circle denotes II Zw 096, see \S\  \ref{zwicky}.}
\label{metallicity residual}
\end{figure}

For our secondary calibration sample of 25 galaxies, we limited the oxygen abundance of the galaxies to \zmetal\ $\leq$ 8.55 as shown in Figure \ref{metallicity residual}.  This figure shows that galaxies at lower oxygen abundances have lower PAH luminosities compared to that expected from our fit between the total PAH luminosity and extinction-corrected \ha\ for the higher metallicity galaxies.  While the metallicity cut between the primary and secondary calibration samples is arbitrary, it defines the location where we start to observe a break in the \lpah/\lfit\ ratios in Figure \ref{metallicity residual}.  The strength of the PAH features is known to decline with decreasing metallicity \citep{Engelbracht2005,Calzetti2007}, which is likely driving the trend.

We fit the secondary calibration sample with a robust linear fit routine (same as the PAH SFR relations) as shown in Figure \ref{metallicity residual} for \zmetal\ $\leq$ 8.55.  We derive metallicity-dependent corrections to the PAH luminosity for each PAH combination as 
\begin{equation} \label{metal correct eq}
\mathrm{log}\ \mathrm{L}_{\mathrm{PAH}, \lambda}^{\mathrm{corr}} = \mathrm{log}\ \mathrm{L}_{\mathrm{PAH}, \lambda} - \mathrm{A[Z} - (\zsun\ + \mathrm{Z}_{0})]
\end{equation}
for \zmetal\ $\leq 8.55$.  We define Z$_0 = \zsun - 8.55 = -0.14$.  We do not have a sufficient number of low metallicity star-forming galaxies to determine if the fit between metallicity and L$_{\mathrm{PAH}, \lambda}$/\lfit\ is linear or more complex.  For galaxies with \zmetal\ $>$ 8.55, we force the ratio of \lpah /\lfit\ to be unity.  We need more super-solar metallicity galaxies to determine if this assumption is accurate.  The metallicity corrections for each combination of PAH features are listed in Table \ref{metallicity corrections}.

\begin{deluxetable}{cc}		
\tablecaption{Metallicity Corrections \vspace{-6pt} 
\label{metallicity corrections}}
\tablecolumns{2}
\tabletypesize{\small}
\tablewidth{0pc}
\setlength{\tabcolsep}{36pt}
\tablehead{
\colhead{PAH Feature(s)} & \colhead{A}
}
\startdata
6.2 + 7.7 + 11.3 & 4.1 $\pm$ 0.3 \\ 
6.2 & 4.0 $\pm$ 0.3 \\ 
7.7 & 4.0 $\pm$ 0.3 \\ 
11.3 & 3.7 $\pm$ 0.3 \\ 
6.2 + 7.7 & 4.1 $\pm$ 0.4 \\ 
6.2 + 11.3 & 3.8 $\pm$ 0.3 \\ 
7.7 + 11.3 & 4.1 $\pm$ 0.3 \\ 
\enddata
\vspace{-6pt}
\tablecomments{Column 1 defines which PAH features were used for the fit (see Figure \ref{metallicity residual}, bottom).  Column 2 is the linear relation coefficients for the metallicity corrections (defined as log L$_{\mathrm{PAH}, \lambda}^{\mathrm{corr}}$ = log L$_{\mathrm{PAH}, \lambda}$ - A[Z - (\zsun\ + Z$_{0}$)], where Z$_0$ = $-$0.14).  Note metallicity corrections are for galaxies with Z~$\leq$~\zsun~+~Z$_{0}$.}
\end{deluxetable}		

\subsection{The Interesting Galaxy:  II Zw 096}
\label{zwicky}

II Zw 096 is a high luminosity galaxy \citep[\lir\ $\sim 10^{12}$~\lsun,][]{Goldader1997,Inami2010,Haan2013} with \zmetal\ = 8.59, which 
implies a metallicity of 0.8~\zsun. It is currently in the process of a merger \citep{Inami2010,Brown2014}. The PAH luminosity is much lower than that expected from our SFR calibration, and this makes II Zw 096 an interesting case study. Our primary calibration sample includes other merging systems \citep[NGC 6090 and possibly SSGSS 18,][]{Brown2014,Battisti2015} that lie on the relations and thus the behavior of II Zw 096 is unique within our sample.

\citet{Inami2010} showed that the huge luminosity of this galaxy does not originate in its nucleus, but is from Region D (67\% of the 24 \micron\ flux)  well removed from the visible traces of the interacting galaxies. The diameter of this source is only a few hundred pc and its luminosity is $\sim 7 \times 10^{11}$~\lsun, so conditions within it are likely to resemble those in the nuclei of local ULIRGs, where it is 
well established that the PAH features are suppressed \citep[e.g.,][]{Rieke2009}.  Additional discussion of the unique characteristics of this object can be found in \citet{Inami2010}; see also Section \ref{SFR comparisons} below.  However, we conclude that the deviation from the PAH SFR calibration is not surprising, given these characteristics.  It nonetheless serves as a warning of rare conditions that can affect the accuracy of the derived SFRs.

\subsection{Photometric Calibrations of the 7.7\micron\ PAH Feature Using Broad-band Photometry}
\label{broadband photocalb}

\begin{figure*}[t!]	
\epsscale{1.1}
\plottwo{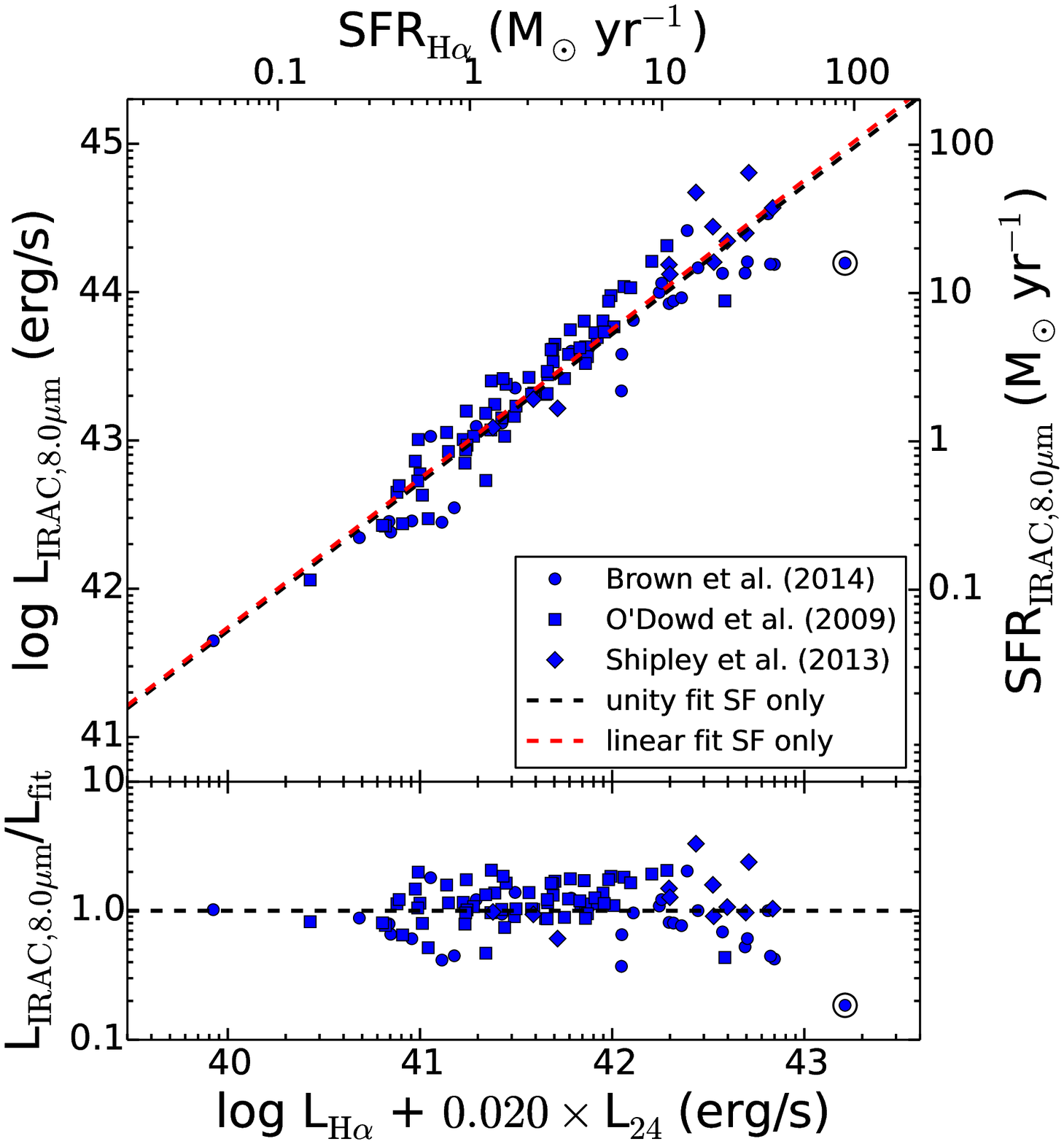}{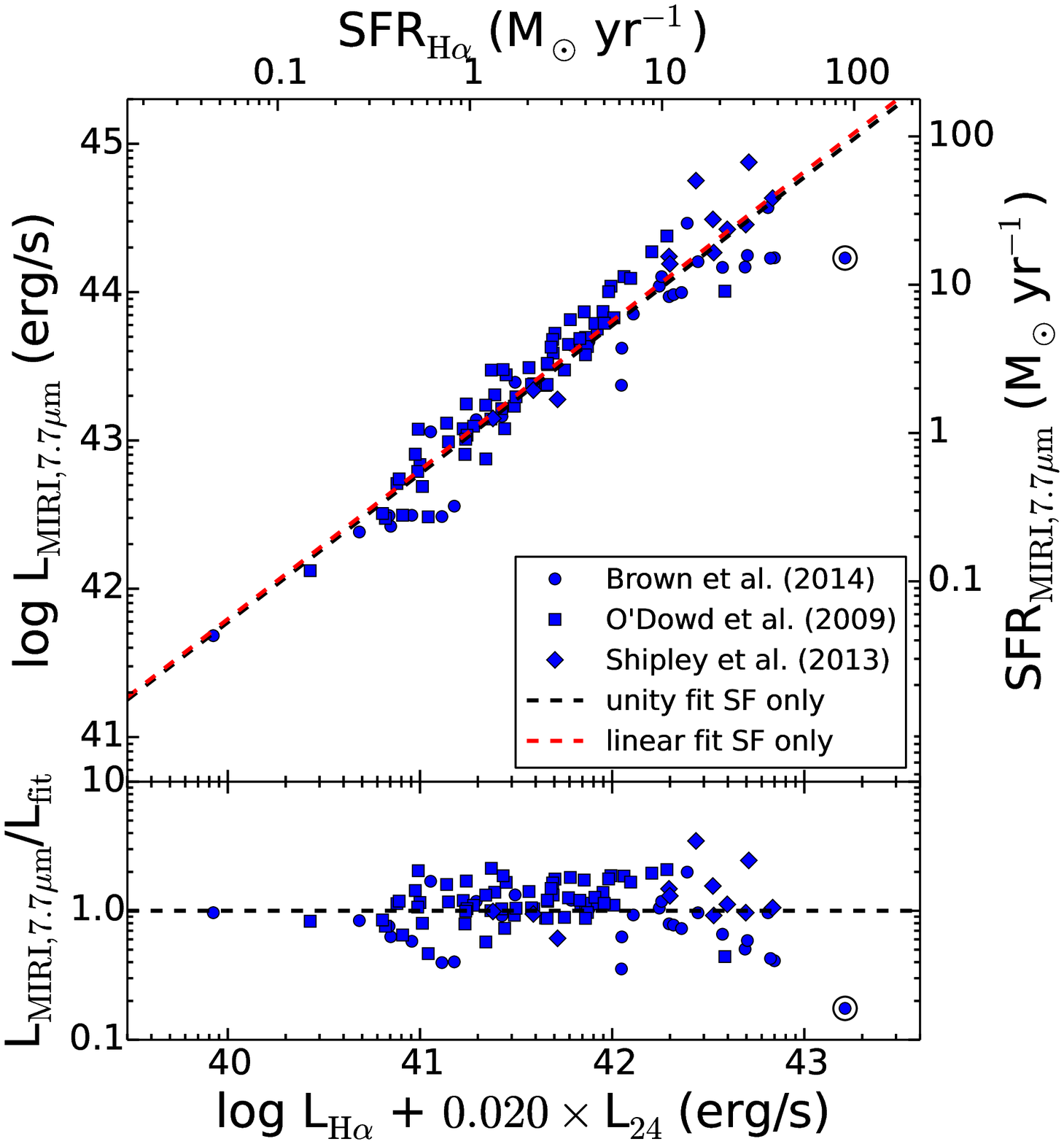}
\caption{Same as Figure \ref{ha24 pah onlySF} but for the broad-band photometric calibrations of the 7.7\micron\ PAH feature.  The \underline{Left} panel is the calibration for the (rest-frame) \spitzer/IRAC 8~\micron\ filter.  The \underline{Right} panel is the calibration for the (rest-frame) \jwst/MIRI 7.7~\micron\ filter.  The derived SFR relations from the two photometric calibrations are given in \S~\ref{broadband photocalb}.
}
\label{photometric calbs}
\end{figure*}

Many galaxies observed with \spitzer\ imaging do not have spectra, and \jwst \ faint imaging will also focus on objects too distant and faint for efficient spectroscopy.  We have therefore performed photometric calibrations assuming rest-frame \spitzer/IRAC 8\micron\ and \jwst/MIRI 7.7\micron\ filters, to determine the accuracy of (rest-frame) mid-IR photometry that covers the PAH features as  a possible SFR tracer.  The comparison of the two photometric bands provides a cross-calibration and also lets us assess systematic effects in the photometry. We estimated broadband flux densities for the filters from the rest-frame spectra of the primary calibration sample of star-forming galaxies.  To measure the rest-frame IRAC and MIRI flux densities, we followed the same procedure as for the 24~\micron\ broadband, rest-frame flux densities (see \S\ \ref{ha contribution}).  Then we followed the same procedure for the SFR calibrations of the PAH features (\S\ \ref{results}) to derive the photometric SFR relations.  Figure \ref{photometric calbs} shows the relation between the synthesized (rest-frame) IRAC 8 \micron\ and MIRI 7.7 \micron\ bands and the extinction-corrected \ha\ luminosity.  The relations are consistent with linear, where the derived relations between the extinction-corrected \ha\ luminosity and the \spitzer/IRAC 8~\micron\ band are
\begin{multline}
\mathrm{log\ SFR\ (M}_{\odot} \mathrm{\ yr}^{-1}) = \\
(-42.99 \pm 0.06) + \mathrm{log\ } \mathrm{L}_{8\micron} (\mathrm{erg\ s}^{-1})
\end{multline}
for the unity relation ($\sigma_\mathrm{MAD} = 0.166$) and
\begin{multline}\label{irac8 linear}
\mathrm{log\ SFR\ (M}_{\odot} \mathrm{\ yr}^{-1}) = (-42.79 \pm 0.07)\ + \\
(0.995 \pm 0.030)\ \mathrm{log\ } \mathrm{L}_{8\micron} (\mathrm{erg\ s}^{-1})
\end{multline}
for the linear relation ($\sigma_\mathrm{MAD} = 0.132$).  The derived relations between the extinction-corrected \ha\ luminosity and the \jwst/MIRI 7.7\micron\ band are
\begin{multline}
\mathrm{log\ SFR\ (M}_{\odot} \mathrm{\ yr}^{-1}) = \\
(-43.05 \pm 0.06) + \mathrm{log\ } \mathrm{L}_{7.7\micron} (\mathrm{erg\ s}^{-1})
\end{multline}
for the unity relation ($\sigma_\mathrm{MAD} = 0.156$) and
\begin{multline}
\mathrm{log\ SFR\ (M}_{\odot} \mathrm{\ yr}^{-1}) = (-42.84 \pm 0.07)\ + \\
(0.995 \pm 0.031)\ \mathrm{log\ } \mathrm{L}_{7.7\micron} (\mathrm{erg\ s}^{-1})
\end{multline}
for the linear relation ($\sigma_\mathrm{MAD} = 0.136$).  The relations for both photometric bands compare very well to the 7.7\micron\ PAH feature calibration with slopes of nearly one and illustrate the ability for mid-IR photometry in bands dominated by  PAHs to give accurate SFRs for samples of star-forming galaxies.

\citet{Yuan2011} show a tight relation between SFR and Akari 9 \micron\ photometry, where they determined the SFRs by fitting templates that simultaneously accounted for the optical-UV and reradiated IR. Their relation is very similar to ours in equation (\ref{irac8 linear}), i.e., converted to our units:
\begin{multline}
\mathrm{log\ SFR}\ (\sfrunits) = (- 42.61)\ + \\
(0.99 \pm 0.03)\ \mathrm{log}\ \mathrm{L}_{9 \micron} (\mathrm{erg\ s}^{-1})
\end{multline}
but with a $\sim 0.2$~dex offset.  They show that this offset is a systematic effect (their Figure 11), probably due to differences in the photometric bands, so the agreement is virtually perfect.

These SFR estimates will not be accurate where an AGN dominates the galaxy emission near 7.7~\micron. Three approaches can be used to eliminate this potential source of error.  First, mid-IR spectroscopy will identify AGN-dominated galaxies through the reduced equivalent widths of the PAH features. Second, such galaxies can be identified with ancillary data, such as deep X-ray imaging. Third, with multi-band mid-IR photometry it should be possible to identify AGN-dominated cases as demonstrated by \citet{Teplitz2011} using \spitzer\ observations. The latter approach is based on identification of AGN from the shape of their mid-IR spectral energy distributions \citep[e.g.,][]{Lacy2004,Lacy2007,Stern2005}.  For example, \citet{Donley2012} showed that AGN can be identified reliably as IRAC ``power-law" sources where AGN have mid-IR spectral behavior of $f_\nu > \lambda^{-2}$.  We discuss application of this to high-redshift galaxies in a later section (\S\ \ref{JWST photometry}).

\section{Previous PAH SFR Calibrations at Low Redshift}
\label{SFR comparisons}

We compare with previous calibrations for PAH features against various SFR indicators. We confine this section to galaxies at z $<$ 0.5. There are additional limitations. First, our calibration has been carried out for galaxies with SFRs up to $\sim$~100~\sfrunits. More-luminous low redshift galaxies show substantial deviations from proportionality between mid-IR and total IR luminosities \citep[e.g.,][]{Rigby2008, Marcillac2008, Rieke2009}, behavior that is not captured in our calibration.  Second,  if there are AGN in the calibration studies of other samples, their presence can affect the results by supplementing the signatures of star-formation or by destroying the PAH carriers. Third, measurement of the strengths of the PAH features depends critically on the treatment of the mid-IR continuum, which varies significantly from study to study \citep[e.g.,][]{HC2009, Treyer2010}.  All the studies under discussion start from the relation reported by \citet{Kennicutt1998}, where we correct to a Kroupa IMF \citep[using the factor of $0.7$ as in][]{Kennicutt2009} and for the larger luminosity obtained from the \spitzer\ data,
\begin{multline}
\mathrm{SFR}_\mathrm{Kroupa}\ (\sfrunits) = \\
2.54\times {10}^{-44} \mathrm{L}_{\mathrm{IR}, 8-1000\micron} (\mathrm{ergs}\thinspace \mathrm{s}^{-1})\thinspace.
\label{eq:eqn2}
\end{multline}

The previous calibrations of the PAH features as a SFR indicator discussed below illustrate these points.  However, given the diverse range of samples and the various methods used to derive previous PAH SFRs, it is remarkable the difference in the calibrations is relatively small (less than a factor of 2 for non-ULIRGs and even the ULIRG samples agree within a factor of a few).  This possibly indicates that the PAH luminosity as a SFR indicator is generally robust regardless of redshift, sample selection, or calibration method (except for very high luminosities).  This behavior points to its accuracy over a broad range of conditions.  Indeed, in \S~\ref{highz galaxies} we show tentative evidence that the PAH-derived SFRs for galaxies with ULIRG luminosities at $1 < z < 3$ agree with those derived from other SFR-tracers.  We expect our sample to provide the most robust calibration of the PAH features, at least over its luminosity range, because we have carefully selected our primary calibration sample to be star-forming galaxies with about solar metallicity over a broad range of luminosities ($10^9 - 10^{12}$~\lsun) and morphologies (from dwarf galaxies to large disk galaxies).

\subsection{\citet{Treyer2010}} 

\citet{Treyer2010} performed an analysis similar to \citet{Kennicutt2009} for various \ha\ to IR luminosity ratios, for galaxies at z $\sim$ 0.1. They found consistent ratios to those of \citet{Kennicutt2009}, especially for the 24~\micron\ luminosity and scatter and derived a new IR luminosity SFR relation \citep[see][for a detailed description of the derivation]{Treyer2010}.

An important test is whether our calibration is reproducible.  \citet{Treyer2010} use the SSGSS sample \citep{ODowd2009}, one of the 
three samples in our full calibration sample.  As in our analysis, they also used PAHFIT to measure PAH luminosities.  The two studies should therefore be in close agreement, but this is somewhat convoluted to demonstrate.  \citet{Treyer2010} give the ratio of the 7.7~\micron\ and 11.3~\micron\ feature strengths to the total IR luminosity, which they base on \spitzer\ photometry. They also define the integrated luminosity from 3 to 1100~\micron, which they describe as being 0.04 dex larger on average than the commonly used definition integrating from 8 to 1000~\micron.  However, \citet{Kennicutt2009} find that basing the \lir\ calculation on \spitzer/MIPS data results in 24\% larger values because the 160~\micron\ band captures cold emission that is not apparent in IRAS measurements (out to 100~\micron).

We equate Equation \ref{eq:eqn2} to those in Equation \ref{unity single SFRs} to obtain the relation between the PAH feature and total IR luminosity and correct to the $3 - 1100$~\micron\ luminosity, the resulting relations are
\begin{equation}
\mathrm{log}\ \lptwo = (-1.27 \pm 0.05) + \mathrm{log}\ \mathrm{L}_{3-1100 \micron}
\label{eq:eqn3} 
\end{equation}
\begin{equation}
\mathrm{log}\ \mathrm{L}_{11.3 \micron}= (-1.84 \pm 0.07) + \mathrm{log}\ \mathrm{L}_{3-1100 \micron}
\label{eq:eqn4} 
\end{equation}
\citet{Treyer2010} obtained normalization values of $-1.204 \pm 0.087$ and $-1.851 \pm 0.071$ in excellent agreement with the relations we derive.

\subsection{\citet{SW2009}}

\citet{SW2009} investigated the PAH luminosity as a SFR indicator for a sample of galaxies at z $<$ 0.5 by calibrating the flux at the peak of the 7.7\micron\ PAH feature against the total IR luminosity using the \citet{Kennicutt1998} IR SFR relation.  Their motivation for using the peak of the emission feature is that the SFR can still be estimated from the observed PAH emission even if the wavelength coverage is limited or the spectra have poor S/N, as might be the case for high redshift sources.  Using the flux at the peak of the PAH features, however, does introduce possible issues with the measurement of the continuum contributions (for example, contribution from an AGN).  \citet{SW2009} limited their sample to objects with EW(6.2\micron) $> 0.4$~\micron\ to exclude likely AGN and they measured a linear relation between the 7.7~\micron\ flux and SFR, with a constant of proportionality smaller by 0.2 dex (with an uncertainty of 0.2 dex) from the one we derive here.

\subsection{\citet{DR2012}}

\citet{DR2012} used SED templates \citep{Rieke2009} to derive a SFR for local galaxies from the 11.3\micron\ PAH feature, which showed evidence to be the most resistant to the influence of an AGN.  We can use our calibration to test their assumption that the 11.3\micron\ feature correctly represents the SFR.  This allows us to have confidence that for galaxies with an AGN (at least up to moderate luminosity AGN), we can accurately derive SFRs from the integrated luminosity of the PAH features, especially for the 11.3\micron\ feature.

From \citet{HK2007}, and taking a nominal ratio of [\ion{Ne}{2}] to [\ion{Ne}{3}] \citep{PS2010}, we find that the typical relation between [\ion{Ne}{2}] and SFR is 
\begin{equation}
\mathrm{log}\ \mathrm{L}_\mathrm{[Ne II]} = -2.88 + \mathrm{log}\ \mathrm{L}_{8-1000 \micron}
\label{eq:eqn5} 
\end{equation}
We put Equation \ref{eq:eqn4} on the same basis as Equation \ref{eq:eqn5} by correcting it to 
the IRAS luminosity parameters:
\begin{equation}
\mathrm{log}\ \mathrm{L}_{11.3 \micron} = -1.70 + \mathrm{log}\ \mathrm{L(IRAS)}_{8-1000 \micron}
\label{eq:eqn6} 
\end{equation}
Comparing, we obtain an estimate that the ratio of fluxes is f([\ion{Ne}{2}]/f(11.3 \micron) $\sim$ 0.07, with errors of perhaps a factor of two. \citet{Farrah2007} find a value of 0.17 for a sample of local ULIRGs, again with an error of a factor of about two. Their value may be overestimated (by a factor of about two) because they subtracted the continuum with a spline fit rather than using an approach similar to that of PAHFIT \citep[e.g.,][]{HC2009, Treyer2010}.  These values are comparable to that of f([\ion{Ne}{2}])/f(11.3 \micron) = 0.12 found around moderate luminosity AGNs by \citet{DR2010}.  They determined the PAH feature strength through PAHFIT, so their result should be on the same scale as our calibration. This value should be dominated by emission associated with star-formation, although a small contribution from the AGN may occur \citep{PS2010}.  In support of this conclusion, \citet{AH2014} find that the 11.3\micron\ PAH feature seems to persist to within $\sim$ 200 pc of a number of AGN.  Thus, our calibration of SFRs from PAH features reinforces the conclusion that the 11.3\micron\ feature is not significantly reduced around moderate-luminosity AGN and can be used as a robust indicator of the SFR in these regions.

\subsection{\citet{Pope2008}}

\citet{Pope2008} derived linear relations between PAH features (6.2\micron, 7.7\micron\ and 11.3\micron) and IR luminosity for a local sample of 22 galaxies with moderate luminosity IR galaxies \citep[$10^{10} < $~\lir/\lsun ~$10^{12}$;][]{Brandl2006}, and high redshift sub-millimeter galaxies (SMGs).  Within the errors, the relations are consistent with ours (except that they indicate a lower value for the flux ratio f[6.2 \micron]/f[7.7 \micron]).

\subsection{\citet{Farrah2007}}

\citet{Farrah2007} derived a PAH SFR from a sample of 53 ULIRGs at z $<$ 0.3, using the 6.2\micron\ and 11.3\micron\ PAH features, which they calibrate against the [\ion{Ne}{2}] $\lambda$12.81\micron\ and [\ion{Ne}{3}] $\lambda$15.55\micron\ emission lines. They focused on these features because they are relatively isolated spectrally and should be relatively easy to measure. They combined both features to mitigate variations in the strengths of individual PAH features between different starburst galaxies.  The calibration of the PAH features with the [\ion{Ne}{2}] $+$ [\ion{Ne}{3}] luminosity resulted in SFR (\sfrunits) = $1.18 \times 10^{-41}$\lpah\ (erg s$^{-1}$), and then we converted it to the Kroupa IMF. The resulting conversion factor is an order of magnitude larger than our calibration.  Part of this difference may arise from the use of spline fitting to remove the continuum, which has been shown to lead to underestimates of the PAH feature strengths by factors of $2 - 3$ \citep[e.g.,][]{HC2009, Treyer2010}, but much of it reflects that the PAH features in local ULIRGs tend to be suppressed.

\section{Application to Distant Galaxies: A Preview for \jwst}
\label{preview jwst}

IR SEDs evolve with redshift so comparisons must be over similar redshift ranges or must make the necessary adjustments \citep[e.g.,][]{HC2009, Papovich2009, Elbaz2010, Finkelstein2011, Kirkpatrick2012, Rujopakarn2013}. We first discuss previous works on this subject; in a following subsection we present results for our demonstration sample of gravitationally lensed, high-redshift galaxies as a case study for galaxies with luminosities and redshifts to be probed with \jwst.

\subsection{Previous studies}

\subsubsection{\citet{Pope2008,Pope2013}, \citet{Shi2009}, \citet{MD2009} \& \citet{Fiolet2010}}

From a sample of 13 SMGs, \citet{Pope2008} demonstrated that eight star-formation-dominated galaxies at z $\sim$ 2 (seven ULIRGs) show linear correlations between \lpah\ and \lir, very similar to those we have derived (the remaining SMGs showed evidence of AGN).  Additional calibrations of the PAH band strengths versus the IR luminosities of high redshift galaxies have been reported by \citet{Shi2009}, \citet{MD2009}, \citet{Fiolet2010} and \citet{Pope2013}.  Combining the 28 galaxies from the \citet{Pope2013} and \citet{Fiolet2010} samples, one can deduce that for $\lir \lsim 3 \times 10^{12}\ \lsun$ galaxies follow the relation between \lpah\ and \lir\ derived for our calibration sample.  However, at greater IR luminosities, the high-redshift samples show lower \lpah/\lir\ ratios.  The difference in this regard with low redshift ULIRGs, which fall substantially below the proportional relationship by 10$^{12}$~\lsun, is shown dramatically in \citet{Pope2013}.  The fact that the high redshift galaxies fit on the low redshift trend (albeit with significant scatter) up to such high luminosities supports other indications that at high redshifts the IR spectral properties even of highly luminous galaxies resemble those of moderate luminosity low redshift galaxies, rather than low redshift galaxies of similarly high luminosity \citep[e.g.,][]{Papovich2009,Muzzin2010,Hwang2010,Finkelstein2011}.  We discuss this behavior in more detail in the following section (\S\ \ref{summary highz}).

\subsubsection{\citet{HC2009}}

\citet{HC2009} derived a PAH SFR relation using the \lpah/\lir\ ratios of seven bright, starburst-dominated galaxies at $0.6 < z < 1.0$, using the \citet{Kennicutt1998} IR SFR calibration. They derive SFR calibrations for each PAH feature (6.2\micron, 7.7\micron\ and 11.3\micron). Compared to our results, their PAH SFR calibrations are larger by 0.15, 0.2, and 0.3 dex for the 6.2\micron, 7.7\micron\ and 11.3\micron\ PAH features, respectively.\footnote{The comparisons include a correction for the 11.3\micron\ PAH luminosity and the SFR, which is misreported in \citet{HC2009} and should be SFR (\sfrunits) = 1.52 $\times$ 10$^{-8}$\lpthree\ (\lsun).}  Their estimated PAH SFRs have uncertainties as large as factors of two \citep[see][for further discussion of issues that contribute to these uncertainties, as well as estimates from previous studies]{HC2009}.  It is difficult to evaluate their results further because they use an independent method to derive the PAH luminosities and measure the mid-IR continuum (and without a cross-calibration to PAHFIT it is not possible for a direct comparison to our results).  They also derived total IR luminosities largely extrapolated from mid-IR flux densities, which may suffer systematics, further complicating any direct comparison.  For the galaxies hosting quasars they find that the SFR estimated from the PAH features as a whole is $3-10$ times less than indicated by the far-IR luminosities, suggesting either the AGN contributes substantially to the far-IR and/or that the AGN destroys the PAH molecules.

\subsubsection{Rujopakarn et al. 2013}

\citet{Rujopakarn2013} focus on the ability of 24 \micron\ photometry to provide accurate estimates of SFRs out to z $\sim$ 3. Thus, at the larger redshifts, they are probing the performance of the 7.7 \micron\ PAH feature in this regard. Particularly for the stacked samples they find very good performance (e.g., their Figure 6) relative to their recommended far-IR SED.  That is, the performance is not significantly redshift-dependent indicating that there is a consistent SFR estimation regardless of the part of the spectrum being sampled from rest wavelengths of 6 to 24 \micron.  \citet{Wuyts2012} and \citet{Berta2013} have independently evaluated the prescription of \citet{Rujopakarn2013} against other star formation metrics and show very good agreement.

\subsection{SFRs of High Redshift Lensed Galaxies}
\label{highz galaxies}

To probe the performance of the PAH bands as SFR indicators at lower luminosities, we now discuss the derived SFRs for the demonstration sample of lensed high-redshift galaxies (\S~\ref{application sample}).  For a comparison, we use the \ha+24\micron\ and \paa\ fluxes from \citet{Rujopakarn2012} to estimate SFRs.  Although they report some measurements in \bra, the signal to noise and the potential for systematic errors in line extraction due to the small equivalent widths of this latter line make it less useful.  \citet{Kennicutt2009} demonstrate the effectiveness of \ha+24\micron\ as an extinction-free SFR indicator, while \paa\ is considered to be exceptionally accurate in this application because of the small extinction and the direct measure of the ionized gas. We have derived its strength relative to the shorter wavelenth H recombination lines assuming Case B recombination at electron temperature T$_e$ = 10,000K and density N$_e$ = 100 cm$^{-3}$ \citep{Osterbrock1989} resulting in
\begin{multline}
\mathrm{log\ SFR}\ (\sfrunits) = \\
(-40.33) + \mathrm{log\ L}_{\paa}\ (\mathrm{erg s}^{-1})
\end{multline}
We compare these determinations with the PAH-derived values using the PAH SFR calibration in Table~\ref{Unity SFRs}.

\begin{figure}[t!]	
\epsscale{1.1}
\plotone{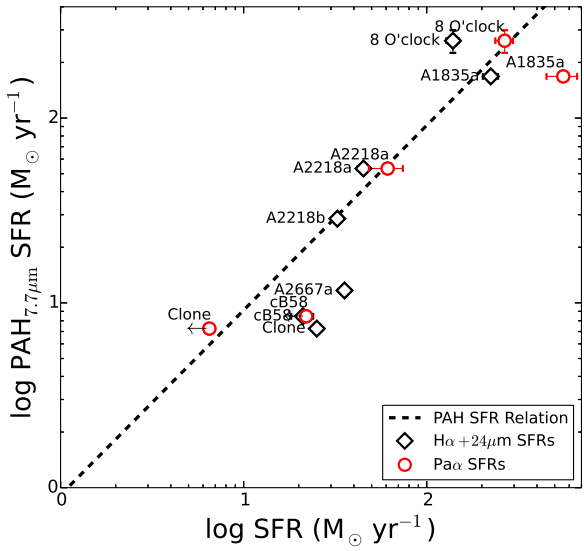}
\caption{The estimated SFRs from the PAH 7.7\micron\ in comparison to the estimated \paa\ (open red circles) and \ha+24\micron\ (open black diamonds) SFRs for the demonstration sample (some of the statistical uncertainties are smaller than the symbol size, primarily for the vertical axis; see \S\  \ref{highz galaxies} for discussion).  The line shown is the unity PAH 7.7\micron\ SFR relation from our primary calibration sample (\S\ \ref{SFR relations}) to the \paa\ and \ha+24\micron\ SFR relations.}
\label{lowz trend highz SFRs}
\end{figure}

The results are summarized in Table~\ref{high-z SFRs} and Figure~\ref{lowz trend highz SFRs} gives a comparison of the various SFR indicators.  We draw two main conclusions.  First, the SFRs derived from the various PAH bands are generally consistent, indicating no significant change in their relative strengths with redshift.  Second, the values derived from the PAH features are generally consistent with those from the other two indicators, typically within 0.3 dex (a factor of two).  Moreover, in all cases the PAH luminosities have smaller uncertainties than the \paa\ measurements.  This is because the PAH features are intrinsically brighter, and they will be among the brightest SFR indicators available for cosmologically distant galaxies.

For A1835a, the presence of an AGN seems likely given the  emission line ratio of log([\ion{N}{2}]/\ha) $\sim -0.25$ \citep{Rujopakarn2012}.  An AGN would contribute to the Pa$\alpha$ line flux without increasing the PAH fluxes, possibly contributing to the factor of about two higher SFR deduced from the line than from the PAH features.  The degree of scatter is not terribly surprising given the  scatter of each of the SFR indicators, plus potential issues such as hidden AGN and systematic errors such as slit losses in the spectra measuring the H recombination lines, and the uncertain corrections for that problem.

\begin{deluxetable*}{cccccccc}		
\tablecaption{Demonstration Sample SFRs \vspace{-6pt} 
\label{high-z SFRs}}
\tablecolumns{8}
\tabletypesize{\small}
\tablewidth{0pc}
\setlength{\tabcolsep}{3pt}
\tablehead{
\colhead{SFR Indicator} & \colhead{A2218b} & \colhead{A2667a} & \colhead{The Clone} & \colhead{A2218a} & \colhead{A1835a} & \colhead{cB58} & \colhead{8 O'clock}
}
\startdata
     &    z = 1.034    &    1.035   &    2.003   &   2.520   &     2.566     &     2.729     &     2.731     \\
\hline
Pa$\alpha$ & $\ldots$ & $\ldots$ & $<$ 6.5 & 61.2 $\pm$ 12.9 & 557. $\pm$ 105. & $<$ 22.0 & 266. $\pm$ 29.4 \\ 
H$\alpha$ + 24$\mu$m & 32.5 $\pm$ 1.3 & 35.5 $\pm$ 0.5 & 25.1 $\pm$ 1.5 & 45.1 $\pm$ 3.2 & 223. $\pm$ 21.2 & 21.1 $\pm$ 2.9 & 139. $\pm$ 10.9 \\ 
6.2 + 7.7 + 11.3 & 27.2 $\pm$ 0.5 & 10.8 $\pm$ 0.3 & 8.3 $\pm$ 0.3 & $\ldots$ & $\ldots$ & $\ldots$ & $\ldots$ \\ 
6.2 & 27.5 $\pm$ 0.3 & 8.1 $\pm$ 0.1 & 15.0 $\pm$ 0.6 & 68.6 $\pm$ 3.4 & 211. $\pm$ 39.4 & 11.1 $\pm$ 0.8 & 302. $\pm$ 20.0 \\ 
7.7 & 28.5 $\pm$ 0.7 & 11.7 $\pm$ 0.4 & 7.3 $\pm$ 0.2 & 53.3 $\pm$ 1.3 & 168. $\pm$ 10.3 & 8.5 $\pm$ 0.4 & 262. $\pm$ 36.7 \\ 
11.3 & 25.3 $\pm$ 1.0 & 11.4 $\pm$ 0.7 & 7.6 $\pm$ 1.2 & $\ldots$ & $\ldots$ & $\ldots$ & $\ldots$ \\ 
6.2 + 7.7 & 27.9 $\pm$ 0.6 & 10.8 $\pm$ 0.3 & 8.6 $\pm$ 0.2 & 55.3 $\pm$ 1.2 & 173. $\pm$ 11.0 & 8.8 $\pm$ 0.4 & 265. $\pm$ 29.6 \\ 
6.2 + 11.3 & 25.1 $\pm$ 0.5 & 9.4 $\pm$ 0.4 & 10.5 $\pm$ 0.7 & $\ldots$ & $\ldots$ & $\ldots$ & $\ldots$ \\ 
7.7 + 11.3 & 27.7 $\pm$ 0.6 & 11.5 $\pm$ 0.4 & 7.3 $\pm$ 0.3 & $\ldots$ & $\ldots$ & $\ldots$ & $\ldots$ \\ 
\enddata
\vspace{-6pt}
\tablecomments{Row 1 in the data gives the redshifts of the galaxies.  Column 1 denotes the SFR indicator that was used to derive the SFR for each galaxy (PAH SFRs were estimated using the unity relations given in Table \ref{Unity SFRs}).  Columns 2$-$8 are the estimated SFRs and uncertainties (at 1$\sigma$) of the flux measurements for each galaxy in the demonstration sample (see \S\ \ref{highz galaxies}).  All SFRs are in units of \sfrunits.}
\end{deluxetable*}		

\label{SFRs z1}

\subsection{The Validity of PAH-estimated SFRs for Very High Luminosity High-Redshift Galaxies}
\label{summary highz}

Locally, there is evidence for a trend of decreasing PAH luminosity with increasing total IR luminosity that appears to occur at an \lir~$\gsim 10^{12}$~\lsun.  This trend appears to be reduced for higher redshift galaxies \citep[e.g.,][]{Papovich2009,Muzzin2010,Finkelstein2011,Rujopakarn2011}.   The simplicity of applying observed frame 24-$\mu$m photometry to determining SFRs at high redshift depends on this trend being largely absent at high redshift, due to the vigorous star formation being more extended and of lower surface density at high redshift (Rujopakarn et al. 2013). We now evaluate the luminosity range over which this behavior can be assumed.  

\begin{figure}[t!]	
\epsscale{1.1}
\plotone{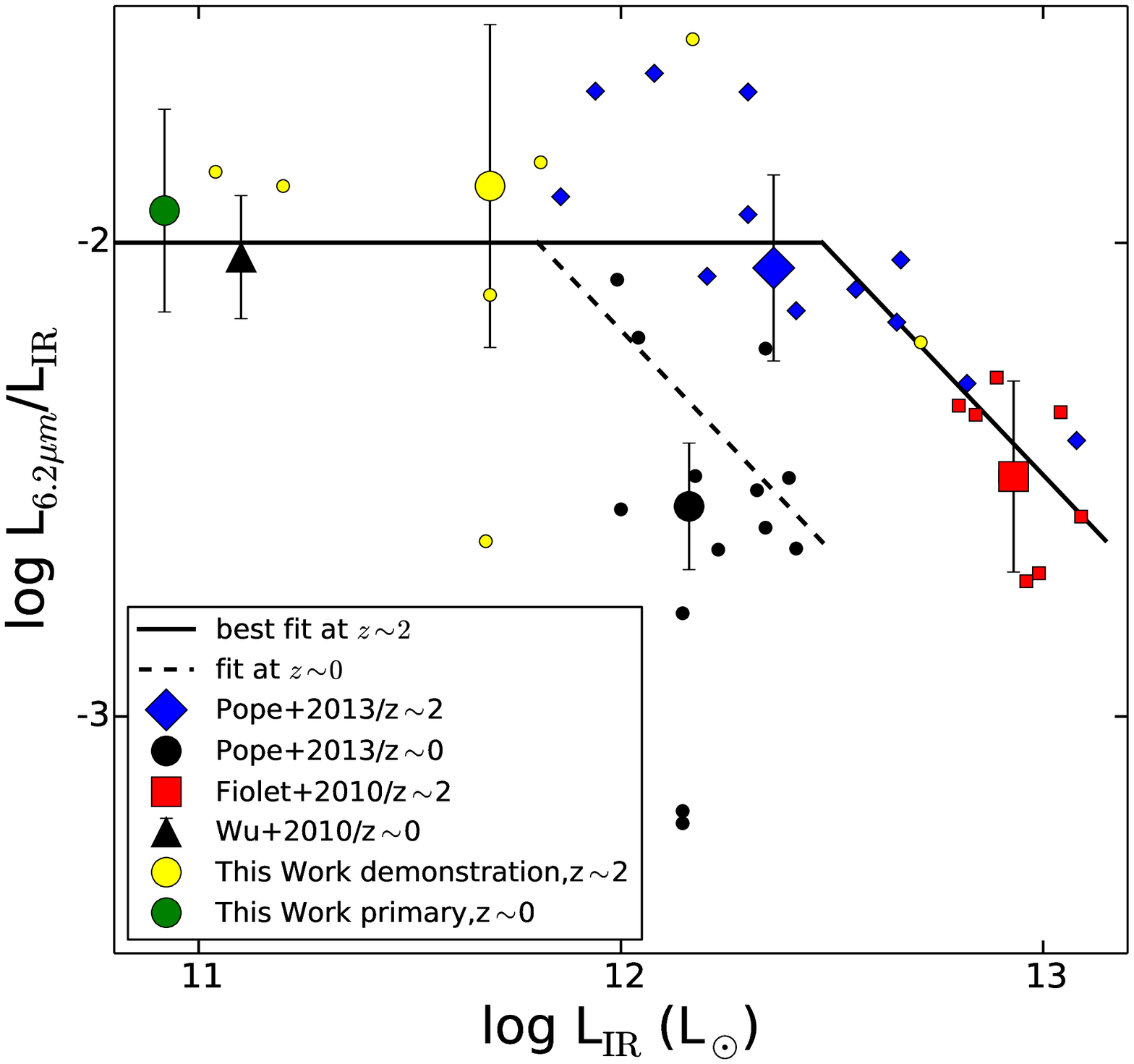}
\caption{Comparison of low-redshift LIRGs and ULIRGs samples to high-redshift (z $\sim$ 2) samples of LIRGs/ULIRGs.  The low-redshift samples are from \citet[black circles]{Pope2013}, \citet[black triangle; median of sample]{Wu2010} and the primary calibration sample (green circle, median of sample).  The high-redshift samples are from \citet[blue diamonds]{Pope2013}, \citet[red squares]{Fiolet2010} and the demonstration sample (yellow circles).  We give a linear fit to the high-redshift ULIRGs \citep{Pope2013,Fiolet2010}, where a trend for decreasing \lpone /\lir\ becomes evident above $\lir /\lsun \approx 3 \times 10^{12}$ (see \S\ \ref{summary highz} for fit parameters) and constant below this luminosity for $z > 1$.  We use the same slope for the low redshift ULIRGs for the trend seen at $z \sim 0$ for $\lir /\lsun \approx 10^{12}$.  Error bars are shown for the medians of the samples from the scatter in the samples ($\sigma_\mathrm{MAD}$).}
\label{summary plot}
\end{figure}

To study this trend at higher redshift, we compared the PAH-to-total IR luminosities using our primary calibration sample and several samples of low-redshift LIRGs and ULIRGs \citep{Pope2013,Wu2010} to our demonstration sample of lensed galaxies and other samples of high-redshift LIRGs and ULIRGs \citep[$z \sim 2$,][]{Pope2013,Fiolet2010}.  Figure \ref{summary plot} gives the results of this relation as the total IR luminosity to the ratio of the luminosity of the 6.2\micron\ PAH feature over the total IR luminosity.  We use the 6.2\micron\ feature as it is available for all samples considered here.  All the samples use PAHFIT as we have done in this work except for the \citet{Pope2013} sample, which has taken this into account by suggesting an increase of 1.7 to the PAH luminosities of their spline fitting method.

Figure \ref{summary plot} shows the medians and scatters ($\sigma_\mathrm{MAD}$) of the samples for the PAH-to-total IR luminosity ratios of high-redshift ULIRGs are more similar to low-redshift LIRGs rather than low-redshift ULIRGs, at least to \lir\ $\approx 3 \times 10^{12}$~\lsun.  This accounts for why the SFRs derived from the PAH features for the high-redshift lensed ULIRGs are consistent with the SFRs derived from \paa\ and \ha+24\micron.  It also indicates that the low-redshift SFR calibration based on PAH luminosities is appropriate to use for high-redshift ULIRGs, even if it is not appropriate for low-redshift ULIRGs, consistent with other recent studies that have demonstrated this result \citep[i.e.][]{Rujopakarn2013}.  However, at even higher total IR luminosities ($\lir/\lsun \gsim 3 \times 10^{12}$), there is evidence that the \lpah/\lir\ ratio declines for the high-redshift samples \citep[e.g.,][]{Fiolet2010}.  We fit this trend seen in Figure \ref{summary plot} as a broken power-law and derive a linear fit above this luminosity using the high redshift samples of \citet{Pope2013} and \citet{Fiolet2010}, which yields 
\begin{multline}
\mathrm{log}\ \lpone/\lir = (9.7 \pm 0.4)\ + \\
(-0.94 \pm 0.24)\ \mathrm{log}\ \lir\ (\lsun)
\end{multline}
for $\lir > 3 \times 10^{12}$ \lsun.
The constant value below this luminosity is log $\lpone /\lir = -2.0 \pm 0.2$.  We use the same slope for the low redshift trend seen at an $\lir \approx 10^{12}\ \lsun$.

\label{SFRs z2}

\subsection{Use of JWST Photometry to Determine SFRs at High Redshift}
\label{JWST photometry}

We conclude that our PAH SFR calibration is accurate for high-redshift ($1 < z < 3$) and for galaxies with SFRs up to $\lsim 300 \sfrunits$ (corresponding to IR luminosities $\lsim 3 \times 10^{12} \lsun$).  The calibration we derive in this paper for galaxies at $z < 0.4$ can be refined quickly with observations from \jwst/MIRI, which will be sensitive to the PAH emission from distant galaxies out to $z \lsim 2$.  Therefore, our calibration will be very useful for studies with \jwst\ of galaxies with dust embedded star-formation over a wide range of SFRs, and over the peak of the SFR density in the Universe \citep{MD2014}.

In cases where mid-IR spectroscopy is available, it will be possible to measure SFRs from PAH features directly.  However, \jwst\ mid-IR spectroscopy will be relatively inefficient as MIRI can observe only one object for each observation.  Efficient SFRs for large numbers of galaxies should be possible from broad-band imaging with MIRI provided the samples can exclude AGN (whose emission will contribute to the mid-IR emission).  Three approaches can be used to eliminate this potential source of error as discussed earlier (\S\ \ref{broadband photocalb}).  However, earlier work by \citet{Donley2008} showed that identification of power-law AGN with IRAC data becomes unreliable by $z \ge 1.75$ because the continua of stellar populations at the relevant rest wavelengths start to mimic power laws \citep[see][their Figure 5]{Donley2008}.  This issue can be avoided by applying the power law method at longer wavelengths. Multi-wavelength observations with \jwst\ can extend the power-law method so it can be applied at all relevant redshifts. One simple way to accomplish this would be to image at 3.56~\micron\ (NIRCam) and with MIRI at 5.6~\micron, 10~\micron, 15~\micron, and 21~\micron\ or a subset of these bands\footnote{We select these bands because they start at the wavelength of the channel 1 IRAC band and are spaced by about a factor of 1.5 in wavelength.}.  To get to equal depth in four shorter-wavelength bands as at 21~\micron\ relative to a power law with an index of $-2$ takes very little extra time (only about 40\% more).  This provides a method completely internal to \jwst\ to identify any possible AGN interlopers that might be contributing too much at rest 8~\micron\ to give valid SFR estimates.

\jwst\ data can also address the issue of whether to apply the local calibration for centrally-concentrated ULIRGs, or whether to apply the calibration for more extended star formation that is dominant at high redshift.  It is unlikely that there is an abrupt transition between the two cases.  Instead, there is probably a mixture of behaviors with the proportion of centrally concentrated cases decreasing with increasing redshift.  This transition appears to be virtually complete by $z = 1$ \citep{Rujopakarn2013}.  Below this redshift, \jwst\ photometry can help select the appropriate template through comparing the behavior of the output of the PAH bands, which are dominant for $6 - 13$ \micron, with that of the dust grains that dominate the emission at $13 - 30$ \micron.  For example, the MIRI photometric bands at 12.8 \micron\ and 21 \micron\ sample these two spectral components separately for $0 < z < 0.6$.  Their relative behavior can be expressed by the ratio of the flux densities in the two bands, i.e., f(21 \micron)/f(12.8 \micron).  To test the behavior of this ratio, we used the log \lir/\lsun\ $= 11.25$ and log \lir/\lsun\ $= 12.25$ templates from \citet{Rieke2009}; the former is representative of the SEDs of high redshift ULIRGs, while the latter is for low redshift ones \citep{Rujopakarn2012}.  We found that the ratio (high redshift over low redshift) is a factor of 1.5 or more greater for the low redshift ULIRG template than for the high-redshift case.  This approach can be extended to $z \sim 1$ by using the 15 \micron\ and 25.5 \micron\ bands in an analogous manner.  Although the sensitivity of the latter band is reduced by the thermal emission of the telescope, an integration time of only an hour would provide sufficient signal to noise (S/N $>$ 20) for $z=1$ LIRGs.  Below the LIRG luminosity range, the evolution in IR SEDs is no longer a significant issue in estimating SFRs.

\section{CONCLUSIONS}
\label{conclusions}

We have studied a sample of 105 star-forming galaxies covering a range of total IR luminosity, \lir\ = L(8-1000~\micron) = $10^9 - 10^{12}$ \lsun\ and redshift $0 < z < 0.4$ to calibrate the luminosity of the PAH features at 6.2\micron\ $-$ 11.3\micron\ as SFR indicators.  The total PAH luminosity (6.2\micron\ + 7.7\micron\ + 11.3\micron\ features) correlates linearly with the SFR as measured by the extinction-corrected \ha\ luminosity \citep[using the sum of the observed \ha\ emission and corresponding 24~\micron\ luminosity as established by][]{Kennicutt2009}, with a tight scatter of 0.14 dex.  The scatter is similar to those seen in comparing other accurate indicators of SFRs, showing that the PAH-derived values are also accurate.

We have provided a new robust SFR calibration using the luminosity emitted from the PAHs at 6.2\micron, 7.7\micron\ and 11.3\micron\ and all possible combinations.  The scatter is smallest, $\sigma_{\mathrm{MAD}}$ = 0.14 dex (i.e., 40\%) whenever the 7.7\micron\ feature is involved in the fit.  This implies this feature may provide the most robust measure of the SFR.  The relation against the 6.2\micron\ feature has the highest scatter, $\sigma_{\mathrm{MAD}}$ = 0.21 dex (i.e., 60\%), which may indicate more variation in galaxies between the luminosity in this feature and the SFR.  For low redshift galaxies ($z < 0.5$), this calibration is valid only up to the high LIRG/ULIRG luminosity range, where IR luminosities at and above this range underestimates the SFR.  We have also demonstrated that in the case of star-forming galaxies, the rest-frame broad-band 7.7~\micron\ emission should also be a robust SFR tracer, if samples of AGN can be excluded.

The PAH SFR calibration shows a dependency on galaxy gas-phase metallicity, where the PAH luminosity of galaxies with Z $\lsim 0.7$~\zsun\ departs from the linear SFR relationship.  We have calibrated a correction to the PAH SFRs using a simple empirical model that depends linearly on metallicity for a sample of 25 galaxies.  Larger samples of low metallicity galaxies are necessary to improve the accuracy of these fits on the dependence of gas-phase metallicity and reduce the scatter seen in the metallicity relations.

We have presented a case study for observations from \jwst\ that will be capable of measuring the PAH luminosities in galaxies to z $\lsim 2$.  Because the PAH features are so bright, our PAH SFR calibration enables an efficient way to measure SFRs in distant galaxies with \jwst\ to SFRs as low as $\sim$10 \sfrunits\ for galaxies that dominate the peak of the SFR density evolution.  We used \spitzer/IRS observations of PAH features in seven lensed star-forming galaxies at $1 < z < 3$ to demonstrate the utility of the PAHs to derive SFRs as accurate as those available from other indicators.  We demonstrate that the SFRs from the PAHs are consistent with those derived from \ha+24\micron\ (and from \paa\ when it is available).

The PAH features may provide the most accurate SFR measurements for distant galaxies, z~$\lsim 2$, with \jwst/MIRI spectroscopy and photometry with much higher signal-to-noise ratios than other SFR indicators.  This is because the PAH luminosities are brighter than most other SFR indicators (\paa\ and \bra) with S/N higher by factors $> 5-10$ in comparable exposure times.  For this reason, the PAH features will be the brightest SFR indicators in high redshift galaxies, and the calibration we provide will have a large utility for studies with \jwst\ of galaxies in the distant universe.  This new SFR indicator will be useful for probing the peak of the SFR density in the universe ($z \lsim 2$) and because PAHs trace star-formation even in galaxies with AGN, this new SFR indicator allows for studies of the co-evolution of star-formation and supermassive blackhole accretion contemporaneously in a galaxy.

\acknowledgements

We thank our colleagues on the NDWFS and AGES teams.  We thank Daniela Calzetti, Darren DePoy, Robert Kennicutt, Jr., Guilane Lagache, Alain Omont, Alexandra Pope, Nicholas Suntzeff and Marie Treyer for comments that helped improve the manuscript.  Support for this work was provided by the NASA Astrophysics Data Analysis Program (ADAP) through grant NNX15AF11G.  Further support for this work was provided to the authors by the George P.\ and Cynthia Woods Mitchell Institute for Fundamental Physics and Astronomy.  This work is based in part on observations and archival data obtained with the \textit{Spitzer Space Telescope}, which is operated by the Jet Propulsion Laboratory, California Institute of Technology under a contract with NASA.  Partial support for this work was provided by NASA through awards 1255094 and 1365085 issued by JPL/Caltech.  This work utilized the PAHFIT IDL tool for decomposing IRS spectra, which J. D. Smith has generously made publicly available \citep{Smith2007}.  This work made use of images and/or data products provided by the NOAO Deep Wide-Field Survey \citep{Jannuzi1999, Jannuzi2004, Dey2004}, which is supported by the National Optical Astronomy Observatory (NOAO). NOAO is operated by AURA, Inc., under a cooperative agreement with the National Science Foundation.

Funding for the SDSS has been provided by the Alfred P. Sloan Foundation, the Participating Institutions, NASA, NSF, the U.S. Department of Energy, the Japanese Monbukagakusho, the Max Planck Society, and the Higher Education Funding Council for England. The SDSS Web site is http://www.sdss.org/. The SDSS is managed by the Astrophysical Research Consortium for the Participating Institutions. The Participating Institutions are the American Museum of Natural History, Astrophysical Institute Potsdam, University of Basel, University of Cambridge, Case Western Reserve University, University of Chicago, Drexel University, Fermilab, the Institute for Advanced Study, the Japan Participation Group, Johns Hopkins University, the Joint Institute for Nuclear Astrophysics, the Kavli Institute for Particle Astrophysics and Cosmology, the Korean Scientist Group, the Chinese Academy of Sciences (LAMOST), Los Alamos National Laboratory, the Max-Planck- Institute for Astronomy (MPIA), the Max-Planck-Institute for Astrophysics (MPA), New Mexico State University, Ohio State University, University of Pittsburgh, University of Portsmouth, Princeton University, the United States Naval Observatory, and the University of Washington.

\bibliography{myrefs}

\begin{thebibliography}{90}
\expandafter\ifx\csname natexlab\endcsname\relax\def\natexlab#1{#1}\fi

\bibitem[{{Allam} {et~al.}(2007){Allam}, {Tucker}, {Lin}, {Diehl}, {Annis},
  {Buckley-Geer}, \& {Frieman}}]{Allam2007}
{Allam}, S.~S., {Tucker}, D.~L., {Lin}, H., {et~al.} 2007, \apjl, 662, L51

\bibitem[{{Alonso-Herrero} {et~al.}(2014){Alonso-Herrero}, {Ramos Almeida},
  {Esquej}, {Roche}, {Hern{\'a}n-Caballero}, {H{\"o}nig},
  {Gonz{\'a}lez-Mart{\'{\i}}n}, {Aretxaga}, {Mason}, {Packham}, {Levenson},
  {Rodr{\'{\i}}guez Espinosa}, {Siebenmorgen}, {Pereira-Santaella},
  {D{\'{\i}}az-Santos}, {Colina}, {Alvarez}, \& {Telesco}}]{AH2014}
{Alonso-Herrero}, A., {Ramos Almeida}, C., {Esquej}, P., {et~al.} 2014, \mnras,
  443, 2766

\bibitem[{{Asplund} {et~al.}(2009){Asplund}, {Grevesse}, {Sauval}, \&
  {Scott}}]{Asplund2009}
{Asplund}, M., {Grevesse}, N., {Sauval}, A.~J., {et~al.} 2009, \araa, 47, 481

\bibitem[{{Battisti} {et~al.}(2015){Battisti}, {Calzetti}, {Johnson}, \&
  {Elbaz}}]{Battisti2015}
{Battisti}, A.~J., {Calzetti}, D., {Johnson}, B.~D., {et~al.} 2015, \apj, 800,
  143

\bibitem[{{Beers} {et~al.}(1990){Beers}, {Flynn}, \& {Gebhardt}}]{Beers1990}
{Beers}, T.~C., {Flynn}, K., \& {Gebhardt}, K. 1990, \aj, 100, 32

\bibitem[{{Bendo} {et~al.}(2008){Bendo}, {Draine}, {Engelbracht}, {Helou},
  {Thornley}, {Bot}, {Buckalew}, {Calzetti}, {Dale}, {Hollenbach}, {Li}, \&
  {Moustakas}}]{Bendo2008}
{Bendo}, G.~J., {Draine}, B.~T., {Engelbracht}, C.~W., {et~al.} 2008, \mnras,
  389, 629

\bibitem[{{Berta} {et~al.}(2013){Berta}, {Lutz}, {Santini}, {Wuyts}, {Rosario},
  {Brisbin}, {Cooray}, {Franceschini}, {Gruppioni}, {Hatziminaoglou}, {Hwang},
  {Le Floc'h}, {Magnelli}, {Nordon}, {Oliver}, {Page}, {Popesso}, {Pozzetti},
  {Pozzi}, {Riguccini}, {Rodighiero}, {Roseboom}, {Scott}, {Symeonidis},
  {Valtchanov}, {Viero}, \& {Wang}}]{Berta2013}
{Berta}, S., {Lutz}, D., {Santini}, P., {et~al.} 2013, \aap, 551, A100

\bibitem[{{Brandl} {et~al.}(2006){Brandl}, {Bernard-Salas}, {Spoon}, {Devost},
  {Sloan}, {Guilles}, {Wu}, {Houck}, {Weedman}, {Armus}, {Appleton}, {Soifer},
  {Charmandaris}, {Hao}, {Higdon}, {Marshall}, \& {Herter}}]{Brandl2006}
{Brandl}, B.~R., {Bernard-Salas}, J., {Spoon}, H.~W.~W., {et~al.} 2006, \apj,
  653, 1129

\bibitem[{{Brinchmann} {et~al.}(2004){Brinchmann}, {Charlot}, {White},
  {Tremonti}, {Kauffmann}, {Heckman}, \& {Brinkmann}}]{Brinchmann2004}
{Brinchmann}, J., {Charlot}, S., {White}, S.~D.~M., {et~al.} 2004, \mnras, 351,
  1151

\bibitem[{{Brown} {et~al.}(2014){Brown}, {Moustakas}, {Smith}, {da Cunha},
  {Jarrett}, {Imanishi}, {Armus}, {Brandl}, \& {Peek}}]{Brown2014}
{Brown}, M.~J.~I., {Moustakas}, J., {Smith}, J.-D.~T., {et~al.} 2014, \apjs,
  212, 18

\bibitem[{{Calzetti} {et~al.}(2007){Calzetti}, {Kennicutt}, {Engelbracht},
  {Leitherer}, {Draine}, {Kewley}, {Moustakas}, {Sosey}, {Dale}, {Gordon},
  {Helou}, {Hollenbach}, {Armus}, {Bendo}, {Bot}, {Buckalew}, {Jarrett}, {Li},
  {Meyer}, {Murphy}, {Prescott}, {Regan}, {Rieke}, {Roussel}, {Sheth}, {Smith},
  {Thornley}, \& {Walter}}]{Calzetti2007}
{Calzetti}, D., {Kennicutt}, R.~C., {Engelbracht}, C.~W., {et~al.} 2007, \apj,
  666, 870

\bibitem[{{Chabrier}(2003)}]{Chabrier2003}
{Chabrier}, G. 2003, \pasp, 115, 763

\bibitem[{{Chary} \& {Elbaz}(2001)}]{CE2001}
{Chary}, R. \& {Elbaz}, D. 2001, \apj, 556, 562

\bibitem[{{Dey} {et~al.}(2004){Dey}, {Jannuzi}, {Brown}, {Tiede}, {Stern},
  {Dawson}, {Spinrad}, \& {NDWFS Team}}]{Dey2004}
{Dey}, A., {Jannuzi}, B.~T., {Brown}, M.~J.~I., {et~al.} 2004, in Bulletin of
  the American Astronomical Society, Vol.~36, American Astronomical Society
  Meeting Abstracts \#204, 746

\bibitem[{{Diamond-Stanic} \& {Rieke}(2010)}]{DR2010}
{Diamond-Stanic}, A.~M. \& {Rieke}, G.~H. 2010, \apj, 724, 140

\bibitem[{{Diamond-Stanic} \& {Rieke}(2012)}]{DR2012}
---. 2012, \apj, 746, 168

\bibitem[{{Donley} {et~al.}(2012){Donley}, {Koekemoer}, {Brusa}, {Capak},
  {Cardamone}, {Civano}, {Ilbert}, {Impey}, {Kartaltepe}, {Miyaji}, {Salvato},
  {Sanders}, {Trump}, \& {Zamorani}}]{Donley2012}
{Donley}, J.~L., {Koekemoer}, A.~M., {Brusa}, M., {et~al.} 2012, \apj, 748, 142

\bibitem[{{Donley} {et~al.}(2008){Donley}, {Rieke}, {P{\'e}rez-Gonz{\'a}lez},
  \& {Barro}}]{Donley2008}
{Donley}, J.~L., {Rieke}, G.~H., {P{\'e}rez-Gonz{\'a}lez}, P.~G., {et~al.}
  2008, \apj, 687, 111

\bibitem[{{Elbaz} {et~al.}(2011){Elbaz}, {Dickinson}, {Hwang},
  {D{\'{\i}}az-Santos}, {Magdis}, {Magnelli}, {Le Borgne}, {Galliano},
  {Pannella}, {Chanial}, {Armus}, {Charmandaris}, {Daddi}, {Aussel}, {Popesso},
  {Kartaltepe}, {Altieri}, {Valtchanov}, {Coia}, {Dannerbauer}, {Dasyra},
  {Leiton}, {Mazzarella}, {Alexander}, {Buat}, {Burgarella}, {Chary}, {Gilli},
  {Ivison}, {Juneau}, {Le Floc'h}, {Lutz}, {Morrison}, {Mullaney}, {Murphy},
  {Pope}, {Scott}, {Brodwin}, {Calzetti}, {Cesarsky}, {Charlot}, {Dole},
  {Eisenhardt}, {Ferguson}, {F{\"o}rster Schreiber}, {Frayer}, {Giavalisco},
  {Huynh}, {Koekemoer}, {Papovich}, {Reddy}, {Surace}, {Teplitz}, {Yun}, \&
  {Wilson}}]{Elbaz2011}
{Elbaz}, D., {Dickinson}, M., {Hwang}, H.~S., {et~al.} 2011, \aap, 533, A119

\bibitem[{{Elbaz} {et~al.}(2010){Elbaz}, {Hwang}, {Magnelli}, {Daddi},
  {Aussel}, {Altieri}, {Amblard}, {Andreani}, {Arumugam}, {Auld}, {Babbedge},
  {Berta}, {Blain}, {Bock}, {Bongiovanni}, {Boselli}, {Buat}, {Burgarella},
  {Castro-Rodriguez}, {Cava}, {Cepa}, {Chanial}, {Chary}, {Cimatti},
  {Clements}, {Conley}, {Conversi}, {Cooray}, {Dickinson}, {Dominguez},
  {Dowell}, {Dunlop}, {Dwek}, {Eales}, {Farrah}, {F{\"o}rster Schreiber},
  {Fox}, {Franceschini}, {Gear}, {Genzel}, {Glenn}, {Griffin}, {Gruppioni},
  {Halpern}, {Hatziminaoglou}, {Ibar}, {Isaak}, {Ivison}, {Lagache}, {Le
  Borgne}, {Le Floc'h}, {Levenson}, {Lu}, {Lutz}, {Madden}, {Maffei}, {Magdis},
  {Mainetti}, {Maiolino}, {Marchetti}, {Mortier}, {Nguyen}, {Nordon},
  {O'Halloran}, {Okumura}, {Oliver}, {Omont}, {Page}, {Panuzzo},
  {Papageorgiou}, {Pearson}, {Perez Fournon}, {P{\'e}rez Garc{\'{\i}}a},
  {Poglitsch}, {Pohlen}, {Popesso}, {Pozzi}, {Rawlings}, {Rigopoulou},
  {Riguccini}, {Rizzo}, {Rodighiero}, {Roseboom}, {Rowan-Robinson},
  {Saintonge}, {Sanchez Portal}, {Santini}, {Sauvage}, {Schulz}, {Scott},
  {Seymour}, {Shao}, {Shupe}, {Smith}, {Stevens}, {Sturm}, {Symeonidis},
  {Tacconi}, {Trichas}, {Tugwell}, {Vaccari}, {Valtchanov}, {Vieira},
  {Vigroux}, {Wang}, {Ward}, {Wright}, {Xu}, \& {Zemcov}}]{Elbaz2010}
{Elbaz}, D., {Hwang}, H.~S., {Magnelli}, B., {et~al.} 2010, \aap, 518, L29

\bibitem[{{Engelbracht} {et~al.}(2005){Engelbracht}, {Gordon}, {Rieke},
  {Werner}, {Dale}, \& {Latter}}]{Engelbracht2005}
{Engelbracht}, C.~W., {Gordon}, K.~D., {Rieke}, G.~H., {et~al.} 2005, \apjl,
  628, L29

\bibitem[{{Farrah} {et~al.}(2007){Farrah}, {Bernard-Salas}, {Spoon}, {Soifer},
  {Armus}, {Brandl}, {Charmandaris}, {Desai}, {Higdon}, {Devost}, \&
  {Houck}}]{Farrah2007}
{Farrah}, D., {Bernard-Salas}, J., {Spoon}, H.~W.~W., {et~al.} 2007, \apj, 667,
  149

\bibitem[{{Finkelstein} {et~al.}(2011){Finkelstein}, {Papovich}, {Finkelstein},
  {Willmer}, {Rigby}, {Rudnick}, {Egami}, {Rieke}, \&
  {Smith}}]{Finkelstein2011}
{Finkelstein}, K.~D., {Papovich}, C., {Finkelstein}, S.~L., {et~al.} 2011,
  \apj, 742, 108

\bibitem[{{Finkelstein} {et~al.}(2009){Finkelstein}, {Papovich}, {Rudnick},
  {Egami}, {Le Floc'h}, {Rieke}, {Rigby}, \& {Willmer}}]{Finkelstein2009}
{Finkelstein}, S.~L., {Papovich}, C., {Rudnick}, G., {et~al.} 2009, \apj, 700,
  376

\bibitem[{{Fiolet} {et~al.}(2010){Fiolet}, {Omont}, {Lagache}, {Bertincourt},
  {Fadda}, {Baker}, {Beelen}, {Berta}, {Boulanger}, {Farrah}, {Kov{\'a}cs},
  {Lonsdale}, {Owen}, {Polletta}, {Shupe}, \& {Yan}}]{Fiolet2010}
{Fiolet}, N., {Omont}, A., {Lagache}, G., {et~al.} 2010, \aap, 524, A33

\bibitem[{{Fu} \& {Stockton}(2009)}]{FS2009}
{Fu}, H. \& {Stockton}, A. 2009, \apj, 696, 1693

\bibitem[{{Fumagalli} {et~al.}(2014){Fumagalli}, {Labb{\'e}}, {Patel}, {Franx},
  {van Dokkum}, {Brammer}, {da Cunha}, {F{\"o}rster Schreiber}, {Kriek},
  {Quadri}, {Rix}, {Wake}, {Whitaker}, {Lundgren}, {Marchesini}, {Maseda},
  {Momcheva}, {Nelson}, {Pacifici}, \& {Skelton}}]{Fumagalli2014}
{Fumagalli}, M., {Labb{\'e}}, I., {Patel}, S.~G., {et~al.} 2014, \apj, 796, 35

\bibitem[{{Goldader} {et~al.}(1997){Goldader}, {Goldader}, {Joseph}, {Doyon},
  \& {Sanders}}]{Goldader1997}
{Goldader}, J.~D., {Goldader}, D.~L., {Joseph}, R.~D., {et~al.} 1997, \aj, 113,
  1569

\bibitem[{{Haan} {et~al.}(2013){Haan}, {Armus}, {Surace}, {Charmandaris},
  {Evans}, {Diaz-Santos}, {Melbourne}, {Mazzarella}, {Howell}, {Stierwalt},
  {Kim}, {Vavilkin}, {Sanders}, {Petric}, {Murphy}, {Braun}, {Bridge}, \&
  {Inami}}]{Haan2013}
{Haan}, S., {Armus}, L., {Surace}, J.~A., {et~al.} 2013, \mnras, 434, 1264

\bibitem[{{Hainline} {et~al.}(2009){Hainline}, {Shapley}, {Kornei}, {Pettini},
  {Buckley-Geer}, {Allam}, \& {Tucker}}]{Hainline2009}
{Hainline}, K.~N., {Shapley}, A.~E., {Kornei}, K.~A., {et~al.} 2009, \apj, 701,
  52

\bibitem[{{Hern{\'a}n-Caballero} {et~al.}(2009){Hern{\'a}n-Caballero},
  {P{\'e}rez-Fournon}, {Hatziminaoglou}, {Afonso-Luis}, {Rowan-Robinson},
  {Rigopoulou}, {Farrah}, {Lonsdale}, {Babbedge}, {Clements}, {Serjeant},
  {Pozzi}, {Vaccari}, {Montenegro-Montes}, {Valtchanov},
  {Gonz{\'a}lez-Solares}, {Oliver}, {Shupe}, {Gruppioni}, {Vila-Vilar{\'o}},
  {Lari}, \& {La Franca}}]{HC2009}
{Hern{\'a}n-Caballero}, A., {P{\'e}rez-Fournon}, I., {Hatziminaoglou}, E.,
  {et~al.} 2009, \mnras, 395, 1695

\bibitem[{{Ho}(2005)}]{Ho2005}
{Ho}, L.~C. 2005, \apj, 629, 680

\bibitem[{{Ho} \& {Keto}(2007)}]{HK2007}
{Ho}, L.~C. \& {Keto}, E. 2007, \apj, 658, 314

\bibitem[{{Hwang} {et~al.}(2010){Hwang}, {Elbaz}, {Magdis}, {Daddi},
  {Symeonidis}, {Altieri}, {Amblard}, {Andreani}, {Arumugam}, {Auld}, {Aussel},
  {Babbedge}, {Berta}, {Blain}, {Bock}, {Bongiovanni}, {Boselli}, {Buat},
  {Burgarella}, {Castro-Rodr{\'{\i}}guez}, {Cava}, {Cepa}, {Chanial}, {Chapin},
  {Chary}, {Cimatti}, {Clements}, {Conley}, {Conversi}, {Cooray},
  {Dannerbauer}, {Dickinson}, {Dominguez}, {Dowell}, {Dunlop}, {Dwek}, {Eales},
  {Farrah}, {Schreiber}, {Fox}, {Franceschini}, {Gear}, {Genzel}, {Glenn},
  {Griffin}, {Gruppioni}, {Halpern}, {Hatziminaoglou}, {Ibar}, {Isaak},
  {Ivison}, {Jeong}, {Lagache}, {Le Borgne}, {Le Floc'h}, {Lee}, {Lee}, {Lee},
  {Levenson}, {Lu}, {Lutz}, {Madden}, {Maffei}, {Magnelli}, {Mainetti},
  {Maiolino}, {Marchetti}, {Mortier}, {Nguyen}, {Nordon}, {O'Halloran},
  {Okumura}, {Oliver}, {Omont}, {Page}, {Panuzzo}, {Papageorgiou}, {Pearson},
  {P{\'e}rez-Fournon}, {Garc{\'{\i}}a}, {Poglitsch}, {Pohlen}, {Popesso},
  {Pozzi}, {Rawlings}, {Rigopoulou}, {Riguccini}, {Rizzo}, {Rodighiero},
  {Roseboom}, {Rowan-Robinson}, {Saintonge}, {Portal}, {Santini}, {Sauvage},
  {Schulz}, {Scott}, {Seymour}, {Shao}, {Shupe}, {Smith}, {Stevens}, {Sturm},
  {Tacconi}, {Trichas}, {Tugwell}, {Vaccari}, {Valtchanov}, {Vieira},
  {Vigroux}, {Wang}, {Ward}, {Wright}, {Xu}, \& {Zemcov}}]{Hwang2010}
{Hwang}, H.~S., {Elbaz}, D., {Magdis}, G., {et~al.} 2010, \mnras, 409, 75

\bibitem[{{Inami} {et~al.}(2010){Inami}, {Armus}, {Surace}, {Mazzarella},
  {Evans}, {Sanders}, {Howell}, {Petric}, {Vavilkin}, {Iwasawa}, {Haan},
  {Murphy}, {Stierwalt}, {Appleton}, {Barnes}, {Bothun}, {Bridge}, {Chan},
  {Charmandaris}, {Frayer}, {Kewley}, {Kim}, {Lord}, {Madore}, {Marshall},
  {Matsuhara}, {Melbourne}, {Rich}, {Schulz}, {Spoon}, {Sturm}, {U},
  {Veilleux}, \& {Xu}}]{Inami2010}
{Inami}, H., {Armus}, L., {Surace}, J.~A., {et~al.} 2010, \aj, 140, 63

\bibitem[{{Jannuzi} \& {Dey}(1999)}]{Jannuzi1999}
{Jannuzi}, B.~T. \& {Dey}, A. 1999, in Astronomical Society of the Pacific
  Conference Series, Vol. 191, Photometric Redshifts and the Detection of High
  Redshift Galaxies, ed. R.~{Weymann}, L.~{Storrie-Lombardi}, M.~{Sawicki}, \&
  R.~{Brunner}, 111

\bibitem[{{Jannuzi} {et~al.}(2004){Jannuzi}, {Dey}, {Brown}, {Ford}, {Hogan},
  {Miller}, {Ryan}, {Tiede}, {Valdes}, \& {NDWFS Team}}]{Jannuzi2004}
{Jannuzi}, B.~T., {Dey}, A., {Brown}, M.~J.~I., {et~al.} 2004, in Bulletin of
  the American Astronomical Society, Vol.~36, American Astronomical Society
  Meeting Abstracts, 1478

\bibitem[{{Kauffmann} {et~al.}(2003){Kauffmann}, {Heckman}, {Tremonti},
  {Brinchmann}, {Charlot}, {White}, {Ridgway}, {Brinkmann}, {Fukugita}, {Hall},
  {Ivezi{\'c}}, {Richards}, \& {Schneider}}]{Kauffmann2003}
{Kauffmann}, G., {Heckman}, T.~M., {Tremonti}, C., {et~al.} 2003, \mnras, 346,
  1055

\bibitem[{{Kennicutt} \& {Evans}(2012)}]{KE2012}
{Kennicutt}, R.~C. \& {Evans}, N.~J. 2012, \araa, 50, 531

\bibitem[{{Kennicutt}(1998)}]{Kennicutt1998}
{Kennicutt}, Jr., R.~C. 1998, \araa, 36, 189

\bibitem[{{Kennicutt} {et~al.}(2007){Kennicutt}, {Calzetti}, {Walter}, {Helou},
  {Hollenbach}, {Armus}, {Bendo}, {Dale}, {Draine}, {Engelbracht}, {Gordon},
  {Prescott}, {Regan}, {Thornley}, {Bot}, {Brinks}, {de Blok}, {de Mello},
  {Meyer}, {Moustakas}, {Murphy}, {Sheth}, \& {Smith}}]{Kennicutt2007}
{Kennicutt}, Jr., R.~C., {Calzetti}, D., {Walter}, F., {et~al.} 2007, \apj,
  671, 333

\bibitem[{{Kennicutt} {et~al.}(2009){Kennicutt}, {Hao}, {Calzetti},
  {Moustakas}, {Dale}, {Bendo}, {Engelbracht}, {Johnson}, \&
  {Lee}}]{Kennicutt2009}
{Kennicutt}, Jr., R.~C., {Hao}, C.-N., {Calzetti}, D., {et~al.} 2009, \apj,
  703, 1672

\bibitem[{{Kewley} {et~al.}(2001){Kewley}, {Dopita}, {Sutherland}, {Heisler},
  \& {Trevena}}]{Kewley2001}
{Kewley}, L.~J., {Dopita}, M.~A., {Sutherland}, R.~S., {et~al.} 2001, \apj,
  556, 121

\bibitem[{{Kirkpatrick} {et~al.}(2012){Kirkpatrick}, {Pope}, {Alexander},
  {Charmandaris}, {Daddi}, {Dickinson}, {Elbaz}, {Gabor}, {Hwang}, {Ivison},
  {Mullaney}, {Pannella}, {Scott}, {Altieri}, {Aussel}, {Bournaud}, {Buat},
  {Coia}, {Dannerbauer}, {Dasyra}, {Kartaltepe}, {Leiton}, {Lin}, {Magdis},
  {Magnelli}, {Morrison}, {Popesso}, \& {Valtchanov}}]{Kirkpatrick2012}
{Kirkpatrick}, A., {Pope}, A., {Alexander}, D.~M., {et~al.} 2012, \apj, 759,
  139

\bibitem[{{Kochanek} {et~al.}(2012){Kochanek}, {Eisenstein}, {Cool},
  {Caldwell}, {Assef}, {Jannuzi}, {Jones}, {Murray}, {Forman}, {Dey}, {Brown},
  {Eisenhardt}, {Gonzalez}, {Green}, \& {Stern}}]{Kochanek2012}
{Kochanek}, C.~S., {Eisenstein}, D.~J., {Cool}, R.~J., {et~al.} 2012, \apjs,
  200, 8

\bibitem[{{Kroupa} \& {Weidner}(2003)}]{Kroupa2003}
{Kroupa}, P. \& {Weidner}, C. 2003, \apj, 598, 1076

\bibitem[{{Lacy} {et~al.}(2007){Lacy}, {Petric}, {Sajina}, {Canalizo},
  {Storrie-Lombardi}, {Armus}, {Fadda}, \& {Marleau}}]{Lacy2007}
{Lacy}, M., {Petric}, A.~O., {Sajina}, A., {et~al.} 2007, \aj, 133, 186

\bibitem[{{Lacy} {et~al.}(2004){Lacy}, {Storrie-Lombardi}, {Sajina},
  {Appleton}, {Armus}, {Chapman}, {Choi}, {Fadda}, {Fang}, {Frayer},
  {Heinrichsen}, {Helou}, {Im}, {Marleau}, {Masci}, {Shupe}, {Soifer},
  {Surace}, {Teplitz}, {Wilson}, \& {Yan}}]{Lacy2004}
{Lacy}, M., {Storrie-Lombardi}, L.~J., {Sajina}, A., {et~al.} 2004, \apjs, 154,
  166

\bibitem[{{Le Floc'h} {et~al.}(2005){Le Floc'h}, {Papovich}, {Dole}, {Bell},
  {Lagache}, {Rieke}, {Egami}, {P{\'e}rez-Gonz{\'a}lez}, {Alonso-Herrero},
  {Rieke}, {Blaylock}, {Engelbracht}, {Gordon}, {Hines}, {Misselt}, {Morrison},
  \& {Mould}}]{LeFloc'h2005}
{Le Floc'h}, E., {Papovich}, C., {Dole}, H., {et~al.} 2005, \apj, 632, 169

\bibitem[{{Lin} {et~al.}(2009){Lin}, {Buckley-Geer}, {Allam}, {Tucker},
  {Diehl}, {Kubik}, {Kubo}, {Annis}, {Frieman}, {Oguri}, \& {Inada}}]{Lin2009}
{Lin}, H., {Buckley-Geer}, E., {Allam}, S.~S., {et~al.} 2009, \apj, 699, 1242

\bibitem[{{Lutz}(2014)}]{Lutz2014}
{Lutz}, D. 2014, \araa, 52, 373

\bibitem[{{Lutz} {et~al.}(2008){Lutz}, {Sturm}, {Tacconi}, {Valiante},
  {Schweitzer}, {Netzer}, {Maiolino}, {Andreani}, {Shemmer}, \&
  {Veilleux}}]{Lutz2008}
{Lutz}, D., {Sturm}, E., {Tacconi}, L.~J., {et~al.} 2008, \apj, 684, 853

\bibitem[{{Madau} \& {Dickinson}(2014)}]{MD2014}
{Madau}, P. \& {Dickinson}, M. 2014, \araa, 52, 415

\bibitem[{{Magnelli} {et~al.}(2009){Magnelli}, {Elbaz}, {Chary}, {Dickinson},
  {Le Borgne}, {Frayer}, \& {Willmer}}]{Magnelli2009}
{Magnelli}, B., {Elbaz}, D., {Chary}, R.~R., {et~al.} 2009, \aap, 496, 57

\bibitem[{{Marcillac} {et~al.}(2008){Marcillac}, {Rieke}, {Papovich},
  {Willmer}, {Weiner}, {Coil}, {Cooper}, {Gerke}, {Woo}, {Newman},
  {Georgakakis}, {Laird}, {Nandra}, {Fazio}, {Huang}, \& {Koo}}]{Marcillac2008}
{Marcillac}, D., {Rieke}, G.~H., {Papovich}, C., {et~al.} 2008, \apj, 675, 1156

\bibitem[{{Men{\'e}ndez-Delmestre} {et~al.}(2009){Men{\'e}ndez-Delmestre},
  {Blain}, {Smail}, {Alexander}, {Chapman}, {Armus}, {Frayer}, {Ivison}, \&
  {Teplitz}}]{MD2009}
{Men{\'e}ndez-Delmestre}, K., {Blain}, A.~W., {Smail}, I., {et~al.} 2009, \apj,
  699, 667

\bibitem[{{Moustakas} \& {Kennicutt}(2006)}]{MK2006}
{Moustakas}, J. \& {Kennicutt}, Jr., R.~C. 2006, \apjs, 164, 81

\bibitem[{{Moustakas} {et~al.}(2010){Moustakas}, {Kennicutt}, {Tremonti},
  {Dale}, {Smith}, \& {Calzetti}}]{Moustakas2010}
{Moustakas}, J., {Kennicutt}, Jr., R.~C., {Tremonti}, C.~A., {et~al.} 2010,
  \apjs, 190, 233

\bibitem[{{Moustakas} {et~al.}(2011){Moustakas}, {Zaritsky}, {Brown}, {Cool},
  {Dey}, {Eisenstein}, {Gonzalez}, {Jannuzi}, {Jones}, {Kochanek}, {Murray}, \&
  {Wild}}]{Moustakas2011}
{Moustakas}, J., {Zaritsky}, D., {Brown}, M., {et~al.} 2011, ArXiv e-prints

\bibitem[{{Murphy} {et~al.}(2011){Murphy}, {Chary}, {Dickinson}, {Pope},
  {Frayer}, \& {Lin}}]{Murphy2011}
{Murphy}, E.~J., {Chary}, R.-R., {Dickinson}, M., {et~al.} 2011, \apj, 732, 126

\bibitem[{{Muzzin} {et~al.}(2010){Muzzin}, {van Dokkum}, {Kriek}, {Labb{\'e}},
  {Cury}, {Marchesini}, \& {Franx}}]{Muzzin2010}
{Muzzin}, A., {van Dokkum}, P., {Kriek}, M., {et~al.} 2010, \apj, 725, 742

\bibitem[{{O'Dowd} {et~al.}(2009){O'Dowd}, {Schiminovich}, {Johnson}, {Treyer},
  {Martin}, {Wyder}, {Charlot}, {Heckman}, {Martins}, {Seibert}, \& {van der
  Hulst}}]{ODowd2009}
{O'Dowd}, M.~J., {Schiminovich}, D., {Johnson}, B.~D., {et~al.} 2009, \apj,
  705, 885

\bibitem[{{Osterbrock}(1989)}]{Osterbrock1989}
{Osterbrock}, D.~E. 1989, {Astrophysics of gaseous nebulae and active galactic
  nuclei} (University Science Books), 84

\bibitem[{{Papovich} {et~al.}(2009){Papovich}, {Rudnick}, {Rigby}, {Willmer},
  {Smith}, {Finkelstein}, {Egami}, \& {Rieke}}]{Papovich2009}
{Papovich}, C., {Rudnick}, G., {Rigby}, J.~R., {et~al.} 2009, \apj, 704, 1506

\bibitem[{{Pereira-Santaella} {et~al.}(2010){Pereira-Santaella},
  {Diamond-Stanic}, {Alonso-Herrero}, \& {Rieke}}]{PS2010}
{Pereira-Santaella}, M., {Diamond-Stanic}, A.~M., {Alonso-Herrero}, A.,
  {et~al.} 2010, \apj, 725, 2270

\bibitem[{{P{\'e}rez-Gonz{\'a}lez} {et~al.}(2005){P{\'e}rez-Gonz{\'a}lez},
  {Rieke}, {Egami}, {Alonso-Herrero}, {Dole}, {Papovich}, {Blaylock}, {Jones},
  {Rieke}, {Rigby}, {Barmby}, {Fazio}, {Huang}, \& {Martin}}]{PG2005}
{P{\'e}rez-Gonz{\'a}lez}, P.~G., {Rieke}, G.~H., {Egami}, E., {et~al.} 2005,
  \apj, 630, 82

\bibitem[{{Pettini} \& {Pagel}(2004)}]{Pettini2004}
{Pettini}, M. \& {Pagel}, B.~E.~J. 2004, \mnras, 348, L59

\bibitem[{{Pope} {et~al.}(2008){Pope}, {Chary}, {Alexander}, {Armus},
  {Dickinson}, {Elbaz}, {Frayer}, {Scott}, \& {Teplitz}}]{Pope2008}
{Pope}, A., {Chary}, R.-R., {Alexander}, D.~M., {et~al.} 2008, \apj, 675, 1171

\bibitem[{{Pope} {et~al.}(2013){Pope}, {Wagg}, {Frayer}, {Armus}, {Chary},
  {Daddi}, {Desai}, {Dickinson}, {Elbaz}, {Gabor}, \& {Kirkpatrick}}]{Pope2013}
{Pope}, A., {Wagg}, J., {Frayer}, D., {et~al.} 2013, \apj, 772, 92

\bibitem[{{Rieke} {et~al.}(2009){Rieke}, {Alonso-Herrero}, {Weiner},
  {P{\'e}rez-Gonz{\'a}lez}, {Blaylock}, {Donley}, \& {Marcillac}}]{Rieke2009}
{Rieke}, G.~H., {Alonso-Herrero}, A., {Weiner}, B.~J., {et~al.} 2009, \apj,
  692, 556

\bibitem[{{Rigby} {et~al.}(2008){Rigby}, {Marcillac}, {Egami}, {Rieke},
  {Richard}, {Kneib}, {Fadda}, {Willmer}, {Borys}, {van der Werf},
  {P{\'e}rez-Gonz{\'a}lez}, {Knudsen}, \& {Papovich}}]{Rigby2008}
{Rigby}, J.~R., {Marcillac}, D., {Egami}, E., {et~al.} 2008, \apj, 675, 262

\bibitem[{{Rosario} {et~al.}(2013){Rosario}, {Trakhtenbrot}, {Lutz}, {Netzer},
  {Trump}, {Silverman}, {Schramm}, {Lusso}, {Berta}, {Bongiorno}, {Brusa},
  {F{\"o}rster-Schreiber}, {Genzel}, {Lilly}, {Magnelli}, {Mainieri},
  {Maiolino}, {Merloni}, {Mignoli}, {Nordon}, {Popesso}, {Salvato}, {Santini},
  {Tacconi}, \& {Zamorani}}]{Rosario2013}
{Rosario}, D.~J., {Trakhtenbrot}, B., {Lutz}, D., {et~al.} 2013, \aap, 560, A72

\bibitem[{{Rujopakarn} {et~al.}(2011){Rujopakarn}, {Rieke}, {Eisenstein}, \&
  {Juneau}}]{Rujopakarn2011}
{Rujopakarn}, W., {Rieke}, G.~H., {Eisenstein}, D.~J., {et~al.} 2011, \apj,
  726, 93

\bibitem[{{Rujopakarn} {et~al.}(2012){Rujopakarn}, {Rieke}, {Papovich},
  {Weiner}, {Rigby}, {Rex}, {Bian}, {Kuhn}, \& {Thompson}}]{Rujopakarn2012}
{Rujopakarn}, W., {Rieke}, G.~H., {Papovich}, C.~J., {et~al.} 2012, \apj, 755,
  168

\bibitem[{{Rujopakarn} {et~al.}(2013){Rujopakarn}, {Rieke}, {Weiner},
  {P{\'e}rez-Gonz{\'a}lez}, {Rex}, {Walth}, \& {Kartaltepe}}]{Rujopakarn2013}
{Rujopakarn}, W., {Rieke}, G.~H., {Weiner}, B.~J., {et~al.} 2013, \apj, 767, 73

\bibitem[{{Sargsyan} \& {Weedman}(2009)}]{SW2009}
{Sargsyan}, L.~A. \& {Weedman}, D.~W. 2009, \apj, 701, 1398

\bibitem[{{Seitz} {et~al.}(1998){Seitz}, {Saglia}, {Bender}, {Hopp}, {Belloni},
  \& {Ziegler}}]{Seitz1998}
{Seitz}, S., {Saglia}, R.~P., {Bender}, R., {et~al.} 1998, \mnras, 298, 945

\bibitem[{{Shi} {et~al.}(2009){Shi}, {Rieke}, {Ogle}, {Jiang}, \&
  {Diamond-Stanic}}]{Shi2009}
{Shi}, Y., {Rieke}, G.~H., {Ogle}, P., {et~al.} 2009, \apj, 703, 1107

\bibitem[{{Shipley} {et~al.}(2013){Shipley}, {Papovich}, {Rieke}, {Dey},
  {Jannuzi}, {Moustakas}, \& {Weiner}}]{Shipley2013}
{Shipley}, H.~V., {Papovich}, C., {Rieke}, G.~H., {et~al.} 2013, \apj, 769, 75

\bibitem[{{Smail} {et~al.}(2005){Smail}, {Smith}, \& {Ivison}}]{Smail2005}
{Smail}, I., {Smith}, G.~P., \& {Ivison}, R.~J. 2005, \apj, 631, 121

\bibitem[{{Smith} {et~al.}(2007){Smith}, {Draine}, {Dale}, {Moustakas},
  {Kennicutt}, {Helou}, {Armus}, {Roussel}, {Sheth}, {Bendo}, {Buckalew},
  {Calzetti}, {Engelbracht}, {Gordon}, {Hollenbach}, {Li}, {Malhotra},
  {Murphy}, \& {Walter}}]{Smith2007}
{Smith}, J.~D.~T., {Draine}, B.~T., {Dale}, D.~A., {et~al.} 2007, \apj, 656,
  770

\bibitem[{{Stern} {et~al.}(2005){Stern}, {Eisenhardt}, {Gorjian}, {Kochanek},
  {Caldwell}, {Eisenstein}, {Brodwin}, {Brown}, {Cool}, {Dey}, {Green},
  {Jannuzi}, {Murray}, {Pahre}, \& {Willner}}]{Stern2005}
{Stern}, D., {Eisenhardt}, P., {Gorjian}, V., {et~al.} 2005, \apj, 631, 163

\bibitem[{{Teplitz} {et~al.}(2011){Teplitz}, {Chary}, {Elbaz}, {Dickinson},
  {Bridge}, {Colbert}, {Le Floc'h}, {Frayer}, {Howell}, {Koo}, {Papovich},
  {Phillips}, {Scarlata}, {Siana}, {Spinrad}, \& {Stern}}]{Teplitz2011}
{Teplitz}, H.~I., {Chary}, R., {Elbaz}, D., {et~al.} 2011, \aj, 141, 1

\bibitem[{{Teplitz} {et~al.}(2000){Teplitz}, {McLean}, {Becklin}, {Figer},
  {Gilbert}, {Graham}, {Larkin}, {Levenson}, \& {Wilcox}}]{Teplitz2000}
{Teplitz}, H.~I., {McLean}, I.~S., {Becklin}, E.~E., {et~al.} 2000, \apjl, 533,
  L65

\bibitem[{{Tremonti} {et~al.}(2004){Tremonti}, {Heckman}, {Kauffmann},
  {Brinchmann}, {Charlot}, {White}, {Seibert}, {Peng}, {Schlegel}, {Uomoto},
  {Fukugita}, \& {Brinkmann}}]{Tremonti2004}
{Tremonti}, C.~A., {Heckman}, T.~M., {Kauffmann}, G., {et~al.} 2004, \apj, 613,
  898

\bibitem[{{Treyer} {et~al.}(2010){Treyer}, {Schiminovich}, {Johnson}, {O'Dowd},
  {Martin}, {Wyder}, {Charlot}, {Heckman}, {Martins}, {Seibert}, \& {van der
  Hulst}}]{Treyer2010}
{Treyer}, M., {Schiminovich}, D., {Johnson}, B.~D., {et~al.} 2010, \apj, 719,
  1191

\bibitem[{{Wu} {et~al.}(2010){Wu}, {Helou}, {Armus}, {Cormier}, {Shi}, {Dale},
  {Dasyra}, {Smith}, {Papovich}, {Draine}, {Rahman}, {Stierwalt}, {Fadda},
  {Lagache}, \& {Wright}}]{Wu2010}
{Wu}, Y., {Helou}, G., {Armus}, L., {et~al.} 2010, \apj, 723, 895

\bibitem[{{Wuyts} {et~al.}(2012){Wuyts}, {Rigby}, {Gladders}, {Gilbank},
  {Sharon}, {Gralla}, \& {Bayliss}}]{Wuyts2012}
{Wuyts}, E., {Rigby}, J.~R., {Gladders}, M.~D., {et~al.} 2012, \apj, 745, 86

\bibitem[{{Xu} {et~al.}(2015){Xu}, {Rieke}, {Egami}, {Pereira}, {Haines}, \&
  {Smith}}]{Xu2015}
{Xu}, L., {Rieke}, G.~H., {Egami}, E., {et~al.} 2015, \apjs, 219, 18

\bibitem[{{Yuan} {et~al.}(2011){Yuan}, {Takeuchi}, {Buat}, {Heinis},
  {Giovannoli}, {Murata}, {Iglesias-P{\'a}ramo}, \& {Burgarella}}]{Yuan2011}
{Yuan}, F.-T., {Takeuchi}, T.~T., {Buat}, V., {et~al.} 2011, \pasj, 63, 1207

\end{thebibliography}
\end{document}